\documentclass[11pt]{article}
\RequirePackage[OT1]{fontenc}
\RequirePackage{amsmath}
\RequirePackage[numbers]{natbib}
\RequirePackage[colorlinks,citecolor=blue,urlcolor=blue]{hyperref}
\usepackage{graphicx}
\usepackage{verbatim}
\usepackage{multirow}
\usepackage{color}
\newcommand{\qed}{{\unskip\nobreak\hfil\penalty50\hskip2em\vadjust{}
            \nobreak\hfil$\Box$\parfillskip=0pt\finalhyphendemerits=0\par}}

\newtheorem{thm}{Theorem}[section] %(If you want theorem number
\newtheorem{lemma}{Lemma}[section] %%    with section number.

\newtheorem{definition}{Definition}[section]

\newcommand{\bed}{\begin{definition}}
\newcommand{\eed}{\end{definition}}
\newcommand{\beq}{\begin{equation}}
\newcommand{\eeq}{\end{equation}}

\newcommand{\eps}{\epsilon}

\newcommand{\bitem}{\begin{itemize}}
\newcommand{\eitem}{\end{itemize}}

\newcommand{\goto}{\rightarrow}

\newcommand{\margmax}{\mathrm{argmax}}

\newcommand{\beqn}{\begin{equation}}
\newcommand{\eeqn}{\end{equation}}
\newcommand{\balign}{\begin{align}}
\newcommand{\ealign}{\end{align}}

\newcommand{\hamm}{\mathrm{Hamm}}
\newcommand{\sgn}{\mathrm{sgn}}

\newcommand{\call}{{\cal I}}

\newcommand{\callJ}{{\cal J}}

\newcommand{\diag}{\mathrm{diag}}

\newcommand{\heta}{\hat{\eta}}

\newcommand{\hsig}{\hat{\Sigma}}

\newcommand{\hs}{\hat{S}}
\newcommand{\hrho}{\hat{\rho}}
\newcommand{\hg}{\hat{\gamma}}
\newcommand{\hell}{\hat{\ell}}

\usepackage{amssymb}

\usepackage{xr}
\begin{document}

\title{Partial Correlation Screening for Estimating Large Precision Matrices, with Applications to Classification}
\author{Shiqiong Huang$^*$, Jiashun Jin$^*$  and Zhigang Yao$^\dag$
}
\date{Carnegie Mellon University$^*$ and National University of Singapore$^\dag$}
\maketitle

\begin{abstract}
Given $n$ samples   $X_1, X_2, \ldots, X_n$  from $N(0, \Sigma)$,  we are interested in  estimating
the $p \times p$ precision matrix $\Omega = \Sigma^{-1}$; we assume
$\Omega$ is sparse in that each row has relatively few nonzeros.

We propose {\it Partial Correlation Screening (PCS)} as a new row-by-row approach.  To estimate the $i$-th row of $\Omega$, $1 \leq i \leq p$, PCS uses a {\it Screen} step and a {\it Clean} step. In the Screen step,
PCS recruits a (small) subset of indices using a stage-wise algorithm,  where in each stage,
the algorithm updates the set of recruited indices by
adding the index $j$ that has the largest empirical partial correlation (in magnitude) with $i$, given the
set of indices recruited so far.
In the Clean step, PCS first re-investigates all recruited indices in hopes of removing false positives, and then uses the resultant set of indices to reconstruct the
$i$-th row of $\Omega$.

PCS is computationally efficient and modest in memory use:  to estimate a  row of $\Omega$, it  only needs a few rows (determined sequentially) of the empirical covariance matrix.
This  enables PCS to execute the estimation of a large precision matrix (e.g., $p=10K$) in
a few minutes, and  open doors to estimating much larger  precision matrices.

We use PCS for classification.  Higher Criticism Thresholding (HCT) is a recent classifier that enjoys optimality,
but to exploit its full potential in practice,  one needs a good estimate of the precision matrix $\Omega$. Combining HCT with any approach to estimating $\Omega$ gives a new classifier: examples include HCT-PCS and HCT-glasso.

We have applied HCT-PCS to two large microarray data sets ($p = 8K$ and $10K$) for classification, where it not only significantly outperforms HCT-glasso, but also is competitive to the Support Vector Machine (SVM) and Random Forest (RF) (for one of the data set, $17.4\%$ improvement over  SVM and $57.8\%$ over  RF). The results suggest
that PCS gives more useful estimates of $\Omega$ than the glasso; we study this carefully and have gained
some interesting insight.

We set up a general theoretical framework and  show that in a broad context, PCS fully recovers the support of $\Omega$ and  HCT-PCS
yields optimal classification behavior.  Our proofs shed interesting light on the behavior of
stage-wise procedures.
\end{abstract}

\vspace{0.15in}\noindent

\vspace{0.15in}\noindent

\vspace{0.15 in}\noindent
{\bf Acknowledgments:} The authors thank Mohammadmahdi R. Yousefi for generosity in sharing 
his data sets. JJ thanks Cun-Hui Zhang and Hui Zou for valuable pointers and discussions. SH and JJ are supported in part by NSF Grant DMS-1208315.

%%%%%%%%%%%%%%
%%%%%%%%%%%%%%
%%%%%%%%%%%%%%
%%%%%%%%%%%%%%

\section{Introduction}
\label{sec:Intro}
\setcounter{equation}{0}

There is always the story of ``four blind men  and the elephant" \cite{Elephant}.
A group of blind men  were asked to  touch an elephant to learn what it is like. Each one touched a different part, but only one part (e.g.,  the tusk, the ear, or the leg).  They then compared notes and learnt that they were in complete disagreement, until the King pointed out to them: ``All of you are right. The reason that every one of you is telling it differently is because each one of you touched the different part of the elephant. So actually the elephant has all the features you mentioned".

There are several  similarities between the elephant tale and the
 problem on estimating large precision matrices;   some  are
obvious, but some are not.
\begin{itemize}
\item Both deal with something enormous: an elephant  or a large matrix.
\item Both encourage {\it parallel computing}: either with a group of blind men or a cluster of computers. Individuals only communicate with a `center' (a king, a master computer), but do not communicate with each other.
\item Both are  {\it modest in memory use}.  If we are only interested in a small part of the elephant (e.g., the tail), we do not need to scan the whole elephant. If we are only interested in a row  of a sparse precision matrix, we don't need to use the {\it whole} empirical covariance matrix.
\end{itemize}
Modesty in memory use is especially important when we only have a modest computing platform (e.g., Matlab on a desktop), where it is easy to hit the RAM limit or memory ceiling.

Given a data matrix $X \in R^{n, p}$. We write
\[
X = [x_1, x_2, \ldots, x_p] = [X_1, X_2, \ldots, X_n]',
\]
where $X_i'$ is the $i$-th row and $x_j$ is the $j$-th column, $1\leq i\leq n, 1\leq j\leq p$.
We assume the rows satisfy
\begin{equation} \label{model1}
X_i  \overset{\text{\emph{iid}}}{\sim} N(0,   \Sigma),    \qquad \Sigma \in R^{p,p}.
\end{equation}
Denote by $\hat{\Sigma}$ by the empirical covariance matrix
\begin{equation} \label{DefineSigmahat}
\hat{\Sigma}(i,j)   = (x_i, x_j)/n.
\end{equation}
The precision matrix
\begin{equation} \label{DefineOmega}
\Omega = \Sigma^{-1},
\end{equation}
is unknown to us but is presumably sparse, in the sense that each row of $\Omega$ has relatively few nonzeros, and the primary interest is to estimate $\Omega$.

Our primary interest is in the `large $n$, really large $p$'  regime \cite{Science}, where  it is challenging to estimate $\Omega$
precisely with real-time computing.

The glasso \cite{glasso} is a well-known approach which estimates $\Omega$ by optimizing the $\ell^1$-penalized objective function of the log-likelihood associated with $\hat{\Sigma}$.  The glasso is not exactly modest in memory use,  and  for large $p$ (e.g.,  $p = 10K$),    the glasso can be unsatisfactorily slow, especially when the tuning parameter is small \cite{glasso}.  Also,  by its design, it is  unclear how to implement the glasso with parallel computing.    This makes the glasso disadvantageous when $p$ is large and resources for parallel computing are available.

Alternatively, we can estimate $\Omega$ row by row.  Such approaches include  but are not limited to Nearest Neighborhood (NN) \cite{Buhlmann}, scaled-lasso (slasso) \cite{slasso}, and  CLIME \cite{CLIME}.
These methods relate the problem of estimating an individual row of  $\Omega$ to  a linear regression model and  apply some variable selection approaches:  NN, slasso and CLIME
apply the lasso, scaled-lasso, and Dantzig Selector correspondingly.
Unfortunately, for $p = 10K$ or larger,  these methods are unsatisfactorily slow,  simply because the lasso, scaled-lasso, and Dantzig Selector are not fast enough to accomplish $10K$ different variable selections in real time.
They are not exactly modest in memory use either: to estimate a row of $\Omega$, they need either the whole matrix of $\hat{\Sigma}$ or  $X$.

We propose {\it Partial Correlation Screening (PCS)} as a new approach to estimating the precision matrix. PCS has the following appealing features.
\begin{itemize}
\item {\it Allowing  for  real-time computing}.  PCS estimates $\Omega$ row by row using a fast screening algorithm, and is able to estimate $\Omega$ for $p = 10K$ or larger with real-time computation  on a modest computing platform.
\item {\it Modesty in memory use}.   To estimate each row of $\Omega$,  PCS does not need the whole matrix of $\hat{\Sigma}$. It only needs the diagonals of $\hat{\Sigma}$ and a few rows of $\hat{\Sigma}$ determined sequentially, provided that $\Omega$ is sufficiently sparse.  This enables us to bypass the RAM limit (of Matlab on a desktop, say)  and to accommodate $\Omega$ with much larger $p$. \end{itemize}

However, we must note that,  practically,  estimating $\Omega$  is {\it rarely} the ultimate goal.   In many applications, the goal is usually to use the estimated $\Omega$  to improve statistical inference, such as classification, inference on the genetic networks, large-scale multiple testing, and so on and so forth.

In this paper, largely motivated by interests in gene microarray,  we  focus  on how to use the estimated precision matrix to improve classification results with microarray data. Table~\ref{table:data} displays two microarray data sets we study in this paper. In each data set, we have samples from two classes (e.g.,   normal versus  diseased), and
each sample is measured over the same set of genes.
The main interest is to use the data set to construct a trained classifier.
%%%%%%%%%%%%%
%%%%%%%%%%%%%
%%%%%%%%%%%%%
\vspace{-.1 in} 
\begin{table}[htb!] 
\caption{Two gene expression microarray data sets.}
\vspace{0.05 in} 
\centering
\scalebox{1}{ 
\begin{tabular}{|l|l||c|c|}
\hline
Data Name & Source  &    $n$ ($\#$ of subjects)  & $p$ ($\#$ of genes)   \\
\hline
Rats & Yousefi et al. (2010)    &  181 &  8491  \\
Liver & Yousefi et al. (2010)   & 157  &10237  \\
\hline
\end{tabular}
\label{table:data}
}
\end{table}

%%%%%%%%%%%%%
%%%%%%%%%%%%%
%%%%%%%%%%%%%
We propose to combine PCS with the recent classifier of Higher Criticism Thresholding (HCT) \cite{DJ08, FJY}, and to build a new classifier HCT-PCS.  In \cite{DJ08,FJY}, they investigated
a two-class classification setting with a Gaussian graphical model.
Assuming samples from two classes share the same sparse precision matrix $\Omega$,  they showed that,  given a reasonably good estimate of $\Omega$,  HCT enjoys optimal classification  behaviors. The challenge, however,  is  to find an algorithm that estimates the precision matrix accurately with real-time computation; this is where PCS comes in.

We apply HCT-PCS to the two data sets above. In these data sets,
the precision matrix is unknown, so it is hard to check whether PCS  is more accurate for estimating $\Omega$ than existing procedures.  However, the class labels are given, which can be used as the  `ground truth' to evaluate the performance of different classifiers.
We find that
\begin{itemize}
\item HCT-PCS significantly outperforms other versions of HCT (say,  HCT-glasso, where $\Omega$
is estimated by the glasso), suggesting that PCS yields more accurate estimates of $\Omega$ than  other approaches (the glasso, say).
\item HCT-PCS is competitive, in both computation time (especially when $n$ is large) and classification errors, to the more popular classifiers of Support Vector Machine (SVM)  \cite{SVM} and Random Forest (RF) \cite{RF}.
\end{itemize}

%%%%%%%%%%%%%%
%%%%%%%%%%%%%%
%%%%%%%%%%%%%%
%%%%%%%%%%%%%%
\subsection{PCS: the idea}
\label{subsec:idea}
We present the key idea of PCS, leaving the formal introduction to Section \ref{subsec:PCS}.  To this end,  we consider an {\it idealized} case  where we are allowed  to access
all `small-size' principal sub-matrices of $\Sigma$ (but not any `large-size' sub-matrices), and study how to use such sub-matrices to reconstruct $\Omega$.
Since any small-size principal sub-matrix of $\Sigma$ can be well-approximated by
the corresponding sub-matrix of $\hat{\Sigma}$ (despite that $\hat{\Sigma}$ as a whole is a bad approximation to $\Sigma$ due to $p \gg n$), once we understand such an idealized case, we know how to deal with the real one.

\begin{figure}[tb!]
    \centering
    \includegraphics[width= 5.7 in]{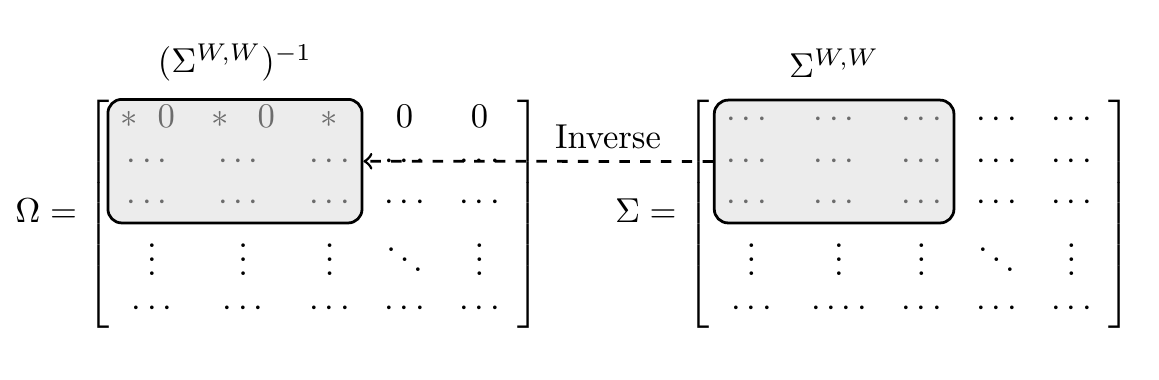}
    \caption{The first row of $\Omega$ only has nonzeros
at column $1, 3, 5$, marked with ``$*"$.  For any subset $W$ such that $\{1, 3, 5\} \subset W$,  the first rows of $\Omega^{W, W}$ and $(\Sigma^{W, W})^{-1}$ are the same.}
    \label{fig:localinv}
\end{figure}

Write $\Omega = (\omega_1, \omega_2, \ldots, \omega_p)$ so that  $\omega_i'$  is the $i$-th row of $\Omega$.
Fixing $1 \leq i \leq p$,  we wish to understand what could be a reasonable approach to reconstructing  $\omega_i'$   using only `small-size' sub-matrices of $\Sigma$.
Define
%%%%%%%%%%%%
%%%%%%%%%%%%
%%%%%%%%%%%%
%%%%%%%%%%%%
\begin{equation} \label{DefineSi}
S^{(i)}(\Omega) =  \{1 \leq j \leq p: \;  \omega_i(j) \neq 0, j\neq i\}.
\end{equation}
Note that $\{i\}\cup S^{(i)}(\Omega)$, not $S^{(i)}(\Omega)$,  is the support of $\omega_i$. Such a notation is  a little bit unconventional; we choose it for simplicity in presentation.
%%%%%%%%%%%
%%%%%%%%%%%
%%%%%%%%%%%
\bed \label{def:submatrix}
For any matrix $A \in R^{n,p}$ and subsets
$\call = \{i_1, i_2,  \ldots, i_{M}\} \subset \{1,   \ldots, n\}$ and
${\cal J}   = \{j_1,   \ldots, j_K\}   \subset \{1,   \ldots, p\}$,
$A^{\call, {\cal J}}$ denotes the $M \times K$ sub-matrix such that  $A^{\call, {\cal J}}(m,k) = A(i_m, j_k)$, $1 \leq m \leq M, 1 \leq k \leq K$ (indices in   either $\call$ or ${\cal J}$ are not necessarily arranged in the ascending order).
\eed
Here is an interesting observation. For any subset $W$  such that
\begin{equation} \label{DefineW}
(\{i\}\cup S^{(i)}(\Omega)) \subset  W \subset \{1, 2, \ldots, p\},
\end{equation}
we can reconstruct  $\omega_i'$  by only knowing a specific row of  $(\Sigma^{W, W})^{-1}$!
%%%%%%%%%%%%
%%%%%%%%%%%%
%%%%%%%%%%%%
%%%%%%%%%%%%
\begin{lemma}   \label{lemma:firstcolumn}
Suppose (\ref{DefineW}) holds, and index $i$ is the $k$-th index in $W$.
The $k$-th row of $\Omega^{W, W}$ coincides with that of $(\Sigma^{W, W})^{-1}$, despite that two matrices are generally unequal.
\end{lemma}
%%%%%%%%%
%%%%%%%%%
%%%%%%%%%
%%%%%%%%%
The proof of Lemam \ref{lemma:firstcolumn}  is elementary so we omit it;  see also Figure \ref{fig:localinv}.
%%%%%%%%%%%%%
%%%%%%%%%%%%%
Lemma \ref{lemma:firstcolumn} motivates a two-step {\it  Screen and Clean} approach (an idea for variable selection that is  applicable in many cases \cite{FanLv, JZZ, KJ, Runze, Wasserman}).
\begin{itemize}
\item In the Screen stage, we identify a subset $S_*^{(i)} = S_*^{(i)}(\Sigma, p)$, in hopes of $S^{(i)}(\Omega) \subset S_*^{(i)}$.
\item In the Clean stage, we reconstruct  $\omega_i'$ from the matrix $\Sigma^{W_*, W_*}$ following the idea in Lemma \ref{lemma:firstcolumn},  where  $W_*=\{i\}\cup S_*^{(i)}$.
\end{itemize}

Seemingly,  the key is how to screen.  Our proposal is to use the {\it partial correlation}, a concept
 closely related to the precision matrix   \cite{BuhlmannBook}.  Consider an (ordered) subset $W \subset \{1, 2, \ldots, p\}$ where $i$ and $j$ are the first and the last indices, respectively.  Let $S = W \setminus \{i, j\}$.
For any vector $Z \sim N(0, \Sigma)$,
the partial correlation between $Z(i)$ and $Z(j)$ given $\{Z(k): k \in S\}$  is defined as
\begin{equation} \label{Definerho}
\rho_{ij}(S) = \frac{ -1  \cdot \mbox{first row last column of $\bigl( \Sigma^{W,W}  \bigr)^{-1}$}}{\bigl[\mbox{product of the first and last diagonals of $\bigl(\Sigma^{W,W}   \bigr)^{-1} $} \bigr]^{1/2}}.
\end{equation}
Note that  $\rho_{ij}(S) = 0$ if and only if the numerator is $0$.
By Lemma \ref{lemma:firstcolumn} above and Lemma~\ref{lem:lowerbound} to be introduced below, we have the following observation:
\[
\mbox{$S^{(i)}(\Omega) \subset  (\{i\} \cup S$)}   \qquad \Longleftrightarrow \qquad \mbox{$\rho_{ij}(S) = 0$ for all $j \notin (\{i\} \cup S)$}.
\]
This observation motivates a stage-wise screening algorithm for choosing $S_*^{(i)}$, where we use the partial correlation to recruit {\it exactly one node} in each step before the algorithm terminates.  Initialize with $S_0^{(i)} = \emptyset$.
\vspace{.075 in}

{\it
Suppose the algorithm has run $(k-1)$ steps and has not yet stopped. Let $S_{k-1}^{(i)} = \{j_1, j_2, \ldots, j_{k-1}\}$ be all the nodes recruited (in that order) by far.  In the $k$-th step, if $\rho_{ij}(S_{k-1}^{(i)}) \neq 0$ for some $j \notin (\{i\} \cup S_k^{(i)})$,  let $j = j_k$ be the index with the largest value of $|\rho_{ij}(S_{k-1}^{(i)})|$, and update with $S_k^{(i)} = S_{k-1}^{(i)} \cup \{j_k\}$. Otherwise, terminates and let  $S_*^{(i)} = S_{k-1}^{(i)}$.
}
\vspace{.075 in}

It is shown in Theorem \ref{thm:idealPCS} that under mild conditions, the algorithm terminates at $\leq C |S^{(i)}(\Omega)|^2$ steps, at which point,   $S^{(i)}(\Omega) \subset S_*^{(i)}$ and $\rho_{ij}(S_*^{(i)}) = 0$ for all $j \notin (\{i\} \cup S_*^{(i)})$.  Letting $W_*=\{i\}\cup S_*^{(i)}$, we can then use  $\Sigma^{W_*, W_*}$ to reconstruct $\omega_i'$, following the connection given in Lemma \ref{lemma:firstcolumn}.

Since all small-size  sub-matrices of $\Sigma$ can be well-approximated by their empirical counterparts in $\hat{\Sigma}$,  the ideas above are readily extendable to the `real case', provided that $|S^{(i)}(\Omega)|$ is sufficiently small. This idea is fleshed out in Section \ref{subsec:PCS}, where PCS is formally introduced.
%%%%%%%%%%%%
%%%%%%%%%%%%
%%%%%%%%%%%%
%%%%%%%%%%%%
\subsection{PCS: the procedure}
\label{subsec:PCS}
From time to time,  especially for analyzing microarray data,  it is desirable
to use the ridge regularization when we invert a principal sub-matrix of $\hat{\Sigma}$ on an {\it as-needed basis}, even when the size of the matrix is small.
Fixing a ridge regularization parameter $\delta > 0$, for any positive definite matrix $W$, define the {\it Ridge Regularized Inverse} by
\begin{equation} \label{RIasneed}
{\cal I}_{\delta}(W) = \left\{
\begin{array}{ll}
W^{-1}, &\qquad \mbox{if all eigenvalues of $W$} \geq  \delta,  \\
(W + \delta I_{|W|})^{-1}, &\qquad \mbox{otherwise},
\end{array}
\right.
\end{equation}
where $I_k$ denotes the $k \times k$ identity matrix (we may drop ``$k$" for simplicity).

For any indices $i,j$ and subset $S \subset \{1,2, \ldots, p\} \setminus \{i,j\}$, let $W = \{i\} \cup S\cup \{j\}$ and suppose $i$ and $j$ are the first and last indices in the subset.  Introduce the {\it regularized empirical partial correlation}  by
%%%%%%%%%%%
%%%%%%%%%%%
%%%%%%%%%%%
\begin{equation} \label{Definehatrho}
\hat{\rho}_{ij}^{(\delta)}(S) =
\frac{ -1  \cdot \mbox{first row last column of ${\cal I}_{\delta}(\hsig^{W,W})$}}{\bigl[\mbox{product of the first and last diagonals of ${\cal I}_{\delta}(\hsig^{W,W})$} \bigr]^{1/2}}.
\end{equation}
Note that if we take $\delta =0$ and replace $\hsig$ by $\Sigma$ everywhere, then $\hrho_{ij}^{(\delta)}(S)$ reduces to the partial correlation $\rho_{ij}(S)$  defined in (\ref{Definerho}).

PCS is specifically designed for very large precision matrices, and we may need to deposit $\hat{\Sigma}$ in the  `data center' instead of the software to bypass the memory ceiling.
A `data center' can be many things:  a hard disk of a laptop, a master machine of a computer cluster, or a large data depository. For example,  suppose we wish to run PCS using Matlab on a  laptop.  For moderately large $p$, we can always load the whole $\hat{\Sigma}$ to Matlab directly. However, for much larger $p$, this becomes impossible, and  depositing $\hat{\Sigma}$ in a `data center' becomes necessary (which then poses great challenges for many procedures, say, glasso).  Fortunately, PCS is able to overcome such a challenge:   to estimate a row of $\Omega$, PCS only needs to load a few rows of the empirical covariance matrix  from the `data center'. See details below.

%%%%%%%%%
%%%%%%%%%
%%%%%%%%%
%%%%%%%%%
\begin{figure}[htb!]
    \centering
    \includegraphics[width = 6 in, height = 2.8 in]{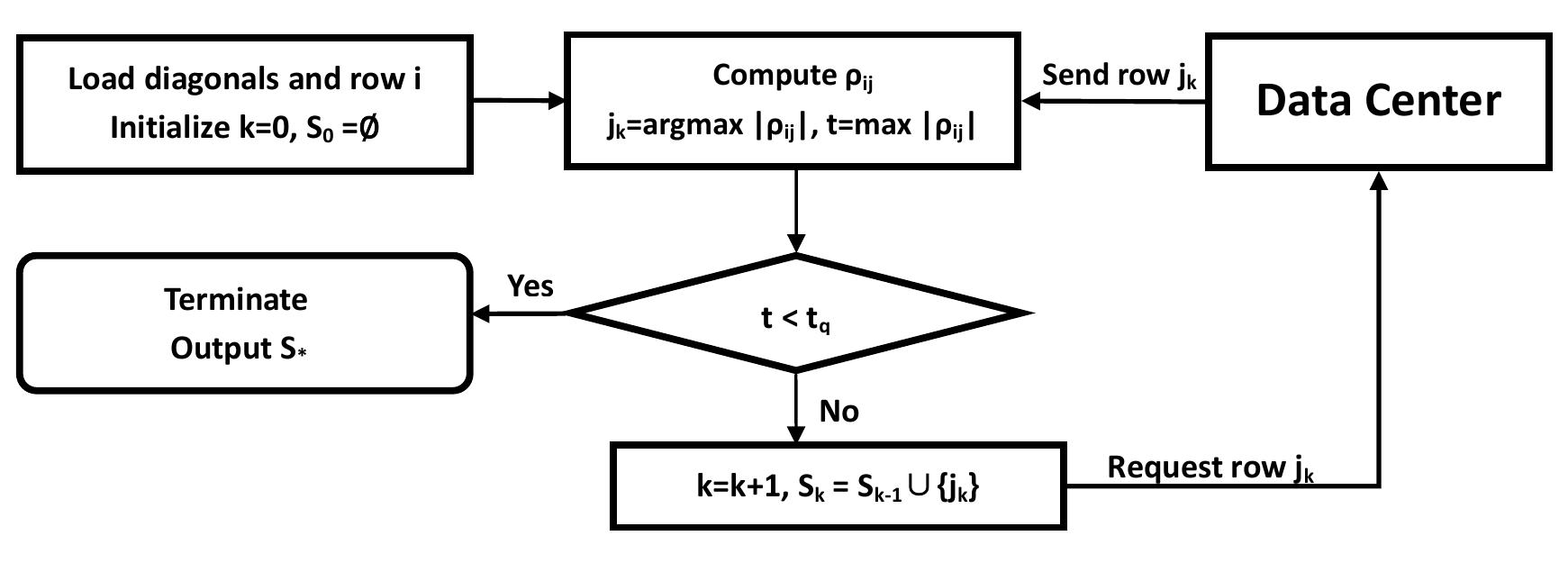}
\caption{Flow chart of PCS (short-hand notations are used for simplicity)}
    \label{fig:flowchart}
\end{figure}

PCS estimates $\Omega$ row by row.  Fixing a tuning parameter  $q > 0$, and set a threshold in the form of
\begin{equation} \label{Definetq}
t_q^* = t^*_q(p,n) =   q \sqrt{2  \log(p)/n}.
\end{equation}
For a small number $\delta > 0$ and an appropriately large integer $L > 0$,
 to estimate the $i$-th row of $\Omega$, $1 \leq i \leq p$,  PCS consists of $4$ steps; see Figure \ref{fig:flowchart}.
 % (we may drop the superscript ``$(i)$" for simplicity).
%%%%%%%%
%%%%%%%%
%%%%%%%%
%%%%%%%%%%%%%
%%%%%%%%%%%%%
%%%%%%%%%%%%%
%%%%%%%%%%%%%
\begin{itemize}
\item {\it Initial step}.  Let $\hat{S}_0^{(i)} =\emptyset$, and load the $i$-th row and the diagonals of  $\hat{\Sigma}$ (from the data center to the software; same below).
\item {\it Screen step}. Suppose the algorithm has not yet terminated at the end of step $(k-1)$, and let   $\hat{S}_{k-1}^{(i)}  = \{j_1, j_2, \ldots, j_{k-1}\}$ be all the nodes recruited so far (in that order).
If
\begin{equation} \label{terminate}
k <  L,  \qquad \mbox{and}  \qquad  \max_{\{j  \notin   (\{i\} \cup  \hat{S}_k^{(i)}) \} }
  |\hrho_{ij}^{(\delta)}(\hat{S}_{k-1}^{(i)})|  \geq   t_q^*,
\end{equation}
let $j = j_k$ be the node satisfying
$j_k  = \margmax_{\{j:  j \notin (\{i\} \cup  \hat{S}_{k-1}^{(i)})\}}  |\hat{\rho}_{ij}^{(\delta)}(\hat{S}_{k-1}^{(i)})|$ (when there are  ties, pick the smallest index).
We load the $j_k$-th row of $\hat{\Sigma}$ to the software and update $\hat{S}_k^{(i)}$ by
$\hat{S}_k^{(i)} =  \hat{S}_{k-1}^{(i)} \cup \{j_k\} = \{j_1, j_2, \ldots, j_k\}$.
Otherwise, the algorithm terminates, and we set $\hat{S}_*^{(i)}  = \hat{S}_*^{(i)}(t, X; p, n)$ as $\hs_{k-1}^{(i)}$, where the indices are arranged in the order they are recruited.
\item {\it Clean step}.  Denote by $\heta'$ the first row of  ${\cal I}_{\delta}(\hat{\Sigma}^{\hat{W}_*, \hat{W}_*})$,  where $\hat{W}_* = \{i\} \cup \hat{S}_*^{(i)}$ for short. Write  $\hat{W}_* = \{i, j_1, j_2, \ldots, j_k\}$ (nodes arranged in that order).
Denote the set of selected nodes after cleaning by
$\hat{S}_{**}^{(i)}  = \hat{S}_{**}^{(i)}(t, X; p, n) = \{j_{\ell}: |\heta(\ell+1)| \geq t_q^*, 1\leq \ell \leq k\}$.
Letting $\hat{W}_{**} = \{i\} \cup \hat{S}_{**}^{(i)}$ (where $i$ is the first node) and writing $A = {\cal I}_{\delta}(\hat{\Sigma}^{\hat{W}_{**}, \hat{W}_{**}})$ for short, we estimate the $i$-th row of $\Omega$ by
\[
\hat{\Omega}^*(i, j) =
\left\{
\begin{array}{ll}
\mbox{first row $\ell$-th column of $A$},  &  \mbox{$j$ is the $\ell$-th node in $\hat{W}_{**}$},  \\
0, & \mbox{$j \notin \hat{W}_{**}$}.
\end{array}
\right.
\]
\item {\it Symmetrization}. $\hat{\Omega}^{pcs}  = [\hat{\Omega}^* + (\hat{\Omega}^*)']/2$.
\end{itemize}
PCS has three tuning parameters $(q, \delta, L)$, but its performance is not sensitive to different choices of  $(\delta, L)$, as long as they are in a reasonable range.  In this paper, we set $(\delta, L) = (.1, 30)$, so essentially PCS only has one tuning parameter $q$.
In practice, how to set  $q$ is generally a difficult problem. Our primary focus on
real data analysis is classification, in which settings we select $q$ by cross validations.  See Section \ref{subsec:realdata} for details.

The computation  cost of PCS is $O(p^2 L^3+np^2)$, where $O(np^2)$ is the cost of obtaining   $\hat{\Sigma}$ from the data matrix $X$, and the $L^3$ term comes from the step of sequentially inverting matrices of sizes $2, 3, \ldots, L+1$.
Also, PCS estimates $\Omega$ row by row and allows for parallel computing. Together, these make PCS a fast algorithm that can have real time computing for
large precision matrices. For example, with $(q, \delta, L) = (.2, .1, 30)$, it takes the PCS only about $5$ and $7.5$ minutes on the rats and the  liver data sets, respectively.

PCS is also modest in memory use: to estimate one row of $\Omega$, PCS  only needs the diagonals and no more than $L$ rows of $\hsig$. This enables PCS to bypass the memory ceiling for very large $p$. Of course,  in such cases, some
communication costs between the software and `data center' are expected, but these seem unavoidable when we hit the memory ceiling. How to design an efficient `communication scheme' is an interesting problem. For reasons of space, we leave this to the future work.

%%%%%%%%%%%%%
%%%%%%%%%%%%%
%%%%%%%%%%%%%
%%%%%%%%%%%%%
%%%%%%%%%%%%%
%%%%%%%%%%
%%%%%%%%%%
%%%%%%%%%%%%%%
%%%%%%%%%%%%%%
%%%%%%%%%%%%%%
\subsection{Applications to classification}
\label{subsec:HCT}
Consider a  classification setting where we have samples $(\tilde{X}_i, Y_i)$, $1 \leq i \leq n$,  from two classes, where $\tilde{X}_i \in R^p$ are the feature vectors and $Y_i \in \{-1, 1\}$ are the class labels. Given a fresh sample $\tilde{X} \in R^p$ where the associated class label $Y \in \{-1, 1\}$ is unknown, the goal is to use $(\tilde{X}_i, Y_i)$
to construct a trained classifier and to use it to   predict $Y$.

Following \cite{FJY},  we model $\tilde{X}_i$ with a Gaussian graphical model, where for two distinct mean vectors $\mu^{\pm} \in R^p$ and a covariance matrix $\Sigma \in R^{p,p}$,
\begin{equation} \label{classificationmodel}
\tilde{X}_i  \sim N(\mu^{\pm}, \Sigma), \qquad \mbox{if $Y_i = \pm 1$, respectively}.
\end{equation}
Similar to that of (\ref{DefineOmega}), we assume the precision matrix $\Omega = \Sigma^{-1}$ is sparse, in the same sense. Additionally, let
$\mu$ be the contrast mean vector:
\begin{equation} \label{Constrastmean}
\mu = \mu^{+} - \mu^{-}.
\end{equation}
We assume $\mu$ is sparse in  that only a small fraction of its entries is nonzero.

We are primarily interested in  classification for microarray data. For the two data sets in Table \ref{table:data},  model (\ref{classificationmodel})   might deviate from the ground truth, but the good thing is that PCS is not tied to model (\ref{model1}) and our proposed classifier works quite well on these data sets; Section \ref{subsec:realdata}.

Higher Criticism Thresholding (HCT) is a recent classifier proposed in \cite{DJ08,FJY}, which adapts Fisher's Linear Discriminant Analysis (LDA) to the modern regime of `large $n$, really large $p$'.
In the idealized case where $\Omega$ is known or can be estimated reasonably well, HCT is shown to have optimal classification behaviors for model (\ref{classificationmodel}).  The question is then how to estimate $\Omega$ accurately with real time computing.

In this paper,  we consider three approaches to estimating $\Omega$: PCS, the glasso \cite{glasso}, and  FoBa. FoBa stands for the classical forward-backward method for variable selection \cite{Seber}, and it has not yet been proposed as an approach to  estimating $\Omega$. However, we can still develop it into such a procedure; we discuss this in details in Section \ref{subsec:foba}.

CLIME and scaled-lasso are not included for comparison, as they are unsatisfactorily slow for large $p$ (e.g., $p = 8K$).  Bickel and Levina \cite{BLT}  proposed to estimate  the precision matrix by the inverse of a thresholded version of the empirical covariance matrix. This method is not included either, for it focuses on the case where $\Sigma$  is sparse (but $\Omega$ may be non-sparse).

To apply PCS, the glasso, or  FoBa,  it is more convenient to start with the empirical
correlation matrix $\hat{R}$ (see below) than with $\hat{\Sigma}$.
Let $n_1$ and $n_2$ be the sample sizes of Class $1$ and Class $2$,
let $\hat{\mu}^{\pm} \in R^p$ be the sample mean vectors  for Class $1$ and Class $2$, respectively,
and let $\hat{s}^{\pm} \in R^p$ be the vectors of sample standard deviations for class $1$ and class $2$, respectively. The pooled standard deviation associated with feature $j$ is then
%%%%%%%%%%
%%%%%%%%%%
%%%%%%%%%%
\begin{equation}\label{shat}
\hat{s}(j)   =    \sqrt{[(n_1-1) (s^+(j))^2+(n_2-1)(s^{-}(j))^2] / (n_1+n_2-2)}.
\end{equation}
For $i = 1, 2, \ldots, n$, let $\hat{\mu}_i^* \in R^p$ be the vectors satisfying $\hat{\mu}_i^* = \hat{\mu}^+$ if  $i \in$ Class $1$ and $\hat{\mu}_i^* = \hat{\mu}^{-}$ otherwise.
The empirical correlation matrix $\hat{R} \in R^{p,p}$ is then
\begin{equation} \label{DefineRhat}
\hat{R}(j,k) = (n\hat{s}(j)\hat{s}(k))^{-1}   \sum_{i  = 1}^n   (\tilde{X}_i(j) - \hat{\mu}_i^*(j)) (\tilde{X}_i(k) - \hat{\mu}_i^*(k)).
\end{equation}
Once $\hat{R}$ is obtained, we apply each of the three methods (PCS, glasso, FoBa) and denote the estimates by $\hat{\Omega}^{pcs}$, $\hat{\Omega}^{glasso}$, and $\hat{\Omega}^{foba}$.

 For $\hat{\Omega}$ being either of the three estimates, the corresponding HCT-classifier (denoted by HCT-PCS, HCT-glasso, and HCT-FoBa) consists of  the following steps for classification.
\begin{itemize}
\item  Let $Z \in R^p$ be the vector of {\it summarizing $t$-scores}:
$Z(j)= (\hat{\mu}^+(j)- \hat{\mu}^{-}(j))/(n_0 \cdot \hat{s}(j))$, $1\leq j\leq p$, where $n_0= (n_1^{-1} + n_2^{-1})^{1/2}$.
\item Normalize $Z$ by $Z^*(j) =
 (Z(j)  - u(j))/d(j)$, where $u(j)$ and $d(j)$ are the mean and standard deviation of different entries of $Z$.
\item Apply the {\it Innovated Transformation} \cite{FJY}:
$\tilde{Z} = \hat{\Omega} Z^*$.
\item {\it Threshold choice by Higher Criticism}. For each $1 \leq j \leq p$,
obtain a  $P$-value by $\pi_j = P(|N(0,1)| \geq  (\hat{\Omega}(j,j))^{-1/2} |\tilde{Z}(j)|)$. Sort
  the $P$-values ascendingly by $\pi_{(1)} < \pi_{(2)} < \ldots < \pi_{(p)}$. Let
$\hat{j}$ be the index among the range $1 \leq j \leq \alpha_0 p$ and that maximizes the so-called HC functional $HC_{p,j} = [j/p - \pi_{(j)}]/\sqrt{(1 - j/p) (j /p)}$ for all $j$ in the range of $1 \leq j \leq \alpha_0 p$ (we usually set $\alpha_0 = .2$, as suggested by \cite{DJ08}).    The HC threshold  $t_p^{HC} = t_p^{HC}(\tilde{Z}, \hat{\Omega},  p, n)$   is   the magnitude of the $\hat{j}$-th largest entry (in magnitude) of $\tilde{Z}$.
\item {\it Assign weights by thresholding}.  Let $w_{HC}(j) = \sgn(\tilde{Z}(j)) \cdot 1\{|\tilde{Z}(j)| \geq t_p^{HC}\}$, $1 \leq j \leq p$. Denote
$w_{HC} = (w_{HC}(1), w_{HC}(2), \ldots, w_{HC}(p))'$.
\item {\it Classification by post-selection LDA}.  We normalize the test feature $\tilde{X}$  by $\tilde{X}^*(j)  = [\tilde{X}(j) - (\hat{\mu}^+(j) + \hat{\mu}^{-}(j))/2]/\hat{s}(j)$,  $1 \leq j \leq p$.
 Let $L_{HC}(\tilde{X}) =  (w_{HC})'  \hat{\Omega} \tilde{X}^*$. We classify $Y = \pm 1$ according to $L_{HC}(\tilde{X}) \gtrless 0$.
\end{itemize}
The rationale behind step $2$ is the phenomenal work by Efron \cite{EmpNull} on empirical null. Efron found that for microarray data, there is a substantial gap between the (marginal) distribution of the  theoretical null and that of the empirical null, and it is desirable to bridge the gap by renormalization.  This step is specifically designed for microarray data, and may not be necessary for other types of data (say, simulated data).  Also, note that when normalizing the test feature $\tilde{X}$, we use $(\hat{\mu}^{\pm}, \hat{s})$ which do not depend on $\tilde{X}$.

%%%%%%%%%%
%%%%%%%%%%
%%%%%%%%%%
%%%%%%%%%%
\subsection{Comparison: classification errors with microarray data}
\label{subsec:realdata}
We consider the two gene microarray data sets  in Table \ref{table:data}.
The original rats data set was collected in a study on gene expressions of live rats in response to different drugs and toxicant, and we use the cleaned version by \cite{liver}. The data set consists of $181$ samples measured on the same set of $8491$ genes, where $61$ samples are labeled by \cite{liver} as toxicant, and the other $120$ as other drugs.  The original liver data set was collected in a study on the hepatocellular carcinoma (HCC),
and we also use the cleaned version by \cite{liver}. The data set consists of $157$ samples measured on the same set of $10,237$ genes,  $82$ of them are tumor samples, and the other $75$ non-tumor.

We consider a total of $6$ different classifiers: naive HCT  (where we pretend that $\Omega$ is diagonal and apply HCT without estimating off-diagonal of $\Omega$; denoted by nHCT), HCT-PCS, HCT-glasso, HCT-FoBa, and two popular classifiers:  Support Vector Machine (SVM) and Random Forest (RF).

Among them, nHCT is tuning free, three methods have one tuning parameter:
$\lambda$ for HCT-glasso, `cost' for SVM, and `number of trees' for RF. The tuning parameter for HCT-glasso (and also those of HCT-PCS and HCT-FoBa) come from the method of estimating the precision matrix. HCT-PCS has three tuning parameters $(\delta, L, q)$, but it is relatively insensitive to $(\delta, L)$. In this paper, we set $(\delta, L) = (0.1, 30)$, so PCS only have one tuning parameter $q$.
For HCT-FoBa, we use the package by \cite{ZhangFoBa}, which has three tuning parameters:
a back fitting parameter (set by the default value of $-.5$ here),  a ridge regression parameter $\delta > 0$ and  a step size parameter $L$. As parameters $(\delta, L)$ have similar roles to $(\delta, L)$ in PCS,  we set them as $(\delta, L) = (0.1, 30)$. The performance of either PCS or FoBa is relatively insensitive to different choices of $(\delta, L)$, as long as they fall in a certain  range.

In our study, we  use two layers of $3$-fold data splitting. To differentiate one from the other, we call them the {\it data-splitting} and {\it cv-splitting}. The former is for comparing classification errors of different methods across different data splitting and it is particularly relevant to evaluating the performance on real data, while the latter is for selecting tuning parameters. The latter is not required for nHCT or HCT-FoBa.
\begin{itemize}
\item {\it Data-splitting}. For each data set, we apply $3$-fold random split to the samples in either of the two classes ($25$ times, independently).
\item {\it Cv-splitting}. For each resultant training set from the data splitting, we apply $3$-fold random split to the samples in either of the two classes (25 times, independently).
\end{itemize}
Sample indices for the $25$ data splitting and sample indices of the $25$ cv-splitting associated with each of the data splitting can be found at  \url{www.stat.cmu.edu/~jiashun/Research/software}.

We now discuss how to set the tuning parameters in HCT-PCS, HCT-glasso, SVM, and RF.
For HCT-PCS, we find the interesting range for $q$ is $0.05\leq q \leq 0.5$.
We discretize this interval evenly with an increment of $0.05$. The increment is sufficiently small,
and a finer grid does not have much difference.
For each data splitting, we determine the best $q$ using $25$ independent cv-splitting,
picking the one that has the smallest ``cv testing error". This $q$ value is then plugged into HCT-PCS
for classification.

For HCT-glasso,  the interesting range for the parameter $\lambda$ is $0.8 \leq \lambda \leq 1$.  Since the algorithm starts with empirical correlation matrix, it is unnecessary to go for $\lambda > 1$ (the resultant estimate would be a scalar times the $p \times p$ identity matrix, a simple result of the KKT condition  \cite{glasso}). On the other hand, it is very time consuming by taking $\lambda < 0.8$.
For example, if we take $\lambda = 0.65, 0.7$, and $0.75$, then on a $12GB$ RAM machine,  it takes the glasso more than a month  for $\lambda = 0.65$,  about a month for  $\lambda = 0.7$, and about $180$ hours  for $\lambda=0.75$ to complete all $25 \times 25$ combinations of data-splitting and cv-splitting, correspondingly.  Similar to that of HCT-PCS, for each data splitting, we take $\lambda \in \{0.8, 0.85, 0.9, 0.95, 1\}$ and use the $25$ cv-splitting to decide the best $\lambda$, which is then plugged into HCT-glasso for classification.

For SVM, we use the package from \url{http://cran.r-project.org/web/packages/e1071/index.html}.   We find the interesting range for the  `cost' parameter is between $0.5$ and $5$, so we take `cost' to be $\{0.5, 1, 1.5, \ldots, 5\}$ and use cv-splitting to pick the best one.   For RF, we use the package downloaded from \url{http://cran.r-project.org/web/packages/randomForest/index.html}. RF has one tuning parameter `number of trees'. We find that the interesting range for  `number of trees' is between $50$ to  $500$, so we take it to be $\{50, 100, \ldots, 500\}$ and use cv-splitting to decide the
best one.

%%%%%%%%%%
%%%%%%%%%%
%%%%%%%%%%
\begin{table}[hbt!]
\caption{Comparison of classification errors (average for $25$  data-splitting). The error rates and their standard deviations (in   brackets) are reported in percentage (e.g.,  $5.7$ means $5.7\%$).}
\vspace{0.05 in} 
\scalebox{1.05}{
\centering
\begin{tabular}{|c|c|c|c|c||c|c|}
\hline
Data   & HCT-PCS  & HCT-FoBa  & HCT-glasso  & nHCT & SVM & RF    \\
\hline
Rats  &  5.7(3.05)  &  8.6(3.36)  & 20.3 (5.10) &  15.1(6.49)   &  6.9(4.02)  &   13.5(3.99)   \\
\hline
Liver &  4.2(3.60)  & 10.0(4.64)  & 20.2 (6.38)  &  10.5(4.30)  & 3.5(2.72)  &   4.3(3.00)  \\
\hline
\end{tabular}
}
\label{table:error}
\end{table}

\begin{figure}[htb!]
    \centering
    \includegraphics[width= 2.8 in]{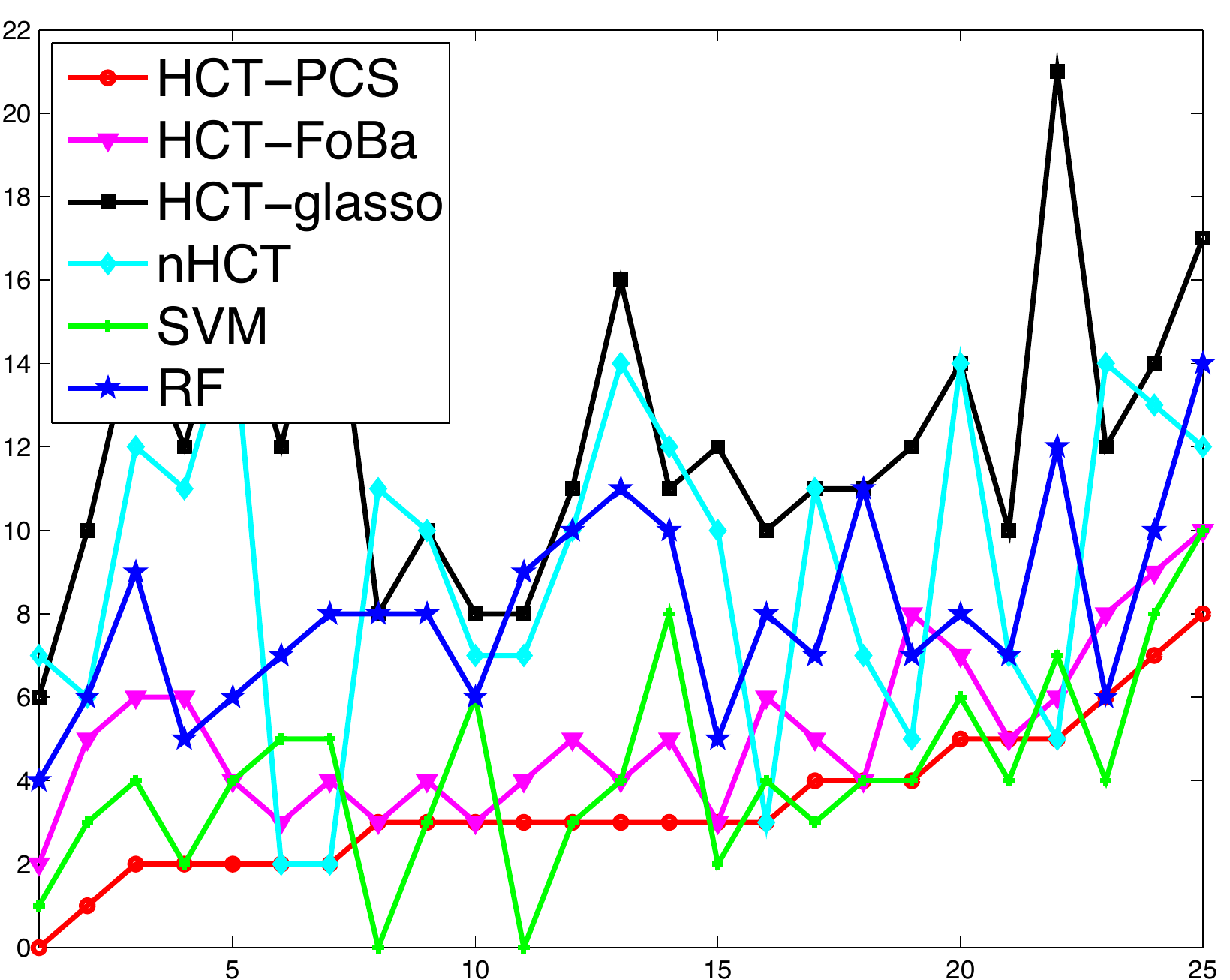}
    \includegraphics[width= 2.8 in]{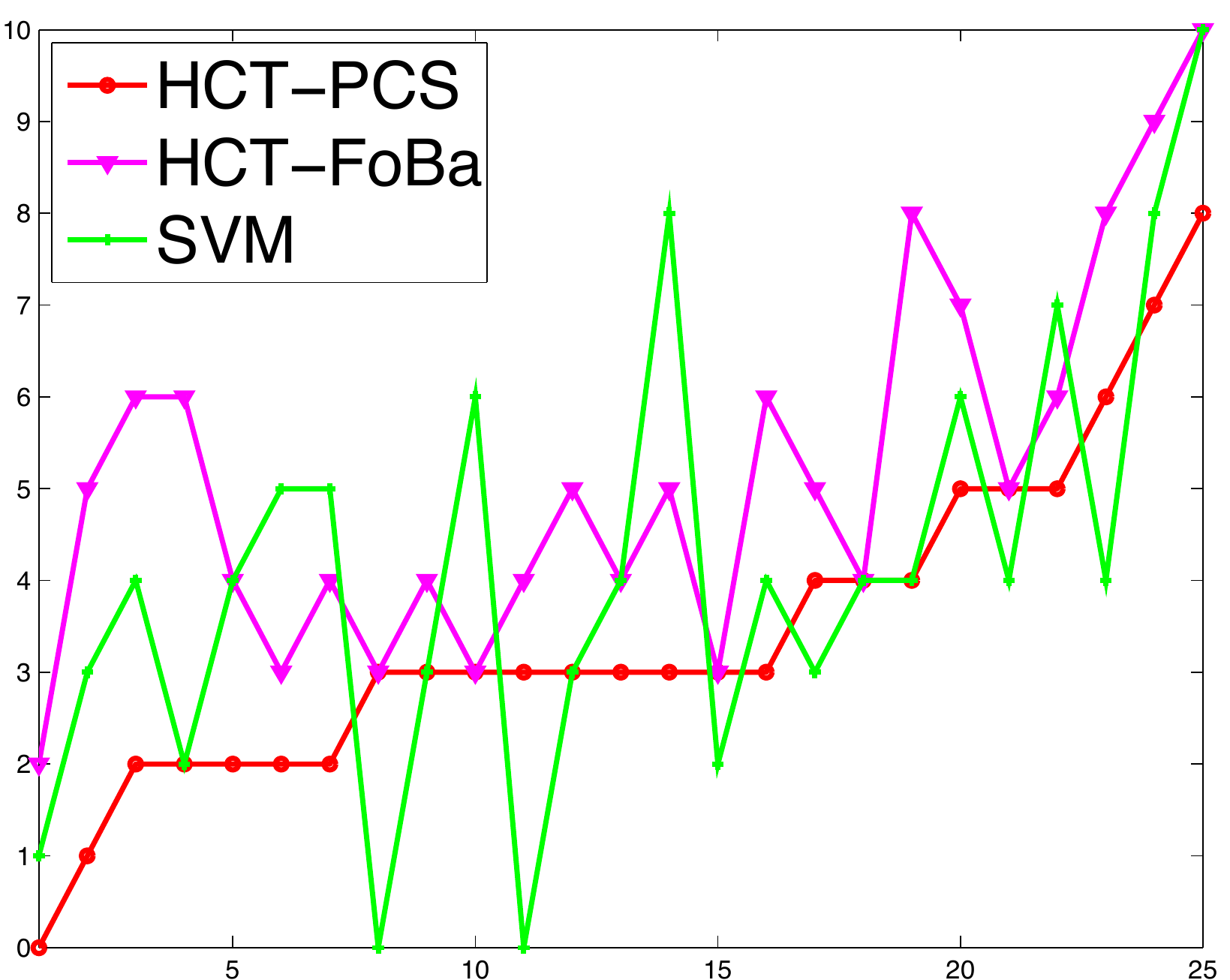}
    \caption{Comparison of number of testing errors for the rats data. Left:  errors (y-axis) of 6 methods for  25 data splittings (x-axis; arranged in a way so that the errors of HCT-PCS increase from left to right).  Right:   the same information but only with HCT-PCS, HCT-FoBa and SVM (for a better view). }
    \label{figure:errorrats}
\end{figure}

\begin{figure}[htb]
     \centering
    \includegraphics[width= 2.8 in]{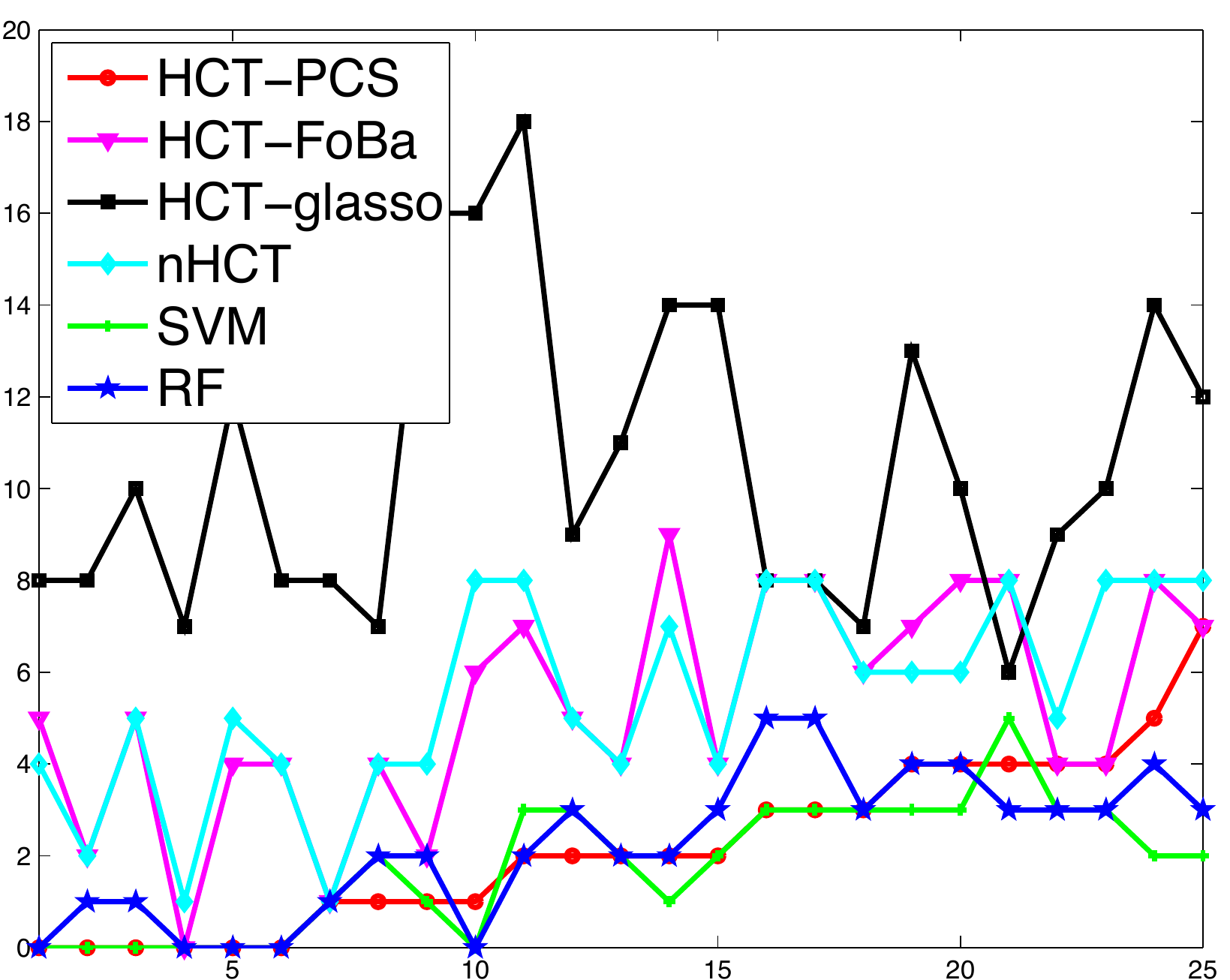}
    \includegraphics[width= 2.8 in]{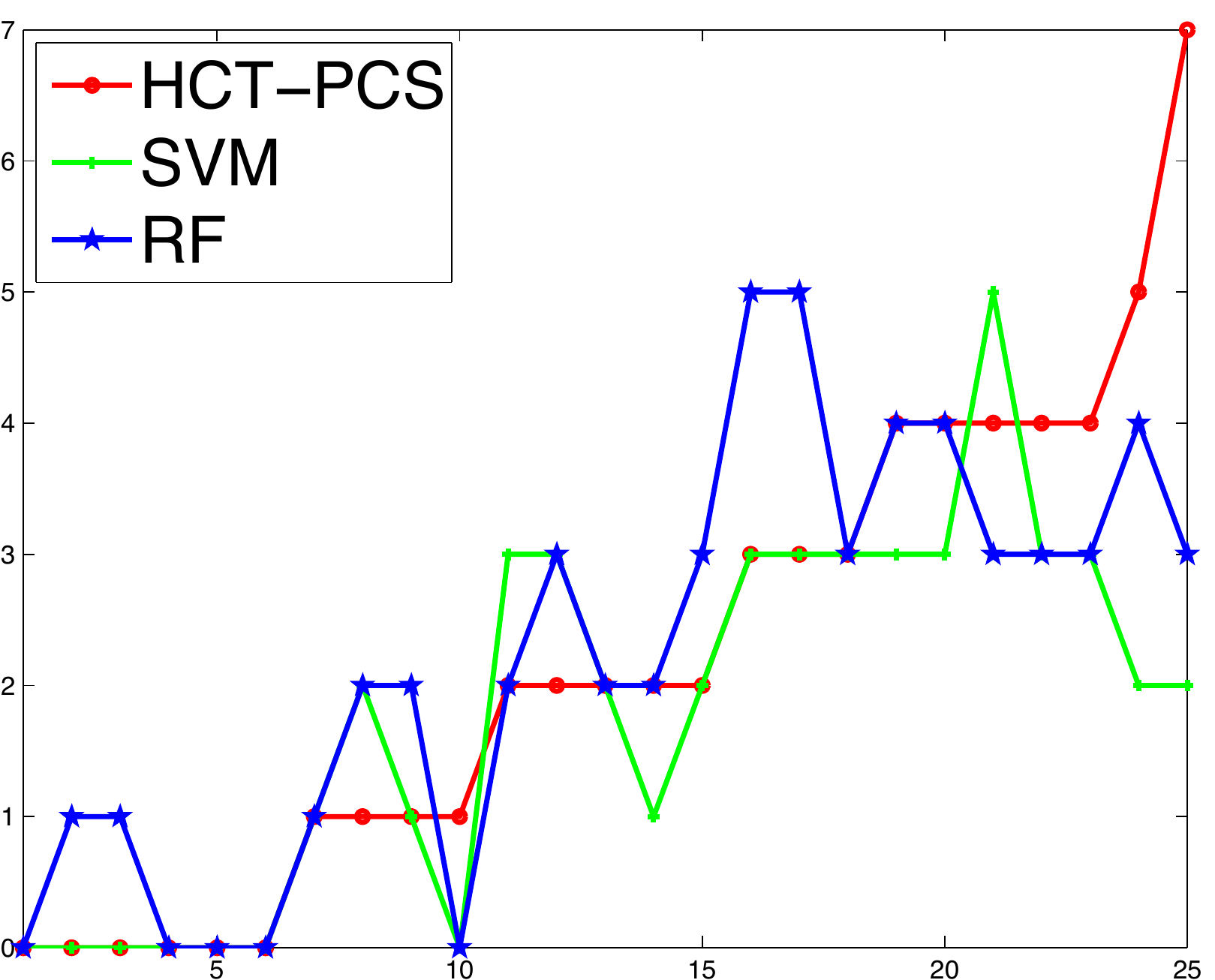}
    \caption{Comparison of number of testing errors for the liver data. Left:  errors (y-axis) of 6 methods at 25 data splittings (x-axis; arranged in a way so that the errors of HCT-PCS increase from left to right). Right:  the same information but only with HCT-PCS, SVM and RF (for a better view).}
    \label{figure:errorliver}
\end{figure}

The average classification (testing) error rates of all $6$ methods across $25$ different data splitting are tabulated in Table \ref{table:error}.  The standard deviations of the error rates are relatively large,
due to the large variability in the data splitting. For more informative comparison, we present the number of testing  errors associated with all $25$ data splittings in Figure~\ref{figure:errorrats} (rats data) and Figure~\ref{figure:errorliver} (liver data), respectively. In these figures, the $25$ data splittings are arranged in a way so that the corresponding errors of HCT-PCS are increasing from left to right.

%%%%%%%%%%%
%%%%%%%%%%%
%%%%%%%%%%%
%%%%%%%%%
%%%%%%%%%
%%%%%%%%%
%%%%%%%%%
From the left panel of Figure \ref{figure:errorrats}, we see that for the rats data, HCT-glasso, nHCT and RF are all above HCT-PCS.  To better show the difference among HCT-PCS, HCT-FoBa and SVM, we further plot the number of  testing errors  in the right panel.   Figure \ref{figure:errorliver} provides the similar information for liver data, with HCT-PCS, SVM and RF being highlighted in the right panel. The results suggest: for rats data, HCT-PCS outperforms all methods with the average errors, including SVM and RF; for liver data, HCT-PCS is slightly inferior to SVM, but still outperforms all other methods; for both data sets,  HCT-PCS significantly outperforms all other HCT-based methods (nHCT,  HCT-glasso, HCT-FoBa), which further suggests PCS gives a better estimate for the precision matrix than the glasso and FoBa.
%%%%%%%%%%%%%
%%%%%%%%%%%%%
%%%%%%%%%%%%%

The computation time is hard to compare, as it depends on many factors such as the data-splitting in use, how professional the code is written, and how capable the user handles the computation. Therefore, the complexity comparison summarized in Table~\ref{tab:complexity} can only be viewed as a qualitative one (for the complexity of the glasso, see \cite{glassocomplexity}). We run all methods using Matlab, with an exception of FoBa, SVM and RF by R, on a workstation with 8 CPU cores and $12GB$ RAM, and the real time elapsed in computation is recorded upon one splitting. Note that for HCT-FoBa, the reported computation time accounts for no cross validation.
\begin{table}[htb!]
\caption{Comparison of computational complexity of all classifiers and running time for rats and liver data at a single data splitting. Here, $n$ is the number of samples, $p$ is the number of variables and $T$ is the number of trees used in RF (assuming $p \geq n$ and $L^3 \leq O(n)$; computation times are based on $25$ cross validations for all but HCT-FoBa and nHT). }
\vspace{0.05 in} 
\centering 
\scalebox{0.915}{
\begin{tabular}{|c|c|c|c|c|c|c|}
\hline
& HCT-PCS  &HCT-FoBa  &   HCT-glasso  &    nHCT     & SVM & RF     \\
 \hline
Complexity   &    $O(n p^2)$   & $O(n p^2)$   &  $O(p^3)$--$O(p^4)$      &  $O(np \log(p))$     & $O(n^2p)$  & $O(T n p\log(n))$       \\
\hline
Time (rats)  & 166.8 min &  8.7 min &  380 min  &  0.11 min   & 7.7 min &  21.5 min  \\
\hline
Time (liver)  & 241.0 min & 12.9 min &  890 min  &  0.12 min  & 7.7 min &  26.3 min  \\
\hline
\end{tabular}}
\label{tab:complexity}
\end{table}

It is noteworthy for much larger $n$ (e.g., $n = 2000$),  SVM becomes much slower, showing a disadvantage of SVM, compared to PCS, FoBa, and the glasso. See Section \ref{sec:Simul} (simulation section) for settings with much larger $n$.

%%%%%%%%%%
%%%%%%%%%%
%%%%%%%%%%%%%%%
%%%%%%%%%%%%%%%
%%%%%%%%%%%%%%%
%%%%%%%%%%

%%%%%%%%%%%%%
%%%%%%%%%%%%%
%%%%%%%%%%%%%
{\bf Remark}. Both PCS and FoBa use ridge regression, but PCS uses ridge regression on an as-needed basis (see \eqref{RIasneed}) and FoBa uses conventional ridge regression at each iteration.  In Table \ref{table:ridge}, we compare the classification error rates of PCS and FoBa for  the cases of with ($\delta = .1$) and without ridge regularization ($\delta = 0$). The results suggest a substantial improvement by using ridge regularization, for both methods. On the other hand, we find that the classification errors for both methods are relatively insensitive to the choice of $\delta$, as long as they fall in an appropriate range. In this paper, we choose $\delta = .1$ for all real data experiments.
%%%%%%%%%
%%%%%%%%%
%%%%%%%%%
\begin{table}[htb!]
\caption{Comparison of classification errors for HCT-PCS and HCT-FoBa for $\delta = .1$ and $\delta = 0$.
The classification errors and their standard deviations (in the brackets) are reported in percentage (e.g.,  $5.7$  means $5.7\%$).
}
\vspace{0.05 in} 
\centering 
\scalebox{0.95}{
\begin{tabular}{|c|c|c|c|c|}
\hline
Data  & HCT-PCS ($\delta = .1$)   & HCT-PCS ($\delta = 0$)  & HCT-FoBa ($\delta = .1$)  &  HCT-FoBa ($\delta = 0$)     \\
\hline
Rats  & 5.7(3.05)   &  13.2(3.57)   &  8.6(3.36)   &  13.7(5.10)   \\
\hline
Liver &  4.2(3.60)   &  12.5(6.08)   & 10.0(4.64)   &  15.1(7.48)  \\
\hline
\end{tabular}}
\label{table:ridge}
\end{table}

%%%%%%%%%%%%%%
%%%%%%%%%%%%%%
%%%%%%%%%%%%%%
\subsection{Comparison with glasso over the estimated $\Omega$}
\label{subsec:glassoomega}
\begin{figure}[htb]
    \centering
    \includegraphics[width= 2.8 in]{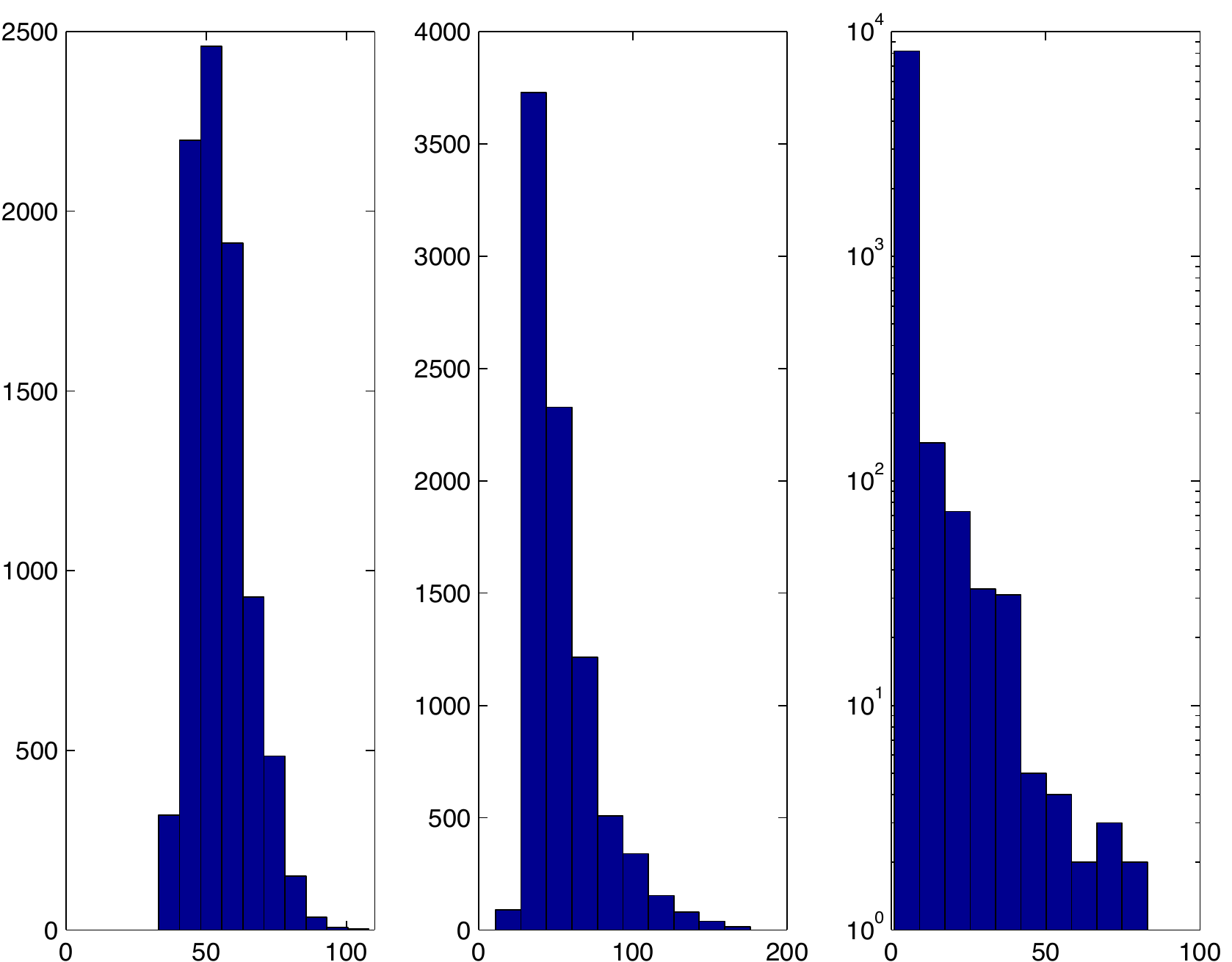}  
    \includegraphics[width= 2.8 in]{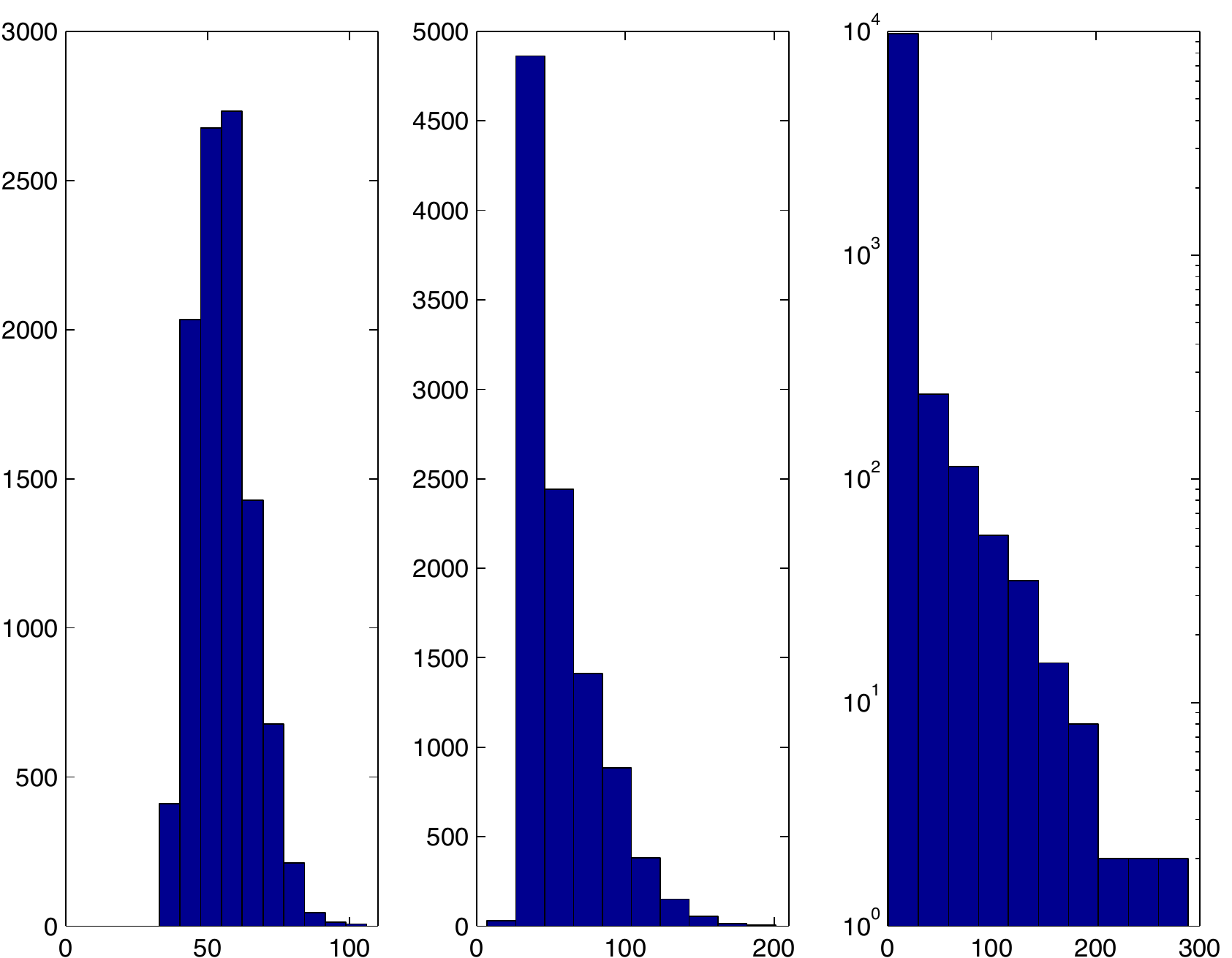}
    \caption{ Panels 1-3:  histograms of the number of nonzeros in different rows of $\hat{\Omega}$ for the rats data by  PCS, FoBa, and the glasso ($y$-axis in Panel 3  is $\log(\mbox{$\#$of nonzeros})$).      For PCS and FoBa, the number of nonzeros range from $33$ to $108$ and from $11$ to $176$, respectively. For the glasso, in $7550$ out of $8491$ rows,
all off-diagonals are estimated as $0$  (maximal number of nonzeros in a row  is $83$).    Panels 4-6: similar but are for the liver data.}
    \label{fig:hist}
\end{figure}

That HCT-PCS significantly outperforms HCT-glasso and HCT-FoBa in classification errors suggests that PCS gives `better' estimations of $\Omega$ than the other two methods. We take a look at the estimated precision matrices by the glasso, PCS, and FoBa on the two microarray data. Figure~\ref{fig:hist} presents the histograms of the number of nonzeros of different rows in
$\hat{\Omega}$ for the three different methods.
For all histograms, we use the whole data set (either the rats or the liver data), without  data splitting. For PCS, we use $(q, \delta, L)=(0.2, 0.1, 30)$. For the glasso, we use $\lambda=0.8$. For FoBa, we use $(\delta, L)=(0.1, 30)$ so that it is consistent with PCS. The histograms look similar when we change the tuning parameters in the appropriate range.
Figure~\ref{fig:hist} reveals very different patterns of the estimated $\Omega$.
\begin{itemize}
\item For the majority of the rows, the glasso estimate is $0$ in all off-diagnal entries,   but for some of the rows, the glasso estimate  can have several  hundreds  of nonzeros.
\item For either of the PCS and the FoBa estimates, the number of nonzeros in each row can be as large as a few ten's, but no smaller than $10$.
\end{itemize}
While the ground truth is unknown, it seems that the estimates by PCS or FoBa make more sense:
it is hard to believe that the off-diagonals of $\Omega$ are all $0$ for most of the rows; it is more likely that in most of the rows, we have at least a few nonzeros. Partially, this explains why the classification errors of the glasso is the largest among the three methods.
It also explains why the naive HCT has unsatisfactory behaviors (recall that in nHCT, we pretend that $\Omega$ is diagonal).

%%%%%%%%%
%%%%%%%%%
%%%%%%%%%
%%%%%%%%%
%%%%%%%%%%
%%%%%%%%%%
\subsection{Comparison with FoBa over the estimated $\Omega$}
\label{subsec:foba}
Forward and Backward regression (FoBa)  is a classical approach to variable selection, which is proposed by
Draper and Smith as early as 1960's \cite{Seber}.  FoBa can be viewed as an extension of the classical
Forward Selection (FS) procedure \cite{Seber},  where
the difference is that FoBa allows for backward elimination, but FS does not.
FS and  FoBa have been studied carefully recently (e.g.,  \cite{StOMP, ZhangGreedy,ZhangFoBa}).

To the best of our knowledge,   FS and FoBa have not yet been proposed as an approach to estimating the precision matrix, but we can always develop them into such an approach as follows. Fix $1 \leq i \leq p$.   Recall that  $X' = [x_1, x_2, \ldots, x_p]$ and that
$\Omega = [\omega_1, \omega_2, \ldots, \omega_p]$.   It is known that we can always associate each row of $\Omega$ with a linear regression model as follows \cite{BuhlmannBook}:
\begin{equation} \label{regression}
x_i  =  (\omega_i(i))^{-1} \sum_{j \neq i} \omega_i(j) x_j + z_i, \qquad  z_i \sim N(0, \sigma^2 \cdot I_n),
\end{equation}
where $\sigma^2 = 1/\omega_i(i)$ and $z_i =  x_i  -   (\omega_i(i))^{-1} \sum_{j \neq i} \omega_i(j) x_j$ is independent of $\{x_j: j \neq i\}$.
We can then apply either FS or FoBa to (\ref{regression}) for each $1\leq i\leq p$, and symmetrize the whole matrix in the same way as the last step of PCS; the resultant procedure is an approach to estimating $\Omega$.
A small gap here is that, for each $1 \leq i \leq p$, FS and FoBa attempt  to estimate the vector $(\omega_i(i))^{-1} \omega_i$, not $\omega_i$ itself (as we desire).

This is  closely related to PCS, but differs in several important ways. Since FoBa is viewed as an improvement over FS, we only compare PCS with FoBa.

The most obvious difference between PCS and FoBa is that, in their `forward selection' steps,  the objective function for recruiting new nodes are different. PCS uses the partial correlation \eqref{Definehatrho}, and FoBa uses the correlation between $x_j$ and the residuals. The following lemma elaborates two objective functions and is proved in Section~\ref{sec:proof}.
%%%%%%%%%%
%%%%%%%%%%
%%%%%%%%%%
\begin{lemma} \label{lemma:objective}
For $i$, $j$, and $S \subset \{1, 2, \ldots, p\}$ such that $i \neq j$, $i, j \notin S$,   and $|S| \leq n-2$,
the objective functions in the `forward selection' steps of PCS  and FoBa associated with $\delta = 0$ are  well-defined with probability $1$, equalling
\begin{equation} \label{grec}
\hat{\rho}_{ij}(S) = x_i' (I - H_S) x_j / \sqrt{x_i' (I - H_S) x_i \cdot  x_j' (I - H_S) x_j},
\end{equation}
and
\begin{equation} \label{FS}
\hat{\rho}_{ij}^*(S) = x_i' (I - H_S) x_j  / \|x_j\|,
\end{equation}
respectively, where $H_S$ is the projection   from $R^n$ to the subspace  $\{x_k: k \in S\}$.
\end{lemma}

PCS and FoBa are also different in philosophy.
It is well-known that FS tends to select ``false variables".  For remedy, FoBa proposes ``immediate backward elimination":   in each step, FoBa is allowed  to add  or remove  one or more variables, in hopes
that whenever we falsely select one or more variables, we can remove them immediately.
 PCS takes a very different strategy.   We recognize that, from a practical perspective, the signals are frequently ``rare and weak" \cite{DJ14,KJ}, meaning that $\Omega$ is sparse and that nonzero entries are relatively small individually.
In such cases,   ``immediate backward elimination" is impossible and we must tolerate many ``false discoveries".
Motivated by this,  PCS employs a Screen and Clean methodology, which attempts
to include all the true nodes while keeping the  ``false discoveries"  as few as possible.
Our results on the two microarray data sets support the ``rare and weak" viewpoint: for example, in Figure~\ref{fig:hist}, the symmetrization step has a significant impact on the histograms of PCS and FoBa for both data sets, which implies that ``false discoveries" are unavoidable.

Though it can be viewed as a method for variable selection, Screen and Clean method has a strong root in the literature of
large-scale multiple testing and in genetics and genomics, where the ``rare and weak" viewpoint is especially appropriate.  In  rare and weak settings, Screen and Clean is more appropriate than other variable selection approaches whose focus is frequently on rare and strong signals. See \cite{DJ14,KJ} for more discussions.

\begin{figure}[htb]
    \centering
    \includegraphics[height  = 2.3 in]{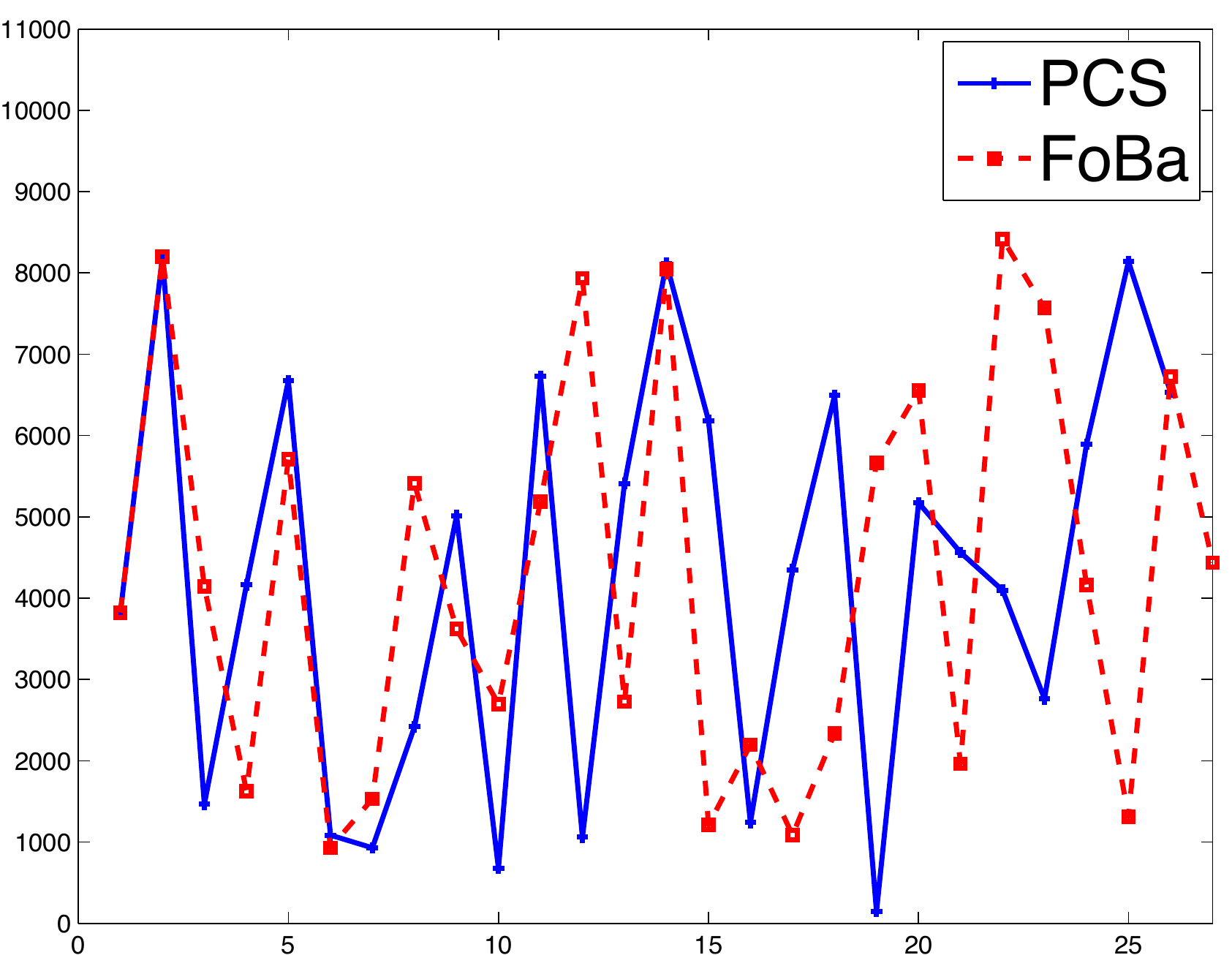}
    \includegraphics[height  = 2.3 in]{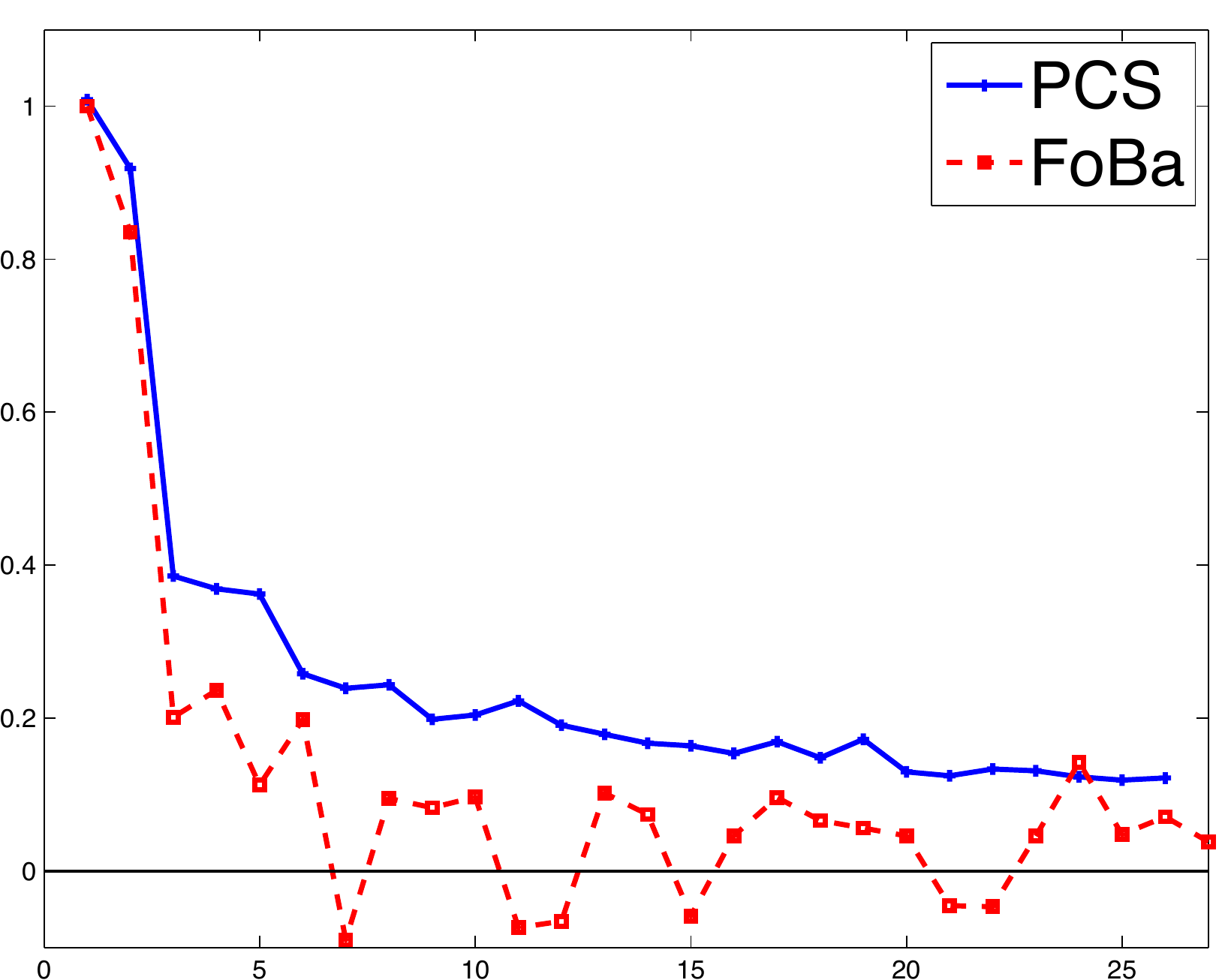}
    \caption{Left: the set of nodes recruited by PCS (solid) and FoBa (dashed) in the forward steps
    (PCS: $26$ nodes;  FoBa:  $27$ nodes).   Right:  objective functions (PCS:  (\ref{grec});  FoBa:  (\ref{FS}))   corresponding to the nodes on the left.  PCS and FoBa only share the first $2$ nodes that have the largest (in magnitude) objective functions for PCS and FoBa, respectively.   }
    \label{fig:pcsfoba}
\end{figure}

In practice, the above differences may lead to noticeable differences between the estimates of $\Omega$ by PCS and FoBa. To illustrate,  we consider the estimation of row $\# 3823$ of $\Omega$ associated with data splitting $\#25$ of the rats data, and compare how the forward selection steps of PCS ($\delta=0.1$)  and  FoBa ($\delta=0.1$) are different from each other.  The cleaning step of PCS and the backward selection of FoBa are omitted for comparison.
%%%%%%%
%%%%%%%
%%%%%%%
%%%%%%%
%%%%%%%
\begin{itemize}
\item \textbf{PCS ($\delta=0.1$).}
In the Screen step, PCS stops at step $26$, and the $26$  recruited nodes are:
3823,  8199,  1466, 4164,  6674,  1087,   931,  2419, 5016,   679,   6726,   1059, 5410,  8116,
6183,  1242,  4348,   6492,   147,  5174,   4561,  4096,  2763, 5894,  8140, and  6532.
\item \textbf{FoBa ($\delta=0.1$).}  We run FoBa for $31$ steps. It turns out that $4$ of the steps are backward steps (one node deleted in each). The $27$ nodes FoBa recruits in each of the forward steps are:   3823, 8199, 4144,  1628, 5707,  931, 1532,  5410,  3620,  2700,  5188,  7933,  2729,
8048,  1212,  2197,  1087,  2337,  5665,  6556,  1962,  8417,  7567,  4164,  1312,  6726,  and  4436.
\end{itemize}
Figure \ref{fig:pcsfoba} displays the two sets of selected nodes (left panel) by PCS and FoBa and their corresponding
coefficients (right panel) given in (\ref{grec}) and (\ref{FS}), respectively.  We see that
the first two recruited nodes by PCS and FoBa are the same, corresponding to large coefficients, either in
(\ref{grec}) and (\ref{FS}). All other nodes recruited by PCS and FoBa are different, corresponding to
comparably smaller coefficients (either in (\ref{grec}) or (\ref{FS})).
This suggests a ``rare and weak" setting where PCS and FoBa differ significantly  from each
other. Also, this provides an interesting angle of explaining why PCS outperforms FoBa in terms of classification error.
 %%%%%%%%%%
 %%%%%%%%%%
 %%%%%%%%%%

%%%%%%%%%%%
%%%%%%%%%%%
%%%%%%%%%%%
\subsection{Summary and contributions}
\label{subsec:summary}
While it is widely accepted that estimating the precision matrices is an interesting problem for high dimensional data analysis,
little attention has been paid to
either the problem of how to develop methods that are practically feasible for very {\it large} precision matrices or the problem of how to integrate the estimated precision matrices for statistical inference.
Motivated by the immediate need for the analysis of microarray data, the main goal of this paper is to find an approach that is executable in real time and also useful in improving statistical inference.

The contribution of this paper is three-fold. First, we propose PCS as a new approach to estimating large sparse precision matrices. PCS estimates the precision matrix row by row. To estimate each row, we develop a stage-wise algorithm which greedily recruits one node at a time using the empirical partial correlations. PCS is computational efficient and modest in memory use. These two features enable PCS to execute accurate estimation of the precession matrices with real-time computing, and also open doors to accommodating much larger precision matrices (e.g., $p \geq 50K$).

Second, we combine PCS with HCT \cite{FJY} for a new classifier  HCT-PCS  and apply it successfully to two microarray data sets.
HCT-PCS is competitive in classification errors, compared to the more popular classifiers of SVM and RF. HCT-PCS is tuning free (given an estimate of $\Omega$), enjoys theoretical optimality \cite{FJY}, and fully exploits the sparsity in both the feature vectors and the precision matrix.
SVM and RF, however, can be unstable with regard to tuning. For example, the tuning parameter in SVM largely relies on training data and structure of the kernel function employed to transform the feature space;
this instability of regularization could end up with non-sparse support vectors \cite{Bi2003, Cawley2010}.
SVM and RF are found faster than HCT-PCS in Section \ref{sec:Intro}, but such  an advantage is much less prominent for larger $n$.

HCT-PCS gives more satisfactory classification results than HCT-glasso,
suggesting that PCS gives `better' or `more useful' estimates for the precision matrix.
The glasso is relatively slow in computation when $p$ is as large as $10K$, especially when
the tuning parameter is small.
For either of two microarray data sets, the glasso estimates are undesirable:
in a majority of rows of $\hat{\Omega}$, all off-diagonals are $0$.
HCT-PCS also gives more satisfactory classification results than HCT-FoBa, and
two main differences between PCS and FoBa are (a) PCS and FoBa use very different objective functions in screening,
(b) FoBa proposes to remove `falsely selected nodes' by immediate backward deletion, while PCS adopts a ``rare and weak signal" view point, and proposes to keep all  `falsely selected nodes' until the end  the Screen step and then remove them in the Clean step.

Last, we justify carefully in Section \ref{sec:main} why and when PCS and HCT-PCS work using a general theoretical framework.  Also, in Section~\ref{subsec:comptheory},  we further compare PCS with other methods theoretically. Our theoretical studies shed interesting new light on the behaviors of stage-wise algorithms.

%%%%%%%%%%%%%%%%
%%%%%%%%%%%%%%%%
%%%%%%%%%%%%%%%%
\subsection{Content and notations}
\label{subsec:content}
The remaining sections are arranged as follows.
Section~\ref{sec:main} presents the main theoretical results. Section~\ref{sec:Simul} presents the
simulations. Section~\ref{sec:Discu} contains discussions and extensions. Section~\ref{sec:proof} contains the proofs of lemmas and theorems.

In this paper, for any vector $a$,   $\|a\|$ denotes the vector $\ell^2$-norm.  For any matrix $A$, $\|A\|$ denotes the
matrix spectral norm, $\|A\|_1$ denotes the matrix $\ell^1$-norm and $\|A\|_{\max}$ denotes the entry-wise max norm. $\lambda_{\max}(A)$ and $\lambda_{\min}(A)$ denote the maximum and minimum eigenvalues of $A$, respectively.
For any matrix $B \in R^{n,p}$ and two subsets $\call, {\cal J}$, $B^{\call,{\callJ}}$ is the same as in Definition~\ref{def:submatrix}.
\begin{comment}
For any matrix $A \in R^{n,p}$ and two subsets
$\call = \{i_1, i_2, \ldots, i_{M}\} \subset \{1, 2, \ldots, n\}$ and
${\cal J}   = \{j_1, j_2, \ldots, j_K\}   \subset \{1, 2, \ldots, p\}$ (where the indices in   either $\call$ or ${\cal J}$ are not necessarily arranged in the ascending order),
denote $A^{\call, {\cal J}}$ by the $M \times K$ sub-matrix such that
$A^{\call, {\cal J}}(m,k) = A(i_m, j_k)$, $1 \leq m \leq M, 1 \leq k \leq K$.
\end{comment}

%%%%%%%%%%%%%
%%%%%%%%%%%%%
%%%%%%%%%%%%%
%%%%%%%%%
%%%%%%%%%
%%%%%%%%%
\section{Main results}  \label{sec:main}
For simplicity, we only study the version of PCS without ridge regularization, and drop the superscript $``(0)"$ by writing   \[
\hrho_{ij}(S) = \hrho_{ij}^{(0)}(S), \qquad \mbox{for any subset $S$, random or non-random}.
\]
We simply set $L=p$, so PCS has only one tuning parameter $q$. In this section,  $C > 0$ is a generic constant which may vary from occasion to occasion.

Theoretically, to characterize the behavior of PCS,  there are two major components:
how PCS behaves in the idealized case where we have access to `small-size' principal
sub-matrices of $\Sigma$ (but not any of the `large-size' sub-matrices), and how to control the
stochastic errors. Below, after some necessary notations, we discuss two components in Sections~\ref{subsec:noiselessPCS}-\ref{subsec:noisePCS}. The main results are presented in the end of Section~\ref{subsec:noisePCS}.

For any positive definite matrix $A$, recall that $\lambda_{\min}(A)$ and $\lambda_{\max}(A)$ denote the smallest and largest eigenvalues, respectively. For any  $1 \leq k \leq p$,
define
\begin{equation} \label{Definemu}
\mu_k^{(1)}(A) = \min_{\{|S| = k  \} } \{ \lambda_{\min}(A^{S,S})\},  \qquad  \mu_k^{(2)}(A)   = \max_{\{ |S| = k  \} } \{ \lambda_{\max}(A^{S,S})\},
\end{equation}
where $S$ is a subset of $\{1, 2, \ldots, p\}$.
Also, for an integer $1 \leq K  \leq p$,
we say that a matrix $A \in R^{p,p}$ is $K$-sparse if each row of $A$ has no more than $K$ nonzero off-diagonals.
Let ${\cal M}_p$ be the set of all $p \times p$ positive definite matrices,   let $0<c_0\leq 1$ be a fixed constant, and
let
\begin{equation} \label{DefineN}
N = N(p, n)=\mbox{the smallest integer that exceeds $n/\log(p)$}.
\end{equation}
We consider the following set of $\Sigma$ (as before, $\Omega$ and $\Sigma$ are tied to each other by $\Omega = \Sigma^{-1}$) denoted by ${\cal M}_p^*(s,c_0) = {\cal M}_p^*(s, c_0; n)$:
\begin{equation}\label{DefinecalM}
{\cal M}_p^*(s,c_0)  =
\{\mbox{$\Sigma \in {\cal M}_p$:  $\Omega$ is $s$-sparse, $\mu^{(1)}_N(\Sigma)\geq c_0$,  $\mu^{(2)}_N(\Sigma)\leq c_0^{-1}$} \}.
\end{equation}

We use $p$ as the driving asymptotic parameter, so $n \goto \infty$ as $p \goto \infty$. We allow $s$ (and other parameters below) to depend on $p$. However,  $c_0$ is a constant not depending on $p$. Recall that
\[
S^{(i)}(\Omega)  = \{1 \leq j \leq p:     j \neq i,  \Omega(i,j) \neq 0\}.
\]
Introduce the {\it minimum signal strength}  by
\[
\tau_p^* = \tau_p^*(\Sigma) = \min_{1 \leq i \leq p} \tau_p(i), \qquad \mbox{where} \quad \tau_p(i) = \tau_p(i; \Sigma) =   \min_{j \in S^{(i)}(\Omega)}  \{|\Omega(i,j)|\}.
\]

%%%%%%%%%%
%%%%%%%%%%
%%%%%%%%%%
We need the following terminology (whenever there is no confusion, we may drop the part ``in row $i$").
\bed
Fix $1 \leq i \leq p$. We call $j$ a signal node (in row $i$)   if $j \neq i$ and  $\Omega(i,j)  \neq 0$, and a noise node (in row $i$) if $j \neq i$ and $\Omega(i,j) = 0$.
\eed
The so-called {\it Lagging Time}  and {\it Energy At Large (EAL)} play a key role in characterizing PCS.
Suppose we apply PCS to estimate the $i$-th row of $\Omega$.

Denote the $k$-th {\it Selecting Time}  for row $i$ by $\hat{m}^{(i)}(k) = \hat{m}^{(i)}(k; X, \Sigma)$,  $1 \leq k \leq |S^{(i)}(\Omega)|$;  this is the index of the stage at which we select a signal node for the $k$-th time. By default, $\hat{m}(0)  = \hat{m}(0; X, \Sigma)  = 0$.
The $k$-th {\it Lagging Time} for row $i$ is then
\[
 \hell^{(i)}(k) = \hell^{(i)}(k; X, \Sigma) =  \hat{m}^{(i)}(k)  - \hat{m}^{(i)}(k-1) -1,  \qquad 1 \leq k \leq |S^{(i)}(\Omega)|.
\]
This is the number of noise nodes PCS recruits between the steps we recruit the $(k-1)$-th and the $k$-th  signal nodes.
Additionally, suppose we are now at the beginning of stage $m$ in the Screen step of PCS, and let $\hs_{m-1}^{(i)}$ be the set of all recruited nodes as before. We say a signal node is ``At Large"  if we have not yet recruited it. The {\it Energy At Large at stage $m$} (for row $i$) is
\[
\hat{E}^{(i)}(m) = \hat{E}^{(i)}(m; X, \Sigma) =  \sum_{j \in (S^{(i)}(\Omega) \setminus \hs_{m-1}^{(i)})} \Omega(i,j)^2,  \qquad 1\leq m\leq p-1.
\]
In the idealized case when we apply PCS to $\Sigma$, Selecting Time,  Lagging Time and EAL reduce to their
non-stochastic counterparts,   denoted correspondingly by
\[
m^{(i)}(k) = m^{(i)}(k; \Sigma),  \;\;\;     \ell^{(i)}(k) = \ell^{(i)}(k; \Sigma), \;\;\;    \mbox{and} \;\;\;  E^{(i)}(m) = E^{(i)}(m;  \Sigma).
\]
Whenever there is no confusion, we may drop the superscript ``$(i)$" for short.

%%%%%%%%%%%%%%%%%
%%%%%%%%%%%%%%%%%
%%%%%%%%%%%%%%%%%
\subsection{Behavior of PCS in the idealized case}
\label{subsec:noiselessPCS}
Consider the idealized case where we have access to all `small-size' principal sub-matrices of $\Sigma$. We wish to investigate how the Screen step of PCS behaves.

Let $\rho_{ij}(S)$ be the partial correlation in \eqref{Definerho}.
In this idealized case, recall that PCS runs as follows.
Initialize with $S_0^{(i)} = \emptyset$.  Suppose the algorithm has run $(m-1)$ steps and has not yet stopped. Let $S_{m-1}^{(i)} = \{j_1, j_2, \ldots, j_{m-1}\}$ be all the nodes recruited (in that order) by far.  At stage $m$, if $\rho_{ij}(S_{m-1}^{(i)}) \neq 0$ for some $j \notin (\{i\} \cup S_{m-1}^{(i)})$,  let $j = j_m$ be the index with the largest value of $|\rho_{ij}(S_{m-1}^{(i)})|$, and update with $S_m^{(i)} = S_{m-1}^{(i)} \cup \{j_m\}$. Otherwise, terminate and let  $S_*^{(i)} = S_{m-1}^{(i)}$.

The key of the analysis lies in  the interesting connection between partial correlations and EAL. The following two lemmas are proved in Section~\ref{sec:proof}.
%%%%%%%%%%%%%%%%%%%%%%%%%
%%%%%%%%%%%%%%%%%%%%%%%%%
\begin{lemma} \label{lem:upperbound}
Fix $p$, $0<c_0\leq 1$,  $1 \leq i, s  \leq p$ and $\Sigma \in {\cal M}_p^*(s, c_0)$.   For each $1\leq k\leq |S^{(i)}(\Omega)|$,
\[
\sum_{m^{(i)}(k-1) < m <m^{(i)}(k) } \rho^2_{ij_m}(S^{(i)}_{m-1}) \leq \big[\mu^{(2)}_{m^{(i)}(k) +s-k}(\Sigma)\big]^2 \sum_{j\in \bigl(S^{(i)}(\Omega)\setminus S^{(i)}_{m^{(i)}(k)-1} \bigr)} \Omega(i,j)^2.
\]
\end{lemma}
%%%%%%%%%%%%%%%%%%%%%%%%%
%%%%%%%%%%%%%%%%%%%%%%%%%
\begin{lemma} \label{lem:lowerbound}
Fix $p$, $0<c_0 <1$,  $1 \leq i, s  \leq p$ and $\Sigma \in {\cal M}_p^*(s,c_0)$.  For each $m\geq 1$,
\[
\sum_{j\in (S^{(i)}(\Omega)\setminus S^{(i)}_{m-1})} \rho^2_{ij}(S^{(i)}_{m-1}) \geq \frac{[\mu^{(1)}_{m+s}(\Sigma)]^3}{\mu^{(2)}_{m+s}(\Sigma)} \sum_{j\in (S^{(i)}(\Omega)\setminus S^{(i)}_{m-1})} \Omega(i,j)^2.
\]
\end{lemma}
Recall that $\Sigma \in {\cal M}_p^*(s, c_0)$, so $\mu^{(1)}_{m+s}(\Sigma)\geq C$ and $\mu^{(2)}_{m+s}(\Sigma)\leq C$.
Fix $1 \leq k \leq |S^{(i)}(\Omega)|$.  Suppose we have recruited $(k-1)$ signals and $(|S^{(i)}(\Omega)| - k + 1)$ ones are at large. The implications of these lemmas are:
\begin{itemize}
\item The sum of squares of all such partial correlations associated with noise nodes we recruit between the $(k-1)$-th and the $k$-th Selecting Times is smaller than a constant $C$ times the EAL associated with the signal nodes that are currently At Large.
\item PCS is a greedy algorithm.  For each noise node we recruit between the $(k-1)$-th  and the $k$-th Selecting Times, the square of the associated partial correlation is no smaller than that of one of the signal nodes At Large,  which in turn is greater than  $C (|S^{(i)}(\Omega)| - k + 1)^{-1}$ times the EAL associated with all signal nodes that are currently At Large.
\item As a result,  the $k$-th Lagging Time satisfies
$\ell^{(i)}(k; \Sigma) \leq C (|S^{(i)}(\Omega)| -  k+1)  \leq C (s - k +1)$, and PCS  must have recruited all true signal nodes in no more than
$C \sum_{k = 1}^s (s - k+1) \leq C s^2$
steps, at which point, all partial correlations are $0$ and the algorithm stops immediately.
\end{itemize}

The above arguments are made precise in the following theorem, the proof of which can be found in Section~\ref{sec:proof}.
%The proof follows directly from the above observations, so we omit it.
%%%%%%%%%%%%%
%%%%%%%%%%%%%
\begin{thm} \label{thm:idealPCS}
Suppose $\Sigma\in \mathcal{M}_p^*(s,c_0)$, and $s^2\log(p)=o(n)$. In the idealized case that we can access all principal sub-matrices of $\Sigma$ with size no more than $N(p,n)$ defined in (\ref{DefineN}), for each row $1\leq i\leq p$, the following holds:
\begin{itemize}
\item At each stage $m$ before all signal nodes are recruited, there exists $j\in (S^{(i)}(\Omega)\setminus S^{(i)}_{m-1})$ such that $|\rho_{ij}(S^{(i)}_{m-1})|\geq C\tau_p^*$, and PCS keeps running.
\item PCS takes no more than $C s^2$ steps to terminate.
\item When PCS terminates, $\rho_{ij}(S^{(i)}_*) = 0$ for all $j \notin (\{i\} \cup S^{(i)}_*)$.
\end{itemize}
\end{thm}

%%%%%%%%%%%
%%%%%%%%%%%
%%%%%%%%%%%
\subsection{Stochastic fluctuations, consistency of PCS}
\label{subsec:noisePCS}
In this section, we aim to extend Theorem \ref{thm:idealPCS} to the real case where we have access to small principal sub-matrices of $\hat{\Sigma}$ instead of $\Sigma$. Recall that in the Screen step of PCS, we use the threshold
\beq \label{threshold}
t_q^* =  t_q^*(p,n) = q\cdot \sqrt{2\log(p)/n}.
\eeq
We hope that there is a $q>0$ such that except for a negligible probability,
\begin{itemize}
\item The algorithm stops at no more than $C s^2$ steps.
\item Suppose we are at stage $m$ of the Screen step of PCS.  If the algorithm has not yet recruited all the signal nodes by stage $(m-1)$, then  there is a $j \notin (\{i\} \cup \hs_{m-1}^{(i)})$ such that
$|\hat{\rho}_{ij}(\hs_{m-1}^{(i)})|  \geq t^*_q(p,n)$.
If the algorithm has recruited all signal nodes by stage $(m-1)$, then for all $j \notin (\{i\} \cup \hs_{m-1}^{(i)})$,
$|\hat{\rho}_{ij}(\hs_{m-1}^{(i)})| < t^*_q(p,n)$.
\end{itemize}
Such a `phase transition' effect ensures PCS to run till all signal nodes are recruited.

The key is to characterize the stochastic fluctuations. Under mild conditions, we can show that except for a probability of $o(p^{-3})$, there is a constant $c_1$ that only depends on $c_0$ in (\ref{DefinecalM}) such that  for each $m\geq 1$,
\beq \label{stochastic}
\max_{j\notin (\{i\}\cup \hs^{(i)}_{m-1})} |\hat{\rho}_{ij}(\hs_{m-1}^{(i)})-\rho_{ij}(\hs_{m-1}^{(i)})| \leq c_1 s\sqrt{2\log(p)/n}.
\eeq
We need the minimum signal strength to be large enough to counter the effect of stochastic fluctuations.  In light of this, we assume
\beq \label{condition1}
\tau_p^* / [s \sqrt{\log(p) / n}] \to\infty,
\eeq
and assume that
\beq \label{threshrange}
2 c_1 s\sqrt{2 \log(p) /n} \leq t_q^*(p,n) \leq    (1/2) c_0^2 \tau_p^*,
\eeq
where $c_0$ is as in \eqref{DefinecalM}.
The constants $2$ and $1/2$ are chose for convenience and can be replaced by any constants $a > 1$ and $b \in (0,1)$, respectively.
When \eqref{condition1}-\eqref{threshrange} hold, we
we are able to derive results similar to those in Lemmas \ref{lem:upperbound}-\ref{lem:lowerbound}, which can then be used to derive the `phase transitional' phenomenon aforementioned.  Roughly saying, with high probability:
if all signal nodes have not yet been recruited by stage $(m-1)$, then the partial correlation associated with the next node to be recruited is at least $C\tau_p^* -c_1s\sqrt{2\log(p)/n}$ which is much larger than the threshold $t_q^*$ and so PCS continues to run. On the other hand,
once all signal nodes are recruited, the partial correlation associated with all remaining nodes fall below  $c_1 s\sqrt{2\log(p)/n}$ which is no larger than $t^*_q/2$, and PCS stops immediately.

The above arguments are made precise in the following theorem, which is the main result of this paper and proved in Section~\ref{sec:proof}.
%%%%%%%%%%%%%%%%%%%%%
%%%%%%%%%%%%%%%%%%%%%
\begin{thm} \label{thm:main}
Fix $1 \leq i \leq p$ and apply the Screen step of PCS to row $i$. Suppose $\Sigma\in \mathcal{M}_p^*(s,c_0)$, $s^2\log(p)=o(n)$, the minimum signal strength $\tau_p^*$ satisfies (\ref{condition1}), and
  the threshold $t_q^*(p,n)$ satisfies (\ref{threshrange}) with the constant  $c_1$  properly large.   With probability at least $1-o(p^{-3})$:
\begin{itemize}
\item At each stage $m$ before all signal nodes are recruited, there exists $j\in (S^{(i)}(\Omega)\setminus \hs^{(i)}_{m-1})$ such that $|\hat{\rho}_{ij}(\hs^{(i)}_{m-1})|\gtrsim c_0^2 \tau_p^*$, and PCS keeps running.
\item PCS takes no more than $C s^2$ steps to terminate.
\item Once PCS recruits the last signal node, it stops immediately, at which point, $|\hrho_{ij}(\hat{S}_*^{(i)})| \leq c_1s \sqrt{2\log(p)/n}$
for all $j  \notin (\{i\} \cup \hat{S}_*^{(i)})$.
\end{itemize}
\end{thm}
An explicit formula for $c_1$ can be worked out but is rather tedious; see the proofs of Theorems \ref{thm:main}-\ref{thm:estimation} for details.
The first two claims of the theorem are still valid if  $t_q^*(p,n) \asymp  s \sqrt{2\log(p)/n}$ but $t_q^*(p,n) \leq c_1 s \sqrt{2\log(p)/n}$.
In such a case, the difference is that, PCS may continue to run for finitely many steps (without immediate termination)  after all signals are recruited.

{\bf Remark}. We can slightly relax the condition \eqref{condition1} by allowing
$\tau_p^*  \sim r  \cdot  s\sqrt{\log(p)/n}$ for some constant $r > 0$.   In this case, there exists a constant $r^*$ that only depends on $c_0$ such that whenever $r>r^*$, we can find constants $\underline{c}=\underline{c}(c_0,r)$ and $\overline{c}=\overline{c}(c_0,r)$, so that Theorem~\ref{thm:main} continues to hold when $\underline{c}\leq q\leq \overline{c}$.
Furthermore, if we only want the first two claims of Theorem~\ref{thm:main} to hold, we do not need the lower bound $\underline{c}$ for $q$.

Theorem \ref{thm:main} discusses  the Screen step of the PCS for individual rows.
The following theorem characterizes properties of the estimator $\hat{\Omega}^{pcs}=\hat{\Omega}^{pcs}(t_q^*, X; p,n)$, and is proved in Section~\ref{sec:proof}.
%%%%%%%%%%%%%%%%%%%%%
%%%%%%%%%%%%%%%%%%%%%
\begin{thm} \label{thm:estimation}
Under conditions of Theorem~\ref{thm:main}, with probability at least $1-o(p^{-2})$,
each row of $\hat{\Omega}^{pcs}$ has the same support as the corresponding row of $\Omega$, and $\|\hat{\Omega}^{pcs}-\Omega\|_{\max}\leq C\sqrt{\log{(p)}/n}$.
\end{thm}
%%%%%%%%%%%%%%%%%%%%%%
%%%%%%%%%%%%%%%%%%%%%
While Theorem \ref{thm:estimation} is for $\|\hat{\Omega}^{pcs}-\Omega\|_{max}$, the results can be extended to accommodate other types of matrix norms (e.g., $\|\hat{\Omega}^{pcs}-\Omega\|_1$).

{\bf Remark}.  In Theorems~\ref{thm:idealPCS}-\ref{thm:main}, we use the lower bound $(s-k)(\tau_p^*)^2$ for the EAL associated with signals that are At Large between the $(k-1)$-th and $k$-th Selecting Time.
Such a bound is not tight, especially when a few smallest
nonzero entries are much smaller than other nonzero entries (in magnitude).
Here is a better bound. Suppose row $i$ has $s$ off-diagonal nonzeros, denoted as $\eta_1, \cdots, \eta_s$. We sort
$\eta_j^2$ in the ascending order: $\eta_{(1)}^2 \leq \eta_{(2)}^2 \leq \ldots \leq \eta_{(s)}^2$.
Then, the EAL is lower bounded by $\sum_{\ell = 1}^{s-k} \eta_{(\ell)}^2$.
Such a bound can help relax the condition \eqref{condition1} for Theorem~\ref{thm:main}, especially when our goal is not to show exact support recovery, but to control the number of signal nodes not recruited in the Screen step.

{\bf Remark}.  We control the stochastic fluctuations \eqref{stochastic} by showing that for each $1\leq i\leq p$ and $m\ll N$, with probability at least $1-o(p^{-3})$,
\beq  \label{specnorm}
\| \hat{\Sigma}^{W,W} - \Sigma^{W,W}\|\leq C\sqrt{m\log(p)/n}, \qquad\mbox{where}\quad W=\{i\}\cup \hs_{m-1}^{(i)}.
\eeq
If we replace $\hs_{m-1}^{(i)}$ by a fixed subset $S$ with $|S| = m-1$, then
by basics in multivariate analysis, the factor $\sqrt{m}$ on the right hand side
can be removed.
In general, if we can find an upper bound for the number of possible realizations of
$\hs_{m-1}^{(i)}$, say, $K(p, m)$, then we can replace $\sqrt{m}$ by $\sqrt{\log(K(p,m))}$.
In \eqref{specnorm},  $K(p, m) = {p \choose m}$  which is the most conservative bound.
How to find a tighter bound for $K(p, m)$ is a difficult problem \cite{ZhangYao}.
We conjecture that in a broad situation, a better bound is possible  so \eqref{specnorm} can be much improved.

At the same time, if we are willing to impose further conditions on $\Sigma$, then such a tighter bound is possible; we investigate this in Section \ref{subsec:PCSalt}.

%%%%%%%%%%%%%%%%%%%
%%%%%%%%%%%%%%%%%%%
\subsection{Consistency of PCS for much weaker signals}
\label{subsec:PCSalt}
In the above results, in order for PCS to be successful, we need $\tau^*_p \gg s\sqrt{\log(p)/n}$. We wish to relax this condition by considering
\beq \label{condition3}
\tau_p^*\geq r \cdot \sqrt{2\log(p)/n}, \qquad \mbox{where $r > 0$ is a fixed constant}.
\eeq
We show  PCS works in such cases if  we put additional conditions
on $\Sigma$.  Let
\[
\kappa^* = \kappa^*(\Sigma) = \max_{1 \leq i \leq p} \kappa(i, \Sigma)
\]
\[
\gamma^* = \gamma^*(\Sigma) = \min_{1 \leq i \leq p} \gamma(i, \Sigma),
\]
where
\[
\kappa(i; \Sigma)    =  \max_{j \notin \{i\} \cup S^{(i)}(\Omega)}  \|(\Sigma^{S^{(i)}(\Omega), S^{(i)}(\Omega)})^{-1} 
\Sigma^{S^{(i)}(\Omega), \{j\}}\|_1
\]
and
\[
\gamma(i;\Sigma)   = \min_{j \notin (\{i\}\cup S^{(i)}(\Omega))} \bigl\{ [ \mbox{first diagonal of $(\Sigma^{\{j\} \cup S^{(i)}(\Omega), \{j\} \cup S^{(i)}(\Omega)})^{-1}$} ]^{1/2} \bigr\};
\]
here,  we always assume $j$ as the first index listed in $\{j\} \cup S^{(i)}(\Omega)$.
The quantity $\kappa^*$ is motivated by a similar quantity in \cite{ZhangGreedy} for linear regressions, and $\gamma^*$ is a normalizing factor which comes from the definition of partial correlations.
Fix a constant $\delta\in (0,1)$. In this sub-section, we assume
\beq \label{condition4}
\kappa^*(\Sigma)/\gamma^*(\Sigma)  \leq 1-\delta, \qquad  \Sigma(i,i)=1, \qquad 1\leq i\leq p;
\eeq
the second assumption is only for simplicity in presentation.
Introduce 
\[
\theta^*=\theta^*(\Sigma)=\min_{1\leq i\leq p}\theta(i,\Sigma),
\]
where 
\[
\theta(i,\Sigma)=\lambda_{\min}(\Sigma^{S^{(i)}(\Omega),S^{(i)}(\Omega)}).
\]
The following theorem is proved in Section~\ref{sec:proof}.
%%%%%%%%%%%%%%%%%%%%%
%%%%%%%%%%%%%%%%%%%%%
\begin{thm} \label{thm:greedy}
Fix $1 \leq i \leq p$ and apply the Screen step of PCS to row $i$.
Suppose $\Sigma\in \mathcal{M}_p^*(s,c_0)$,  (\ref{condition4}) holds for some $\delta\in (0,1)$, $s^2\log(p)=o(n)$, the minimum signal strength $\tau_p^*$ satisfies (\ref{condition3}) with
$r \geq \sqrt{5\Omega(i,i)}[\theta^*(\Sigma)]^{-1}$ $\max\big\{ \sqrt{\theta^*(\Sigma)} + 2\delta^{-1},  2\sqrt{\Omega(i,i)}\big\}$,
and the threshold $t_q^*(p,n)$ satisfies $\sqrt{5}<q\leq \theta^*(\Sigma)r/\Omega(i,i)$. With probability at least $1-o(p^{-3})$,
\begin{itemize}
\item Before all signal nodes are recruited, PCS keeps running and recruits a signal node at each step.
\item PCS takes exactly $|S^{(i)}(\Omega)|$ steps to terminate.
\item When PCS stops, $|\hat{\rho}_{ij}(\hs^{(i)}_*)|\leq \sqrt{10\log(p)/n}$ for all $j \notin (\{i\} \cup \hs^{(i)}_*)$.
\end{itemize}
\end{thm}
%%%%%%%%%%%%%%%%%%%%%%
%%%%%%%%%%%%%%%%%%%%%
By Theorem~\ref{thm:greedy}, the claim of   Theorem \ref{thm:estimation} continues to hold, the proof of which is
straightforward so we omit it.
In Theorem, \ref{thm:greedy},  we require $r \geq \sqrt{5\Omega(i,i)}[\theta^*(\Sigma)]^{-1} (\sqrt{\theta^*(\Sigma)} + 2\delta^{-1})$ and $r \geq   2 \sqrt{5\Omega(i,i)}[\theta^*(\Sigma)]^{-1} \sqrt{\Omega(i,i)}\big\}$. The first condition
ensures  that PCS always recruits signal nodes before termination. The second one ensures the existence of a threshold by which PCS terminates immediately once all signals are recruited.
%%Regarding that PCS has a Clean step, the second term can be largely removed.

%%%%%%%%%%%%%%%%%%%
%%%%%%%%%%%%%%%%%%%
\subsection{Optimal classification phase diagram  by HCT-PCS}
Combe back to model (\ref{classificationmodel}) where
$\tilde{X}_i \sim N(\mu^{\pm}, \Sigma)$,  $\Omega = \Sigma^{-1}$,  if $Y_i = \pm 1$, respectively.
In this model, the optimality of HCT was justified carefully in \cite{DJ08, FJY} in theory.
At the heart of the theoretical framework is the notion of {\it classification phase diagram}.
Call the two-dimensional space calibrating the signal sparsity (fraction of nonzeros in the contrast mean vector $(\mu^+-\mu^-)$) and signal strength (minimum magnitudes of the nonzero contrast mean entries) the {\it phase space}. The phase diagram is a partition of the phase space into three sub-regions, where
successful classification is {\it relatively easy}, {\it possible but relatively hard}, and {\it impossible}  simply because the signals are too rare and weak.

We say a trained classifier achieves the optimal phase diagram if it partitions the phase space in exactly the same way as the optimal classifier does. It was shown in \cite[Theorm 1.1-1.3]{FJY} that HCT achieves the optimal phase diagram (with some additional regularity conditions) provided that
\begin{itemize}
\item $\Omega$ is $s_p$-sparse, where $s_p \leq L_p$.
\item $\Omega$ is known, or can be estimated by $\hat{\Omega}$ such that $\|\hat{\Omega}  - \Omega\|_{\max} \leq L_p / \sqrt{n}$.
\end{itemize}
Here, $L_p > 0$ is a generic multi-$\log(p)$ term such that for any constant $c > 0$, $L_p p^{-c} \goto 0$ and $L_p p^{c} \goto \infty$.

We now consider HCT-PCS. By results in Sections~\ref{subsec:noisePCS}-\ref{subsec:PCSalt}, we have shown
\beq
\|\hat{\Omega}^{pcs}-\Omega\|_{\max}\leq C\sqrt{\log(p)/n}.
\eeq
Therefore, HCT-PCS achieves the optimal phase diagrams in classification,  provided that $s_p \leq L_p$.  See   \cite{FJY} for details.

Note that the condition on $s_p$ is relatively strict here. For much larger $s_p$ (e.g., $s_p  = p^{\vartheta}$ for some constant $0<\vartheta <1$), it remains unknown which procedures achieve the optimal phase diagram, even when $\Omega$ is known.

%%%%%%%%%%%%%%%
%%%%%%%%%%%%%%%
\subsection{Comparisons with other methods}
\label{subsec:comptheory}
There are
some existing theoretical results on exact support recovery of the precision matrix,
including but are not limited to those on the glasso \cite{BinYu},
CLIME \cite{CLIME}, and scaled-lasso \cite{slasso}.

For exact support recovery,  the glasso requires the  so-called  ``Incoherent Conditions" (IC) \cite{BinYu}.
The IC condition is relatively restrictive, which can be illustrated by the following simple example.
Suppose $p$ is divisible by $3$, and
$\Omega$ is block-wise diagonal where each diagonal
block is a symmetric matrix $D \in R^{3,3}$ satisfying $D(1,1) = D(2,2)  = 1, D(1,2) = 0, D(1,3)=a, D(2,3) = b$, and $D(3,3) = c$, where
$c^2 > a^2 + b^2$ so $D$ is positive definite.
In this example, the IC condition imposes a restriction $|a|+|b|+2|ab|< 1$.

The conditions required for CLIME to achieve the exact support recovery is given in
\cite{CLIME}, which in our notations can be roughly translated into
that the minimum off-diagonals of $\Omega$ are no smaller than $C \|\Omega\|_1 /\sqrt{n}$.
Such a condition overlaps with ours in many cases. It also covers some cases we do not cover,
but the other way around is also true.  As for scaled-lasso, note that the primary interest in \cite{slasso} is on
the convergence in terms of the matrix spectral norm,  where conditions for exact support recovery
are not given.

Note that the largest advantage of PCS is that, it allows for real-time computing for very large matrices, and has nice results in real data analysis.

The method in \cite{BLT} and FoBa \cite{ZhangGreedy, ZhangFoBa} are also related. However, the main results of \cite{BLT} is on the case where  $\Sigma$ is sparse. Since the primary interest here is on the case where $\Omega$ is sparse, their
results do not directly apply. The results in \cite{ZhangGreedy, ZhangFoBa} are on variable selection, and have not yet been adapted to precision matrix estimation. Recall that  in Section \ref{subsec:foba},  we have already carefully compared PCS with FoBa, from the perspective of  real data applications:  PCS is different from FoBa in  philosophy, method and implementation, and yields much better classification results.

The results (numerical and theoretical) presented in this paper suggest
that PCS is an interesting procedure and is worthy of future exploration. In particular, we believe that, with some technical advancements in proofs, the conditions required for the success of PCS can be largely weakened.

%%%%%%%%%%%%
%%%%%%%%%%%%
%%%%%%%%%%%%
\section{Simulations}
\label{sec:Simul}
We conducted simulation studies under various data set configurations, to assess the behavior of PCS in estimating the precision matrix and its performance in classification.  The first experiment consists of three sub-experiments, where we  compare PCS with other methods including the glasso and FoBa in estimating the precision matrix. In the second experiment, we focus on classification and compare HCT-PCS with other classifiers including HCT-Foba, HCT-glasso, nHT, SVM and RF.

%%%%%%%%%%%
%%%%%%%%%%%
%%%%%%%%%%%
\subsection{Experiment 1 (precision matrix estimation)}
\label{subsec:Simul1}
Experiment $1$ consists of three sub-experiments, 1a--1c.
In each sub-experiment, we generate samples $X_i \overset{iid}{\sim}(0,\Omega^{-1})$, $i=1,2,\ldots,n$, where $\Omega \in R^{p,p}$, for $10$ repetitions. For any $\hat{\Omega}$, an estimate of $\Omega$, we measure the
performance by the average errors across $10$ different repetitions. We use four different error measures:  spectrum norm, Frobenius norm,  and the matrix $\ell^1$-norm of  $(\hat{\Omega} -  \Omega)$, and the matrix Hamming distance between $\hat{\Omega}$ and $\Omega$.  The spectral norm, Frobenius norm, and the matrix $\ell^1$-norm are as in textbooks. The matrix Hamming distance between $\hat{\Omega}$ and $\Omega$ is
\begin{equation} \label{Defineell0}
\hamm_p(\hat{\Omega}, \Omega) = \frac{1}{p}  \sum_{1 \leq i, j \leq p} 1\{\sgn(|\hat{\Omega}(i,j)|)  \neq \sgn(|\Omega(i,j)|)\},
\end{equation}
where $\sgn(x) = 1$ if $x > 0$ and $\sgn(x) = 0$ if $x = 0$.  Alternatively, we can replace
the factor $p^{-1}$ by $1$ or $p^{-2}$, but the resultant values would be either too large or too small; the current one is the best   for presentations.

In these experiments, matrix singularity is not as extreme as in the microarray data, so we use PCS and FoBa without the ridge regularization.

For PCS, we take the tuning parameter $L$ to be $15$ in experiments 1a--1b for  the algorithm generally stops after 10 steps due to the simple structure of $\Omega$. In experiment 1c, we use $L=30$ because $\Omega$ is more complex. For tuning parameter $q$, we test $q$ from 0.5 to 3 with increment of 0.5 in experiment 1a and 1b. We find that for $q\leq 0.5$ or $q\geq 2.5$, the errors are higher, while the errors remain similar for $1\leq q \leq2$, so we use $q=1.5$. In experiment 1c, we test $q$ from 0.25 to 2 with increment of 0.25.  We find that for $q\leq 0.25$ or $q\geq 1.5$, the errors are higher than $0.5\leq q \leq 1.25$, so we use $q=0.75$.  For FoBa, we set $L$ the same as in PCS.

For the glasso, we set  the tuning parameter  $\lambda$ as  $0.5$. To finish all $10$ repetitions,  it takes about $10$  hours for experiments
1a-1b  and more than $24$ hours  for experiment 1c, and so
we do not consider $\lambda$ smaller than $0.5$.

We now describe $\Omega$ in three sub-experiments. For  experiments 1a and 1c, we set   $(p,n) = (5000,1000),  (2000,1000), (1000,500)$.  For experiment 1b, we set $(p,n) =   (4500,1000)$, $(3000,1000)$, $(1500,500)$ so that $p$ is divisible by $3$.

{\it Experiment 1a}:  $\Omega(i,j) = 1\{i = j\} + \rho \cdot 1\{ |i - j| = 1\}$, $\rho=0.4$, $1 \leq i, j \leq p$.
    Here, the IC condition (see Section~\ref{subsec:comptheory}) for the glasso holds, but no longer holds if we increase $\rho$ slightly.

{\it Experiment 1b}: $\Omega$ is a block-wise diagonal matrix, and
each diagonal block is a $3 \times 3$ symmetric matrix $A$ satisfying $A(1,1) = A(2,2) = A(3,3) =1$, $A(1,2) = 0$,  $A(1,3) = 0.5$, and $A(2,3) = 0.7$.  This matrix $\Omega$ is positive definite but does not satisfy the IC condition.

{\it Experiment 1c}: We generate  $\Omega$ as follows. First, we generate a $p \times p$  Wigner matrix $W$ \cite{Vershynin} (the symmetric matrix with $0$ on all the diagonals and iid $\mathrm{Bernoulli}(\eps)$ random variables for entries on the upper triangle; here we take $\eps=.01$). Next, we let $\Omega^* = .5 W + \vartheta I_p$, where $I_p$ is the  $p \times p$ identity matrix and $\vartheta=\vartheta(W)$ is such that the conditional number of $\Omega^*$ (the ratio of the maximal and the minimal singular values) is $p$. Last, we scale $\Omega^*$ to have unit diagonals and let $\Omega$ be the resultant matrix.

%%%%%%%%%%%%%
%%%%%%%%%%%%%
%%%%%%%%%%%%%
%%%%%%%%%%%%%
\begin{table}[htb]
\caption{Estimation errors (with standard deviations in brackets) for Experiment 1a.}
\vspace{0.05 in} 
\centering 
\scalebox{0.88}{
\begin{tabular}{ccccccccccc}
\hline
%\hline
%\multicolumn{9}{c}{Experience 1a} \\
\hline
    & & \multicolumn{3}{c}{Spectrum norm} &
      & \multicolumn{3}{c}{Matrix $\ell^1$-norm} \\
      \cline{3-5} \cline{7-9}
$p$ & $n$ & PCS & glasso & FoBa && PCS & glasso & FoBa\\
\hline
5000 & 1000 & 0.27(0.021) & 1.19(0.003) & 0.78(0.014)    &&  0.34(0.033) & 1.23(0.003) & 2.45(0.070) \\
2000 & 1000 & 0.26(0.027) & 1.18(0.003) & 0.70(0.018)    &&  0.34(0.035) & 1.23(0.005) & 2.18(0.107)\\
1000 & 500  & 0.34(0.033) & 1.19(0.003) & 1.14(0.025)    &&  0.45(0.051) & 1.24(0.004) & 3.24(0.174)\\
\hline
\hline
    & & \multicolumn{3}{c}{Frobenius norm} &
      & \multicolumn{3}{c}{Matrix Hamming distance} \\
      \cline{3-5} \cline{7-9}
$p$ & $n$ & PCS & glasso & FoBa && PCS & glasso & FoBa\\
\hline
5000 & 1000 & 4.39(0.057) & 49.00(0.013) & 23.08(0.067) && 0.00(0.000) & 0.00(0.001) & 24.93(0.021)\\ %  15.20(4.733) & 124652(106.995)\\
2000 & 1000 & 2.79(0.036) & 30.99(0.012) & 13.03(0.037) && 0.00(0.000) & 0.00(0.002) & 24.43(0.030) \\% 5.00(3.916) & 48860.60(60.289) \\
1000 & 500  & 2.83(0.099) & 21.91(0.019) & 14.89(0.094) && 0.00(0.000) & 0.04(0.011) & 24.12(0.038)\\% 43.60(11.306) & 24116.80(37.750)\\
\hline
\end{tabular}}
\label{table:sim1}
\end{table}
%%%%%%%%%
%%%%%%%%%
%%%%%%%%%
\begin{table}[htb]
\caption{Estimation errors (with standard deviations in brackets) for Experiment 1b.}
\vspace{0.05 in} 
\scalebox{0.88}{
\begin{tabular}{ccccccccccc}
\hline
%\hline
%\multicolumn{9}{c}{Experience 1b} \\
\hline
    & & \multicolumn{3}{c}{Spectrum norm} &
      & \multicolumn{3}{c}{Matrix $\ell^1$-norm} \\
      \cline{3-5} \cline{7-9}
$p$ & $n$ & PCS & glasso & FoBa && PCS & glasso & FoBa\\
\hline
4500 & 1000 & 0.30(0.022) & 1.23(0.004)  & 0.73(0.017)  && 0.36(0.035) & 1.49(0.005) & 2.13(0.043) \\
3000 & 1000 & 0.29(0.027) & 1.22(0.004)  & 0.70(0.018)  && 0.35(0.039) & 1.49(0.006) & 2.01(0.047) \\
1500 & 500  & 0.36(0.018) & 1.23(0.003)  & 1.15(0.022)  && 0.43(0.028) & 1.51(0.006) & 3.22(0.239) \\
\hline
\hline
    & & \multicolumn{3}{c}{Frobenius norm} &
      & \multicolumn{3}{c}{Matrix Hamming distance} \\
      \cline{3-5} \cline{7-9}
$p$ & $n$ & PCS & glasso & FoBa && PCS & glasso & FoBa\\
\hline
4500 & 1000 & 3.99(0.059) & 49.83(0.009) & 19.47(0.069)  && 0.00(0.000) & 0.69(0.002)& 26.90(0.019)\\%3101.1(10.482) &  121056.2(85.637)\\
3000 & 1000 & 3.25(0.029) & 40.69(0.008) & 14.96(0.059)  && 0.00(0.000) & 0.68(0.003) & 26.78(0.019) \\%2046.8(9.531)  121056.2(85.637) & 80346(56.757)\\
1500 & 500  & 3.30(0.092) & 28.76(0.013) & 17.95(0.118)  && 0.00(0.000) & 1.47(0.044) & 26.58(0.035)\\%& 2205.6(65.901) & 39867.6(52.871)\\
\hline
\end{tabular}}
\label{table:sim2}
\end{table}

The results for three experiments are summarized in Tables \ref{table:sim1}, \ref{table:sim2}, and \ref{table:sim3}, correspondingly,  in terms of four error measures aforementioned.
For experiments 1a--1b, it suggests that (a) PCS outperforms the glasso and FoBa in all four different error measures, especially in terms of the Hamming distance, where PCS has $0$ Hamming distance in all cases (and thus exact support recovery of $\Omega$);
(b) the glasso and FoBa
have similar performance in terms of the $\ell^1$-norm and Hamming distance,
but the glasso is significantly inferior to FoBa in terms of
the spectral norm and Frobenius norm.
For  experiment 1c, Table \ref{table:sim3} shows that the glasso  is not that competitive to FoBa as in the previous two experiments, while PCS still has a dominant advantage over the glasso  and FoBa when both $p$ and $n$ get larger, especially in terms of the Hamming distance.
%%%%%%%%%%%
%%%%%%%%%%%
%%%%%%%%%%%
\begin{table}[htb]
\caption{Estimation errors (with standard deviations in brackets) for Experiment 1c.}
\vspace{0.05 in} 
\centering 
\scalebox{0.85}{
\begin{tabular}{ccccccccccc}
\hline
%\hline
%\multicolumn{9}{c}{Experiment 1c($\alpha=0.01,\rho=0.5$)} \\
\hline
    & & \multicolumn{3}{c}{Spectrum norm} &
      & \multicolumn{3}{c}{Matrix $\ell^1$-norm} \\
      \cline{3-5} \cline{7-9}
$p$ & $n$ & PCS & glasso & FoBa && PCS & glasso & FoBa\\
\hline
5000 & 1000 &  2.56(0.009) & 4.13(0.002) & 4.47(0.000)  && 5.05(0.149) &13.50(0.675) & 6.29(0.056) \\
2000 & 1000 &  0.53(0.009) & 2.79(0.001) & 3.11(0.000)  && 1.84(0.097) & 7.25(0.221) & 5.42(0.013) \\
1000 & 500  &  1.00(0.068) & 2.13(0.001) & 2.41(0.001)  && 3.25(0.305) & 13.27(0.249)& 4.11(0.009) \\
\hline
\hline
    & & \multicolumn{3}{c}{Frobenius norm} &
      & \multicolumn{3}{c}{Matrix Hamming distance} \\
      \cline{3-5} \cline{7-9}
$p$ & $n$ & PCS & glasso & FoBa && PCS & glasso & FoBa\\
\hline
5000 & 1000 & 35.46(0.068) & 55.41(0.025) & 72.18(0.004) &&35.22(0.103) & 69.68(0.668) & 55.80(0.037)\\%176080(516.296)&348401.9(3339.144)& 279016.4(185.462) \\
2000 & 1000 & 10.61(0.054) & 33.88(0.020) & 43.63(0.003) && 5.89(0.067) & 34.60(0.295) & 25.94(0.077)\\%11778(133.179)  & 69191(589.005)  & 51878.8(153.800)\\
1000 & 500  & 10.65(0.074) & 22.34(0.013) & 29.23(0.003) && 8.29(0.085) & 14.70(0.093) & 15.27(0.132)\\%8291(84.985)   & 14699.4(92.679)  & 15268(131.626)\\
\hline
\end{tabular}}
\label{table:sim3}
\end{table}

%%%%%%%%%%%%
%%%%%%%%%%%%
%%%%%%%%%%%%
\subsection{Experiment 2 (classification)}
\label{subsec:Simul2}

In this experiment, we  take $\Omega$ to be the tri-diagonal matrix as in experiment 1a, calibrated by the parameter $\rho$.  Also, following \cite{FJY},  we consider the most challenging ``rare and weak" setting
where the contrast mean vector $\mu$ only has a small fraction of nonzeros and the nonzeros are
individually small.
In detail,  let $\nu_a$ be the point mass at $a$. For two numbers $(\eps_p, \tau_p)$ that may depend on $p$,   we generate the scaled vector $\sqrt{n} \mu$
from the mixture of two point masses:
$\sqrt{n} \mu(j) \stackrel{iid}{\sim}  (1 - \eps_p) \nu_0 + \eps_p   \nu_{\tau_p}$.

In this experiment, we take $(p, n, \rho,  \eps_p, \tau_p)=(5000, 1000, 0.4, 0.1, 3.5 )$. For $(\mu, \Omega)$ generated as above,  the simulation contains the following main steps:
\begin{enumerate}
\item  Generate $n$ samples $(\tilde{X}_i, Y_i)$, $1 \leq i \leq n$,  by letting $Y_i = 1$ for $i \leq n/2$ and $Y_i = -1$  for $i > n/2$, and $\tilde{X}_i \sim N(Y_i  \cdot \mu, \Omega^{-1})$.
\item  Split the $n$ samples into training and test sets by following exactly the same procedure in Section 1.4. The only difference is that we use $10$ data splitting and $10$ cv-splitting here.
\item  Use the training set to build all classifiers (HCT-PCS, HCT-FoBa, HCT-glasso,  nHCT, SVM and RF), apply them to the test set, and then record the test errors.
\end{enumerate}
%%%%%%%%%%
%%%%%%%%%%
%%%%%%%%%%
\begin{table}[hb!]
\caption{Comparison of classification errors for Experiment 2 (based on 10 independent data-splitting).
Numbers in each cell are the percentages (e.g., $11.08$ means  $11.08\%$)} 
\vspace{0.05 in}
\centering  
\scalebox{1}{
\begin{tabular}{c|cccccc}
\hline
\hline
       & HCT-PCS  &  HCT-Foba  &  HCT-glasso & nHCT  &    SVM    & RF     \\
 \hline
average error  &  11.08  & 12.11  & 43.67  & 32.02 &    20.03 & 35.51       \\
`best' error          &  8.73   & 10.54  & 37.95  & 21.99 &    18.98 & 31.93       \\
\hline
\end{tabular}
}
\label{table:simclass}
\end{table}

The results are summarized in Table \ref{table:simclass} in terms of both the average error across $10$ data splitting and the minimum error in $10$ data splitting. It suggests that HCT-PCS outperforms other HC-based classifiers; in particular, HCT-PCS significantly outperforms nHCT and HCT-glasso. In addition, both SVM and RF are less competitive compared to HCT-PCS. This is consistent with the theoretical results in \cite{FJY},  where it was shown that given a sufficiently accurate estimate of $\Omega$, the HCT classifier has the optimal classification behavior in the ``rare and weak" settings associated with the sparse Gaussian graphical model \eqref{classificationmodel}.

%%%%%%%%%%%%%%
%%%%%%%%%%%%%%
%%%%%%%%%%%%%%
%%%%%%%%%%%%%%
%%%%%%%%%%%%%%
\section{Discussions and extensions} \label{sec:Discu}
This paper is closely related to areas such as  precision matrix estimation, classification, variable selection, and inference on ``rare and weak"  signals, and
has many possible directions for extensions. Below, we mention some of such possibilities.

The precision matrix is a quantity that is useful in many settings.
It can be either the direct quantity of interest (e.g., genetic regulatory networks),  or a quantity that can be used to
improve the results of inferences. Examples include  classical methods of Hotelling's $\chi^2$-test,
discriminant analysis, post-selection inference for linear regressions and the recent work on Innovated Higher Criticism \cite{HJ09}.
In these examples, a good estimate of the precision matrix could largely improve the results of the
inferences. The proposed approach is especially useful for it allows real-time computation for very large
precision matrices.

The theoretical results in the paper can be extended in various directions.
For example, in this paper, we assume $\Omega$ is strictly sparse in the sense that in each row, most of the entries are exactly $0$. Such an assumption can  be  largely relaxed.
 Also, the theoretical results presented in this paper focus on when it is possible to obtain
exact support recovery. The results are extendable to the cases
where we wish to measure the loss by matrix spectral norm or matrix $\ell^2$-norm.
In particular, we mention that if the ultimate goal is for classification, it is not
necessary to fully recover the support of the precision matrix.
A more interesting problem (but more difficult) is  to  study how the
estimation errors in the precision matrix affect the classification results.

PCS needs a threshold tuning parameter $q$ (it also uses the ridge parameter $\delta$
and a maximal step size parameter $L$, which we usually set by $(\delta, L) = (.1, 30)$; PCS
is relatively insensitive to the choices of $(\delta, L)$).
When we use PCS for classification, we determine $q$ by cross validation,
which increases the computation costs by many times.
The same drawback applies to other classifiers, such as HCT-FoBa, HCT-glasso, SVM, and RF.

From both a theoretical and practical perspective,  we wish to have a trained classifier that is tuning free. In Donoho and Jin \cite{DJ08}, we propose
HCT as a tuning free classifier that enjoys optimality, but  unfortunately the method is only applicable to the case where $\Omega$ is known.
How to develop a tuning free optimal classifier for  the case where $\Omega$ is unknown is a very interesting problem. For reasons of space, we leave to future work.

Intellectually, the idea of HCT classification is closely related to \cite{FAIR, ROAD, Ingster}, but
is in different ways, especially on the data-driven threshold choice and studies on the phase diagrams.
The work is also closely related to other development on Higher Criticism. See for example
\cite{Plan, DJ14, JK,  Wellner,  Chen}.

While the primary interest in this paper is on microarray data,  the idea here
 can be extended to other types of data (e.g., the SNP data).
Modern SNP data sets may have many more features (e.g., $p = 250K$)
than a typical microarray data set. While the sheer large size poses great challenges for computation,  we must note that in many of such studies on SNP,
the (population) covariance matrix among different SNPs is banded.
Such a nice feature can help to substantially reduce the computational burden.
How to extend PCS to analysis of such data sets is therefore of great interest.

%%%%%%%%%%
%%%%%%%%%%
%%%%%%%%%%
%%%%%%%%%%
\section{Proofs}  \label{sec:proof}

In this section, we first present some elementary lemmas on basic Random Matrix Theory in Section \ref{subsec:stoch}, and then give the proofs for Lemmas \ref{lemma:objective}, \ref{lem:upperbound}-\ref{lem:lowerbound} and Theorem \ref{thm:idealPCS}-\ref{thm:greedy}. The proof of Lemma \ref{lemma:firstcolumn} is elementary so we omit it. Throughout this section,   $C > 0$ denotes a generic positive constant the value of which may vary from occasion to occasion.
%%%%%%%%%%
%%%%%%%%%%
%%%%%%%%%%

\subsection{Upper bound for stochastic errors} \label{subsec:stoch}

We present some results about controlling the stochastic terms in $(\hat{\Sigma}-\Sigma)$. The key is the following lemma, which is the direct results of  \cite[Remark 5.40]{Vershynin}.
%%%%%%%%%%%%%%%%%%%%%
%%%%%%%%%%%%%%%%%%%%%
\begin{lemma} \label{lem:vershynin}
Fix $p\geq 1$ and $\Sigma\in R^{p,p}$ which is positive definite. Let $A$ be an $n \times k$ random matrix, where each row of $A$ is an independent copy of $N(0,\Sigma)$. There are universal constants $\tilde{c}, c>0$, not depending on $\Sigma$, such that for every $x\geq 0$,  with probability at least $1 - 2\exp(- cx^2)$,
\[
\|(1/p)A'A - \Sigma\|\leq \max(\delta,\delta^2)\cdot \|\Sigma\|, \qquad \delta = \tilde{c} \sqrt{k/n} + x/\sqrt{n}.
\]
\end{lemma}
%%%%%%%%%%%%%%%%%%%%%%%%%
%%%%%%%%%%%%%%%%%%%%%%%%%
The following lemma is frequently used in the proofs. We recall that $\mu^{(1)}_k(A)$ and $\mu^{(2)}_k(\Sigma)$ are defined in \eqref{Definemu}.
%%%%%%%%%%%%%%%%%%%%%%%%%
\begin{lemma} \label{lemma:stoch}
Fix $p\geq 1$ and $\Sigma\in R^{p,p}$ which is positive definite. There exist universal constants $c>0$, not depending on $\Sigma$, such that for each fixed $1\leq i\leq p$ and $1\leq k<n$, with probability at least $1 - 2p^{-4}$,
\begin{itemize}
\item[(a)]  $|\hat{\Sigma}(i,i) - \Sigma(i,i)| \leq c \sqrt{\log(p)/n}$.
\item[(b)]  $\max_{|U|\leq k}\|\hat{\Sigma}^{U,U} - \Sigma^{U,U}\| \leq \mu^{(2)}_k(\Sigma)\cdot c \sqrt{k \log(p)/n}$.
\item[(c)]  $\max_{|U|\leq k}\|(\hat{\Sigma}^{U,U})^{-1} - (\Sigma^{U,U})^{-1}\| \leq [\mu^{(1)}_k(\Sigma)]^{-2} \mu^{(2)}_k(\Sigma) \cdot c \sqrt{k \log(p)/n}$.
\end{itemize}
\end{lemma}
%%%%%%%%%%%%%%%%%%%%%%%%%
%%%%%%%%%%%%%%%%%%%%%%%%%
\noindent{\it Proof}.
(a) is elementary, (b) follows from Lemma~\ref{lem:vershynin} with $x=\sqrt{c^{-1}\log(p^{k+4})}$ and the probability union bound. To show (c), we note that all size-$k$ submatrices are invertible with probability $1$ \cite{Edelman}. Therefore, (c) comes from combining (b) and the equality that $(\hat{\Sigma}^{U,U})^{-1} - (\Sigma^{U,U})^{-1}=(\Sigma^{U,U})^{-1}(\Sigma^{U,U}-\hat{\Sigma}^{U,U})(\hat{\Sigma}^{U,U})^{-1}$.
\qed

By Lemma~\ref{lemma:stoch}, for $\Sigma\in \mathcal{M}_p^*$, there is a constant $C>0$ that depends on $c_0$ such that for each $1\leq k\leq N(p,n)$, with probability at least $1-O(p^{-4})$,
\[
\max_{|U|\leq k}\|\hat{\Sigma}^{U,U} - \Sigma^{U,U}\|\leq C\sqrt{k\log(p)/n},
\]
and
\[
\max_{|U|\leq k}\|\hat{(\Sigma}^{U,U})^{-1} - (\Sigma^{U,U})^{-1}\|\leq C\sqrt{k\log(p)/n}.
\]

%%%%%%%%%%
%%%%%%%%%%

\subsection{Proof of Lemma~\ref{lemma:objective}}
The proofs are similar, so we only show the first one (associated with PCS).   Fix $1 \leq i,j \leq p$ and a subset $S \subset \{1, 2, \ldots, p\}$  such that $i \neq j$ and $i, j \notin S$.
Let $\tilde{S} = \{i, j\} \cup S$ be the ordered set  such that $i$ and $j$ are the first and second indices, respectively.
Since $|\tilde{S}| \leq n$, by \cite{Edelman},  the matrix $\hat{\Sigma}^{S, S}$ is non-singular with probability $1$,
so $\hrho_{ij}(S)$ is well-defined.

Now, for short, let
\[
A = \left(
\begin{array}{cc}
\hat{a}_{11}& \hat{a}_{12} \\
\hat{a}_{21} & \hat{a}_{22}
\end{array}
\right) = \mbox{$2\times 2$ submatrix of $(\hsig^{\tilde{S}, \tilde{S}})^{-1}$ on the top left}.
\]
By the way it is defined,
\begin{equation} \label{objpf1}
\hrho_{ij}(S) =  -  \hat{a}_{12} / \sqrt{\hat{a}_{11} \hat{a}_{22}}.
\end{equation}
At the same time, by basic algebra \cite{Seber},  it is known that
\begin{equation} \label{objpf2}
A =
\left(
\begin{array}{cc}
\hg_{ii}(S) & \hg_{ij}(S) \\
\hg_{ji}(S) & \hg_{jj}(S)
\end{array}
\right)^{-1},
\end{equation}
where $\hg_{ii}(S)$, $\hg_{jj}(S)$, $\hg_{ij}(S)$, and $\hg_{ji}(S)$ are random variables defined by
\begin{equation}    \label{Definegamma}
\left(
\begin{array}{cc}
\hg_{ii}(S) & \hg_{ij}(S) \\
\hg_{ji}(S) & \hg_{jj}(S)
\end{array}
\right)  =  \hsig^{\{i,j\}, \{i,j\}} - \hsig^{\{i,j\}, S} (\hsig^{S,S})^{-1}   (\hsig^{\{i,j\}, S})'.
\end{equation}
In our notation,  $\hsig= (1/n) X'X$, where $X$ is the $n \times p$ data matrix.
Write $S = \{j_1, j_2, \ldots, j_k\}$,  where $k = |S|$.  For short, let $X_S$ be the $n \times k$ matrix
where the $m$-th column is the $j_m$-th column of $X$, $1 \leq m \leq k$. It is seen that
\[
H_S = X_S (X_S' X_S)^{-1} X_S',
\]
Rewrite the right hand side of (\ref{Definegamma}) by
\begin{align*}
&(1/n)  \biggl[\left(
\begin{array}{cc}
(x_i, x_i),  & (x_i, x_j)  \\
(x_j, x_i),   & (x_j, x_j)
\end{array}
\right)    -  \left(
\begin{array}{cc}
x_i'  \\
x_j'
\end{array}
\right)  X_S (X_S X_S')^{-1} X_S'  \cdot (x_i, x_j) \biggr]     \nonumber     \\
=
&  (1/n)  \left(
\begin{array}{cc}
x_i' (I - H_S) x_i,    & x_i' (I - H_S) x_j  \\
x_j' (I - H_S) x_i,    & x_j' (I - H_S) x_j
\end{array}
\right).
\end{align*}
It follows that
\begin{equation} \label{obj52}
\left(
\begin{array}{cc}
\hg_{ii}(S) & \hg_{ij}(S) \\
\hg_{ji}(S) & \hg_{jj}(S)
\end{array}
\right)   =  (1/n)  \left(
\begin{array}{cc}
x_i' (I - H_S) x_i,    & x_i' (I - H_S) x_j  \\
x_j' (I - H_S) x_i,    & x_j' (I - H_S) x_j
\end{array}
\right).
\end{equation}
Now, for any positive definite $2 \times 2$ matrix $D$, write
\[
D = \left(
\begin{array}{cc}
d_{11} & d_{12}  \\
d_{21}  & d_{22}
\end{array}
\right), \qquad  D^{-1} = \left(
\begin{array}{cc}
h_{11} & h_{12}  \\
h_{21}  & h_{22}
\end{array}
\right).
\]
It is known that
\begin{equation} \label{objpf3}
d_{12}/\sqrt{d_11 d_{22}} = - h_{12} /\sqrt{h_{11} h_{22}}.
\end{equation}
Combining (\ref{objpf1})-(\ref{objpf2}) and applying (\ref{objpf3}),
\[
\hrho_{ij}(S) = \hg_{ij}(S) / \sqrt{\hg_{ii}(S) \hg_{jj}(S)},
\]
and the claim follows from (\ref{obj52}). \qed

%%%%%%%%%%
%%%%%%%%%%
\subsection{Proof of Lemma~\ref{lem:upperbound}}

Fix $i$ and write for short $S_0=S^{(i)}(\Omega)$ and $\omega'$ as the $i$-th row of $\Omega$. For each $m$, we define  $U_m=\{i, j_1,\cdots,j_m\}$. Then
\begin{equation} \label{rho_m1}
\rho_{ij_m}(S^{(i)}_{m-1}) = \frac{-1\cdot [\mbox{first row last column of }(\Sigma^{U_m, U_m})^{-1}]}{[\mbox{product of the first and last diagonals of }(\Sigma^{U_m,U_m})^{-1}]^{1/2}}.
\end{equation}

We need some notations to simplify the matrix $(\Sigma^{U_m,U_m})^{-1}$. Fix $k\geq 1$. Introduce the set $V=\{i, j_1,\cdots,j_{m(k)}\}\cup (S_0\setminus S^{(i)}_{m(k)})$, where $i$ is the first index and $j_m$ is the $(m+1)$-th index in the set, $1\leq m\leq m(k)$. Let $A=\Sigma^{V,V}$. For each $m\geq 1$, we partition $A$ into blocks corresponding to the first $(m+1)$-th indices and the remaining ones
\[
A = \begin{pmatrix}
A^{(m)}_{11} & A^{(m)}_{12}\\
A^{(m)}_{21} & A^{(m)}_{22}
\end{pmatrix},
\]
so that $\Sigma^{U_m, U_m} = A_{11}^{(m)}$. For notation simplicity, we shall omit all the superscripts and write  $A_{11}^{(m)}$ as $A_{11}$.
Using the matrix inverse formula,
\begin{equation} \label{inverse}
A^{-1} =  \begin{pmatrix}
A_{11}^{-1} +  B_{12} B_{22}^{-1} B_{21} & - B_{12} B_{22}^{-1}  \\
- B_{22}^{-1} B_{21} & B_{22}^{-1}
\end{pmatrix},
\end{equation}
where $B_{22} = A_{22} - A_{21} A_{11}^{-1} A_{12}$, $B_{12} = A_{11}^{-1} A_{12}$ and $B_{21} = A_{21} A_{11}^{-1}$.

Now, we show the claim. Since $V \supset S_0$, Lemma 1.1 implies that the first row of $A^{-1}$ is equal to  $\omega'$  restricted to $V$. Combining this with \eqref{inverse}, for each $m(k-1)< m< m(k)$,
\[
\mbox{first row last column of } [A_{11}^{-1} +  B_{12} B_{22}^{-1} B_{21}] = \omega(j_m)=0,
\]
and
\[
-1 \cdot \mbox{first row of } B_{12}B_{22}^{-1} = (\omega^{V\setminus U_m})'.
\]
Also, by definition, $B_{21}=A_{21} A_{11}^{-1}$. Combining the above,
\beq \label{firstlast}
\mbox{first row last column of } A_{11}^{-1} = (\omega^{V\setminus U_m})' A_{21} \cdot \mbox{last column of } A_{11}^{-1}.
\eeq

To simplify \eqref{firstlast}, we introduce a vector $\eta_k\in R^{|V|}$ such that $\eta_k(j+1)=0$ for $0\leq j< m(k)$ and $\eta_k(j+1)=(\omega^V)(j+1)$ for $m(k)\leq j\leq |V|-1$. For each $m(k-1)< m< m(k)$, let $J_m=\{1,\cdots,m+1\}$ and $J_m^c=\{m+2,\cdots,|V|\}$. Then $\omega^{V\setminus U_m}=(\omega^V)^{J_m^c}=(\eta_k)^{J_m^c}$, $A_{21}=A^{J_m^c, J_m}$ and $(\eta_k)^{J_m}=0$. It follows that
\begin{equation} \label{temp1}
(\omega^{V\setminus U_m})' A_{21} = (\eta_k'A)^{J_m}.
\end{equation}
Moreover, let $A=LL'$ be the Cholesky decomposition of $A$, where $L$ is a lower triangular matrix with positive diagonals. By basics of Cholesky decomposition, for $L_m=L^{J_m,J_m}$, $A_{11}=A^{J_m,J_m}=L_mL_m'$ is the Cholesky decomposition of $A_{11}$, and $L_m$ satisfies $L_m^{-1}=(L^{-1})^{J_m,J_m}$.
Therefore,
\begin{equation} \label{temp2}
A_{11}^{-1} = [(L^{-1})']^{J_m, J_m}(L^{-1})^{J_m, J_m}.
\end{equation}
The nice thing about \eqref{temp1}-\eqref{temp2} is that on the right hand sides, $(\eta_k, A, L)$ only depend on $k$ but not $m$ (while on the left hand sides, $A_{21}$ and $A_{11}$ depend on $m$). This allows us to stack the expressions for different $m$.

We plug \eqref{temp1}-\eqref{temp2} into \eqref{firstlast} and use the fact that $L^{-1}$ is also lower triangular. It yields
\begin{align*}
\mbox{first row last column of } & A_{11}^{-1}
=   [\eta_k'A (L^{-1})']^{J_m} \cdot \mbox{last column of } (L^{-1})^{J_m, J_m}   \\
= &  L^{-1}(m+1,m+1) \cdot q(m+1), \qquad q \equiv L^{-1}A'\eta_k.
\end{align*}
This gives the numerator of \eqref{rho_m1}.
For the denominator, by   \eqref{inverse} and \eqref{temp2},
(a)  the first diagonal of $A_{11}^{-1} \geq A^{-1}(1,1) \geq \lambda_{\max}^{-1}(A)$, and (b)
the last diagonal of $A_{11}^{-1} = [L^{-1}(m+1,m+1)]^2$.
Combining the above with \eqref{rho_m1}, we have
\begin{equation} \label{rhoupper}
\rho_{ij_m}^2(S^{(i)}_{m-1}) \leq \lambda_{\max}(A) \cdot q^2(m+1).
\end{equation}

Since \eqref{rhoupper} holds for each $m(k-1)<m<m(k)$, we stack the results for all $m$ and obtain
$\sum_{m(k-1)<m<m(k)} \rho_{ij_m}^2(S^{(i)}_{m-1}) \leq \lambda_{\max}(A)\cdot \|q\|^2 \leq \lambda^2_{\max}(A) \cdot \|\eta_k\|^2$.
Here, the last inequality is due to $A=L'L$ and $q=L^{-1}A'\eta=L'\eta$.
The claim then follows by noting that $\|\eta_k\|^2=\sum_{j\in (S_0\setminus S^{(i)}_{m(k)-1})}\omega^2(j_m)$ and that $A$ is a principal submatrix of $\Sigma$ with size $\leq m(k)+s-k$.
\qed

%%%%%%%%%%
%%%%%%%%%%
\subsection{Proof of Lemma~\ref{lem:lowerbound}}

Fixing $1\leq i\leq p$ and $m\geq 1$, we adopt the notations $S_0$, $\omega'$ and $U_m$ as in the proof of Lemma~\ref{lem:upperbound}. Let $W=S_0\setminus S^{(i)}_{m-1}$. For each $j\in W$, let $V_j=\{i, j_1,\cdots, j_{m-1}, j\}$ and suppose $i$ and $j$ are the first and last indices in the set, respectively. By definition,
\begin{equation} \label{rho_m}
\rho_{ij}(S^{(i)}_{m-1}) = \frac{-1\cdot [\mbox{first row last column of }(\Sigma^{V_j, V_j})^{-1}]}{[\mbox{product of  first and last diagonals of }(\Sigma^{V_j,V_j})^{-1}]^{1/2}}.
\end{equation}
Write $\Sigma_m=\Sigma^{U_{m-1}, U_{m-1}}$ and $\eta_j=\Sigma^{U_{m-1}, \{j\}}$ for short. By basic algebra,
\[
(\Sigma^{V_j,V_j})^{-1}
= \begin{pmatrix}
\Sigma_m^{-1} +  A & - [\Sigma(j,j)-\eta_j'\Sigma_m^{-1}\eta_j]^{-1}\eta_j'\Sigma_m^{-1}  \\
-[\Sigma(j,j)-\eta_j'\Sigma_m^{-1}\eta_j]^{-1}\Sigma_m^{-1}\eta_j  & [\Sigma(j,j)-\eta_j'\Sigma_m^{-1}\eta_j]^{-1}
\end{pmatrix},
\]
where $A$ is a positive-definite matrix. It follows that
\begin{align*}
& \mbox{first row last column of } (\Sigma^{V_j,V_j})^{-1} = -1 \cdot [\Sigma(j,j)-\eta_j'\Sigma_m^{-1}\eta_j]^{-1}e_j'\Sigma_m^{-1}\eta_j,\cr
& \mbox{last diagonal of } (\Sigma^{V_j,V_j})^{-1} = [\Sigma(j,j)-\eta_j'\Sigma_m^{-1}\eta_j]^{-1},
\end{align*}
where $e_1=(1,0,\cdots,0)'$. As a result,
\begin{equation}
\sum_{j\in W}\rho^2_{ij}(S^{(i)}_{m-1}) = \sum_{j\in W}\frac{[\Sigma(j,j)-\eta_j'\Sigma_m^{-1}\eta_j]^{-1}}{(\Sigma^{V_j,V_j})^{-1}(1,1)}(e_1'\Sigma_m^{-1}\eta_j)^2,
\end{equation}
Let
$D = \Sigma^{W,W} - \Sigma^{W, U_m}\Sigma_m^{-1}\Sigma^{U_m, W}$ and $H = \Sigma_m^{-1}\Sigma^{U_m, W}$.
Then
\begin{equation} \label{lowerbound1}
\sum_{j\in W}\rho^2_{ij}(S^{(i)}_{m-1}) \geq \frac{1}{\max_{j\in W}(\Sigma^{V_j,V_j})^{-1}(1,1)} \|e_1'H[\diag(D)]^{-1/2}\|^2.
\end{equation}

Below, we make a connection between $\omega$ and the right hand side of \eqref{lowerbound1}. Introduce set $V=\{i,j_1,\cdots,j_{m-1}\}\cup W$ such that $i$ is the first index and $j_k$ is the $(k+1)$-th index, $1\leq k\leq m-1$. Using the matrix inverse formula,
\[
(\Sigma^{V,V})^{-1} =
\begin{pmatrix}
\Sigma_m & \Sigma^{U_{m-1}, W}\\
\Sigma^{W, U_{m-1}} & \Sigma^{W,W}
\end{pmatrix}^{-1}
= \begin{pmatrix}
\Sigma_m^{-1} + HD^{-1}H & - HD^{-1} \\
-D^{-1}H'  & D^{-1}
\end{pmatrix}.
\]
Since $V\supset S_0$, by Lemma 1, the first row of $(\Sigma^{V,V})^{-1}$ coincides with $(\omega^V)'$. In particular,
\begin{equation} \label{LBtemp1}
e_1'HD^{-1} = -(\omega^W)'.
\end{equation}
Furthermore, since $V_j\subset V$,
\begin{equation} \label{LBtemp2}
(\Sigma^{V_j,V_j})^{-1}(1,1)\leq (\Sigma^{V,V})^{-1}(1,1) \leq \lambda_{\min}^{-1}(\Sigma^{V,V}).
\end{equation}
Plugging \eqref{LBtemp1}-\eqref{LBtemp2} into \eqref{lowerbound1} gives that
$\sum_{j\in W}\rho^2_{ij}(S^{(i)}_{m-1}) \geq \lambda_{\min}(\Sigma^{V,V}) \cdot $  $ \|[\diag(D)]^{-1/2}D\omega^W\|^2$.
Note that $\|[\diag(D)]^{-1/2}D\omega^W\|^2\geq \lambda_{\max}^{-1}(D)\lambda^2_{\min}(D)\cdot \|\omega^W\|^2$. Moreover, the eigenvalues of $D$ are between $\lambda_{\min}(\Sigma^{V,V})$ and $\lambda_{\max}(\Sigma^{V,V})$. Therefore,
$\sum_{j\in W}\rho^2_{ij}(S^{(i)}_{m-1}) \geq \frac{\lambda^3_{\min}(\Sigma^{V,V})}{\lambda_{\max}(\Sigma^{V,V})} \cdot \|\omega^W\|^2$.
The claim follows by noting that $W=S_0\setminus S^{(i)}_{m-1}$ and that the size of $\Sigma^{V,V}$ is $|W|+m\leq m+s$.
\qed
\subsection{Proof of Theorem~\ref{thm:idealPCS}}
Write for short $S_0 = S^{(i)}(\Omega)$, $s_0 = |S_0|$ and $m^{(i)}(k)=m(k)$, $1\leq k\leq s_0$.
The key is to show that for all $1 \leq k \leq  s_0$,  the $k$-th Lagging Time $\ell^{(i)}(k; \Sigma)$
satisfies
\begin{equation} \label{thm2.1pf1}
\ell^{(i)}(k;  \Sigma) \leq  2 c_0^{-6} (s_0 - k +1),
\end{equation}
where $c_0$ is as in ${\cal M}_p^*(s, c_0)$.
Once (\ref{thm2.1pf1}) is proved, the second claim follows by basic algebra,
the third claim follows directly from Lemma~\ref{lemma:firstcolumn}.
As for the first claim, suppose that at the end of stage $(m-1)$, we have recruited $k$ signal nodes, $k < s_0$.
By Lemma~\ref{lem:lowerbound},
\[
\sum_{j \in (S_0 \setminus S_{m-1}^{(i)})} \rho_{ij}^2(S_{m-1}^{(i)})   \geq  c_0^{4} (s_0 - k) (\tau_p^*)^2,
\]
where for the constant $c_0^{4}$, we have used the definition of ${\cal M}_p^*(s, c_0)$ and that the algorithm terminates in no  more than $Cs^2 \ll N(p,n)$ steps.  It follows that there is a  $j \in (S_0 \setminus S_{m-1}^{(i)})$ such that
$|\rho_{ij}(S^{(i)}_{m-1})| \geq c_0^2 \tau_p^*$. Since PCS is a greedy algorithm, recruiting the node with the largest partial correlation in each step, the claim follows.

We now show (\ref{thm2.1pf1}). Our strategy is to use `Reductio ad absurdum'.   Suppose  (\ref{thm2.1pf1}) does not hold. Let $k_0$ be the smallest integer such that
\begin{equation} \label{thm2.1pf2}
\ell^{(i)}(k_0;  \Sigma) > 2  c_0^{-6} (s_0 - k_0 +1).
\end{equation}
Let $m_0$ be the integer such that $m_0\leq 2c_0^{-6}(s_0 - k_0 +1)<m_0+1$.  Note that by (\ref{thm2.1pf2}), we do not recruit any signal nodes in steps $m(k_0-1) +1, m(k_0-1) + 2, \ldots, m(k_0-1) + m_0$, and also that
\begin{equation} \label{thm2.1pf3}
m(k_0-1) + m_0 \leq (k_0-1) + 2 c_0^{-6} \sum_{j  = 1}^{k_0} (s_0 - j +1) \leq Cs^2,
\end{equation}
where the right hand side $\leq N(p,n)$ by the assumption of $s^2 \log(p) = o(n)$.
One one hand, by Lemma~\ref{lem:upperbound} and (\ref{thm2.1pf3}),
\begin{equation} \label{thm2.1pf4}
\sum_{m =  m(k_0-1)  + 1}^{m(k_0-1) + m_0} \rho_{ij_m}^2(S_{m-1}^{(i)}) \leq c_0^{-2} E^{(i)}(m(k_0)),
\end{equation}
where $E^{(i)}(m) = \sum_{j \in (S_0 \setminus S^{(i)}_{m-1})} \Omega(i,j)^2$ denotes the EAL at stage $m$.
On the other hand, by Lemma~\ref{lem:lowerbound}, for any $m\in \{ m(k_0-1)+1,\cdots, m(k_0-1) + m_0\}$,
\[
\sum_{j \in (S_0 \setminus S_{m-1}^{(i)})} \rho_{ij}^2(S_{m-1}^{(i)})   \geq  c_0^{4} E^{(i)}(m)=c_0^{4} E^{(i)}(m(k_0)),
\]
where the last equality is because PCS does not recruit any signal nodes in steps $m(k_0-1) +1, m(k_0-1) + 2, \ldots, m(k_0-1) + m_0$. It follows that there is a $j \in (S_0 \setminus S_{m-1}^{(i)})$ such that
\[
\rho_{ij}^2(S_{m-1}^{(i)}) \geq (s_0 - k_0+1)^{-1} c_0^{4} E^{(i)}(m(k_0)).
\]
Since PCS is a greedy algorithm, recruiting the node with the largest partial correlation in each step,
\beq  \label{thm2.1pf5}
\rho_{ij_m}^2(S_{m-1}^{(i)})\geq (s_0 - k_0+1)^{-1} c_0^{4} E^{(i)}(m(k_0)),
\eeq
for any $m=m(k_0-1)+1,\cdots,m(k_0-1)+m_0$. Inserting \eqref{thm2.1pf5} into (\ref{thm2.1pf4}) gives $m_0\leq c_0^{-6}(s_0-k_0+1)$. However, $m_0 \geq 2c_0^{-6}(s_0 - k_0 +1)-1$ by definition, and a contradiction follows. This concludes the proof. \qed

%%%%%%%%%%%%%%%
%%%%%%%%%%%%%%%
\subsection{Proof of Theorem~\ref{thm:main}}

Write for short $S_0 = S^{(i)}(\Omega)$, $s_0 = |S_0|$ and $\hat{m}^{(i)}(k)=\hat{m}(k)$, $1\leq k\leq s_0$. Recall that $\hat{\ell}^{(i)}(k; X, \Sigma)$ is the $k$-th Lagging Time, $1 \leq k \leq s_0$.
Define $\hat{k}$ to be the smallest integer in $\{1, 2, \ldots, s_0\}$ such that
\[
\hat{\ell}^{(i)}(k; X, \Sigma) > 4 c_0^{-6}(s_0 - k -1);
\]
If no such integer exists, then we let $\hat{k} = \infty$.

To show Theorem~\ref{thm:main},  the key is to show that
\begin{equation} \label{Thm2.2pf0}
P(\hat{k} = \infty)  \geq 1 - o(p^{-3}),
\end{equation}
so with overwhelming probabilities,
\[
\hat{\ell}^{(i)}(k; X, \Sigma)  \leq 4 c_0^{-6}(s_0 - k +1), \qquad k = 1, 2, \ldots, s_0.
\]
Write
\[
P(\hat{k} < \infty) = \sum_{k = 1}^{s_0} P(\hat{k} = k).
\]
To show (\ref{Thm2.2pf0}), it is sufficient to show for any $k_0 \in \{1, 2, \ldots, s_0\}$,
\begin{equation} \label{Thm2.2pf1}
P(\hat{k} = k_0) =  O(p^{-4}),
\end{equation}

Now, we show \eqref{Thm2.2pf1}. For any $k_0$ such that $1 \leq k_0 \leq s_0$, similar to that in the proof of  Theorem~\ref{thm:idealPCS}, let $m_0$ be the integer such that
\[
m_0 \leq 4 c_0^{-4} (s_0 - k_0 +1)<m_0+1.
\]
On one hand, the following lemma extends Lemma~\ref{lem:upperbound} and is proved in Section~\ref{subsubsec:upperbound2}.
%%%%%%%%%%%%%%%%%%%%%
%%%%%%%%%%%%%%%%%%%%%
\begin{lemma} \label{lem:upperbound2}
Suppose conditions of Theorem~\ref{thm:idealPCS} hold. Fix $k_0$ such that $1 \leq k_0 \leq s_0$.
There is an event $B_1$ such that $P(B_1^c) = O(p^{-4})$ and that over the event $B_1 \cap \{\hat{k} = k_0\}$,
\[
\sum_{m = \hat{m}(k_0 -1) +1}^{\hat{m}(k_0 -1) +   m_0}   \hat{\rho}_{ij_m}^2(\hat{S}_{m-1}^{(i)}) \lesssim 2c_0^{-2} \sum_{j \in \bigl(S_0 \setminus \hat{S}_{\hat{m}(k_0-1)}^{(i)}\bigr)}  \Omega(i,j)^2  + C s^2 \log(p)/n,
\]
where $C>0$ is a constant that only depends on $c_0$.
\end{lemma}
Recall that in our notations, $\sum_{j\in(S_0\setminus \hs^{(i)}_{m-1})}\Omega(i,j)^2\equiv \hat{E}^{(i)}(m)$, which is the EAL at stage $m$. Also, by our conditions, the minimum signal strength
\beq  \label{Thm2.2signal}
\tau_p^* / [s \sqrt{\log(p)/n}] \goto \infty.
\eeq
Therefore, it follows that over the event $B_1 \cap \{\hat{k} = k_0\}$,
\begin{equation} \label{Thm2.2pf3}
\sum_{m = \hat{m}(k_0 -1) +1}^{\hat{m}(k_0 -1) +   m_0}   \hat{\rho}_{ij_m}^2(\hat{S}_{m-1}^{(i)}) \lesssim  2c_0^{-2}\hat{E}^{(i)}(\hat{m}(k_0 -1)+1).
\end{equation}
On the other hand, we also have the following lemma, which extends Lemma~\ref{lem:upperbound} and is proved in Section~\ref{subsubsec:lowerbound2}.
%%%%%%%%%%%%%%%%%%%%%%
%%%%%%%%%%%%%%%%%%%%%%
\begin{lemma} \label{lem:lowerbound2}
Suppose conditions of Theorem~\ref{thm:idealPCS} hold.
Fix $k_0$ such that $1 \leq k_0 \leq s_0$.
There is an event $B_2$ such that $P(B_2^c)= O(p^{-4})$ and that over the event $B_2 \cap \{\hat{k} = k_0\}$, for any $m$ satisfying $\hat{m}(k_0-1)<m\leq \hat{m}(k_0-1)+m_0$,
\[
\sum_{j \in (S_0 \setminus \hat{S}^{(i)}_{m-1})}   \hat{\rho}_{ij}^2(\hat{S}^{(i)}_{m-1}) \gtrsim c_0^{4} \sum_{j \in (S_0 \setminus\hat{S}_{m-1}^{(i)})}  \Omega(i,j)^2 -  C s^2 \log(p)/n,
\]
where $C>0$ is a constant that only depends on $c_0$.
\end{lemma}
Combining Lemma~\ref{lem:lowerbound2} and \eqref{Thm2.2signal}, over the event $B_2\cap \{\hat{k}=k_0\}$,  for each $m\in \{\hat{m}(k_0-1)+1,\cdots,\hat{m}(k_0-1)+m_0\}$,
\[
\sum_{j \in (S_0 \setminus \hat{S}^{(i)}_{m-1})}   \hat{\rho}_{ij}^2(\hat{S}^{(i)}_{m-1})\gtrsim c_0^4 \hat{E}^{(i)}(m)
= c_0^{4} \hat{E}^{(i)}(\hat{m}(k_0-1)+1),
\]
where the last inequality is because PCS does not recruit any signal nodes in steps $\hat{m}(k_0-1)+1,\cdots, \hat{m}(k_0-1)+m_0$ on the event $\{\hat{k}=k_0\}$. Therefore, there is a $j\in (S_0\setminus \hs^{(i)}_{m-1})$ such that
\[
\hat{\rho}_{ij}^2(\hat{S}_{m-1}^{(i)}) \gtrsim (s_0-k_0+1)^{-1} c_0^{4} \hat{E}^{(i)}(\hat{m}(k_0-1)+1).
\]
PCS is a greedy algorithm, recruiting the node with the largest partial correlation in each step. It follows that over the event $B_2\cap \{\hat{k}=k_0\}$, for each $m\in \{\hat{m}(k_0-1)+1,\cdots, \hat{m}(k_0-1)+m_0\}$,
\begin{equation} \label{Thm2.2pf4}
\hat{\rho}_{ij_m}^2(\hat{S}_{m-1}^{(i)}) \gtrsim (s_0-k_0+1)^{-1} c_0^{4} \hat{E}^{(i)}(\hat{m}(k_0-1)+1).
\end{equation}
Combining \eqref{Thm2.2pf3}-\eqref{Thm2.2pf4},
\[
m_0\lesssim 2c_0^{-6}(s_0-k+1),
\]
which yields a contradiction. In other words, we have shown that
\begin{equation} \label{Thm2.2pf5}
P(B_1 \cap B_2 \cap \{\hat{k} = k_0\}) = 0,
\end{equation}
and \eqref{Thm2.2pf1} follows.

We now proceed to show Theorem \ref{thm:main}.
Consider the second claim first. By (\ref{Thm2.2pf1}),  $P(\{\hat{k} =  \infty\}) \geq 1 - o(p^{-3})$, and over the event $\{\hat{k} = \infty\}$, PCS stops in no more than
\begin{equation} \label{Thm2.2pf6}
s_0 +  4c_0^{-6}\sum_{k = 1}^{s_0} (s_0 - k -1) \leq C s_0^2
\end{equation}
steps, and the second claim follows directly.

Consider the other two claims.
The following lemma is proved in Section~\ref{subsubsec:noiselevel}.
\begin{lemma} \label{lem:noiselevel}
Suppose conditions of Theorem~\ref{thm:main} hold. There is a constant $c_1>0$ that only depends on $c_0$, and an event $B_3$ with $P(B_3^c)=o(p^{-3})$, such that for each $1\leq m\leq 4c_0^{-6}s^2$, over the event $B_3 \cap \{\hat{k} = \infty\} \cap \{ S_0 \subset \hat{S}_m^{(i)}\}$,
\[
|\hat{\rho}_{ij}(\hat{S}_{m}^{(i)})| \leq c_1s \sqrt{2\log(p)/n}, \qquad \mbox{for all } j\notin (\{i\}\cup \hs^{(i)}_m).
\]
\end{lemma}
Combining Lemmas~\ref{lem:lowerbound2}-\ref{lem:noiselevel},
\begin{itemize}
\item over the event $B_3 \cap \{\hat{k} = \infty\} \cap \{S_0 \subset \hat{S}_m^{(i)} \}$,  for all $j \notin (\{i\}\cup \hat{S}_m^{(i)})$,
\[
|\hat{\rho}_{ij}(\hat{S}_{m})| \leq c_1s \sqrt{2\log(p)/n};
\]
\item over the event $B_3 \cap \{\hat{k} = \infty\} \cap \{S_0 \not \subset  \hat{S}_m^{(i)} \}$, there is a $j \in (S_0 \setminus \hat{S}_m^{(i)})$ such that
\[
|\hat{\rho}_{ij}(\hat{S}_m^{(i)})| \gtrsim c_0^{2}  \bigg(\frac{\hat{E}^{(i)}(m+1)}{|S_0\setminus \hat{S}_m^{(i)}|}\bigg)^{1/2} \geq C \tau_p^*.
\]
\end{itemize}
Since $c_1s\sqrt{2\log(p)/n}\leq t_q^*/2 \ll \tau_p^*$, combining these gives the first and the last claim.  \qed

\subsubsection{Proof of Lemma~\ref{lem:upperbound2}} \label{subsubsec:upperbound2}

Write for short  $m^* = m(k_0 - 1)$ and
\[
V = \{i, j_1, j_2, \ldots, j_{m^*}\}, \qquad W = \{j_{m^*+1},  j_{m^* + 2}, \ldots, j_{m^* + m_0}\},
\]
where $j_1, j_2, \ldots, j_{m^*}$ are the recruited nodes in the first $m^*$ steps, and $j_{m^*+1},  j_{m^* + 2}, \ldots, j_{m^* + m_0}$ are the nodes recruited in the next $m_0$ steps, both are
ordered sets where the indices are arranged in that order.  At the same time,
let
\[
Q =  S_0 \setminus \hat{S}_{m^*}^{(i)}
\]
be all the signal nodes (arranged in the ascending order for convenience)  that have not yet been recruited in the first $m^*$ steps. Note that over the event we consider, $W$ does not contain any signal node, so
\[
Q = S_0  \setminus \hat{S}_{m}^{(i)}, \qquad \mbox{for any
$m$ such that $m^*+1 \leq m \leq   m^*+ m_0$}.
\]

Throughout this section, $V \cup W$ is the ordered set where all nodes in $V$ are arranged before those of $W$, and nodes in $V$ and $W$ are arranged according to their original order aforementioned. Similar rules apply to $W \cup V$,  $V \cup W \cup Q$,
etc.. Note that $V \cup W$ is not the same as $W \cup V$ for indices are arranged in different orders.
Introduce the following short-hand notations:
\begin{align*}
x(V) &= \mbox{the first diagonal of $(\hsig^{V, V})^{-1}$},    \\
x(V,W) &= \mbox{the first diagonal of $(\hsig^{V \cup W, V \cup W})^{-1}$},   \\
x(V,W,Q) &= \mbox{the first diagonal of $(\hsig^{V \cup W \cup Q, V \cup W \cup Q})^{-1}$},   \\
x(V,Q,W) &= \mbox{the first diagonal of $(\hsig^{V \cup Q \cup  W, V \cup Q \cup  W})^{-1}$}.
\end{align*}
The proof for the lemma contains two parts. In the first part, we show that
\begin{equation} \label{ubpf1}
\sum_{m = m_* +1}^{m_* +   m_0}   \hat{\rho}_{ij_m}^2(\hat{S}_{m-1}^{(i)})  \leq \hsig(i,i) \cdot  [x(V,W) - x(V)].
\end{equation}
In the second part, we analyze $[x(V,W) - x(V)]$ and completes the proof.

Consider the first part, where the key is to use Cholesky factorization \cite{Horn}.  To this end, we introduce a short hand notation. For any matrix $D \in R^{m,m}$ and $1 \leq k \leq m$, let
\[
D(1:k,1:k)
\]
denote  the sub-matrix of $D$ consisting of the first $k$ rows and $k$ columns of $D$. Note that $D = D(1:m,1:m)$.
Denote for short $\hat{D}$ by the $(m^* + m_0 + 1) \times (m^* + m_0 + 1)$ matrix (the one extra comes from the first index,  $i$)
\[
\hat{D} =  \hsig^{V \cup W, V\cup W},
\]
and let
\[
\hat{D} = L L'
\]
be the Cholesky factorization (unique provided the diagonals of $L$ are positive); note that $L$ is lower triangular.
Denote $U$ by the inverse of $L$:
\[
L = U^{-1}.
\]
By basic algebra, $U$ is a lower triangular $(m^* + m_0 + 1) \times (m^* + m_0 + 1)$ matrix. The following facts are noteworthy. For any $k$ such that $1 \leq k \leq m^* + m_0 + 1$,  we have
\[
[L(1:k,1:k)]^{-1}  = U(1:k,1:k),
\]
and
\[
[\hat{D}(1:k, 1:k)]^{-1}  = U'(1:k,1:k) \cdot  U(1:k,1:k);
\]
especially, $\hat{D} = U'U$.

In our notations, if we write for short $D_{m+1} = \hat{D}(1:m+1, 1:m+1)$, then
\[
\hrho_{ij_m}(\hat{S}_{m-1}^{(i)}) = \frac{\mbox{first row last column of $D_{m+1}^{-1}$}}{\sqrt{\mbox{product of first and last diagonals of $D_{m+1}^{-1}$}}},
\]
We collect some basic facts.
\begin{itemize}
\item The first row last column of $D_{m+1}^{-1}$ is $U(m+1,m+1) \cdot U(m+1,1)$.
\item The first diagonal of $D_{m+1}^{-1}$ is no smaller than $\hsig(i,i)^{-1}$.
\item The last diagonal of $D_{m+1}^{-1}$ is $U(m+1,m+1)^2$.
\item $\sum_{m = 0}^{m^*} U(m+1,1)^2 = x(V)$.
\item $\sum_{m=0}^{m^* + m_0} U(m+1,1)^2 = x(V,W)$.
\end{itemize}
Combining these
\[
\sum_{m = m_* +1}^{m_* +   m_0}   \hat{\rho}_{ij_m}^2(\hat{S}_{m-1}^{(i)})  \leq \hsig(i,i) \sum_{m = m^* +1}^{m^* + m_0}  U(1, m+1)^2 = \hsig(1,1)[x(V,W) - x(V)],
\]
and the claim follows.

Consider the second part.   Note that $x(V,Q,W) = x(V,W,Q)$.  We write
\begin{eqnarray}
&& x(V,W)  -  x(V)  \label{ubpf2}\\
&=& [x(V,Q) - x(V)]  - [x(V,W,Q) - x(V,W)] + [x(V,W,Q) - x(V,Q)] \nonumber\\
&=& [x(V,Q) - x(V)]  - [x(V,W,Q) - x(V,W)] +  [x(V,Q,W) - x(V,Q)].\nonumber
\end{eqnarray}
We now analyze the three terms on the right hand side. The analysis is similar, so we only discuss the first one in detail.
According to the partition of indices in $V \cup Q$  to those in $V$ and those in $Q$, we write
\[
A  = \Sigma^{V \cup Q, V \cup Q} =
\left(
\begin{array}{cc}
A_{11}   & A_{12} \\
A_{21}  &   A_{22}
\end{array}
\right),  \;\;\;   \hat{A}  = \hsig^{V \cup Q, V \cup Q} =
\left(
\begin{array}{cc}
\hat{A}_{11}   & \hat{A}_{12} \\
\hat{A}_{21} & \hat{A}_{22}
\end{array}
\right).
\]
By basic algebra,  we have
\begin{equation} \label{inverse1}
A^{-1} =  \begin{pmatrix}
A_{11}^{-1} +  B_{12} B_{22}^{-1} B_{21} & - B_{12} B_{22}^{-1}  \\
- B_{22}^{-1} B_{21} & B_{22}^{-1}
\end{pmatrix},
\end{equation}
and
\begin{equation} \label{inverse2}
\hat{A}^{-1} =  \begin{pmatrix}
\hat{A}_{11}^{-1} +  \hat{B}_{12} \hat{B}_{22}^{-1} \hat{B}_{21} & - \hat{B}_{12} B_{22}^{-1}  \\
- \hat{B}_{22}^{-1} \hat{B}_{21} & \hat{B}_{22}^{-1}
\end{pmatrix},
\end{equation}
where $B_{22} = A_{22} - A_{21} A_{11}^{-1} A_{12}$, $B_{12} = A_{11}^{-1} A_{12}$ and $B_{21} = A_{21} A_{11}^{-1}$, and $\hat{B}_{12}$, $\hat{B}_{21}$, and $\hat{B}_{22}$ are defined similarly.
Moreover, denote the first row of
\[
 - \hat{B}_{12} B_{22}^{-1}  \qquad \mbox{and}  \qquad  - B_{12} B_{22}^{-1}
\]
by $\heta'$ and $\eta$, respectively.
The following facts follow from definitions,  basic algebra, and Lemma \ref{lemma:firstcolumn}.
\begin{itemize}
\item $x(V,Q) - x(V) = \heta' \hat{B}_{22} \heta$.
\item $\eta(k) = \Omega(i, j_{m^*+m_0+k}) \neq 0$, $1 \leq k \leq |Q|$; so $\|\eta\|^2=\sum_{j\in(S_0\setminus \hs^{(i)}_{m^*})}\Omega(i,j)^2$, which is the EAL on the right hand side of the claim.
\end{itemize}
It follows that
\begin{equation}  \label{ubpf51}
x(V,Q) - x(V) \leq \| \hat{B}_{22}\| \|\heta\|^2  \leq 2 \| \hat{B}_{22} \| (\|\eta\|^2 + \|\heta - \eta\|^2),
\end{equation}
where $\|\hat{B}_{22}\|\leq \|B_{22}\|+\|\hat{A}^{-1}-A^{-1}\|\leq c_0^{-1} + \|\hat{A}^{-1}-A^{-1}\|$, by $m^*+|Q|+1\leq N(p,n)$ and the regularity condition imposed on $\mathcal{M}^*_p(s,c_0)$. At the same time,
\begin{equation} \label{ubpf52}
\| \heta - \eta \|^2  \leq  \|(\hat{A}^{-1} - A^{-1})^2\|  \leq (\|\hat{A}^{-1} \| \cdot  \|A^{-1}\| \cdot  \|\hat{A} - A\| )^2.
\end{equation}
Since over the event we consider, $m^* + |Q| + 1 \leq C s^2$, by Lemma~\ref{lemma:stoch},
with probability at least $1 - o(p^{-4})$,
\begin{equation} \label{ubpf53}
\|\hat{A} - A\|^2 \leq  C(m^*+|Q|+1)\log(p)/n.
\end{equation}
Combining (\ref{ubpf51})-(\ref{ubpf53}) gives that with probability at least $1 - O(p^{-4})$,
\begin{equation} \label{ubpf61}
0\leq x(V,Q) - x(V) \lesssim 2 c_0^{-1} \|\eta\|^2 + Cs^2 \log(p)/n,
\end{equation}
Similarly, we have
\begin{equation} \label{ubpf62}
0\leq x(V,W,Q) - x(V,W) \lesssim 2 c_0^{-1} \|\eta\|^2 + Cs^2 \log(p)/n ,
\end{equation}
and
\begin{equation} \label{ubpf63}
0\leq x(V,Q,W) - x(V,Q)  \leq C s^2\log(p)/n.
\end{equation}
The right hand side does not have the $\|\eta\|^2$ for that the associated ``$\eta$" vector is the vector of $0$, by a direct use of Lemma \ref{lemma:firstcolumn}.
We inserting (\ref{ubpf61})-(\ref{ubpf63}) into (\ref{ubpf2}), and further combine it with \eqref{ubpf1}.   The claim follows directly by noting that with probability at least $1-O(p^{-4})$, $\hat{\Sigma}(i,i)\leq \Sigma(i,i)+C\sqrt{\log(p)/n}\lesssim c_0^{-1}$.  \qed

%%%%%%%%%%%%%%
%%%%%%%%%%%%%%
\subsubsection{Proof of Lemma~\ref{lem:lowerbound2}} \label{subsubsec:lowerbound2}
Write for short $\omega'$ as the $i$-th row of $\Omega$. Recall that $\{j_1,j_2,\cdots\}$ is the sequence of nodes recruited in the Screen step of PCS. Fix $m$ such that $m(k_0-1)<m\leq m(k_0)+m_0$. Let
\[
V=\{i,j_1,\cdots,j_{m-1}\}\cup (S_0\setminus \hs^{(i)}_{m-1}), \qquad W=S_0\setminus \hs^{(i)}_{m-1},
\]
where we assumed the indices in $V$ are listed in the above order. Define the vector $\tilde{\omega}\in R^p$ such that
\[
\tilde{\omega}(j) =\left\{
\begin{array}{lcl}
0, && \mbox{$j\notin V$},\\
(\hat{\Sigma}^{V,V})^{-1}(1,\ell), && \mbox{$j$ is the $\ell$-th node in $V$}.
\end{array}\right.
\]
Due to similar calculations to those in the proof of Lemma~\ref{lem:lowerbound}, we obtain
\beq \label{rho2-temp1}
\sum_{j\in (S_0\setminus \hs^{(i)}_{m-1})}\hat{\rho}^2_{ij}(\hs^{(i)}) \geq \frac{\lambda^3_{\min}(\hat{\Sigma}^{V,V})}{\lambda_{\max}(\hat{\Sigma}^{V,V})} \cdot \|\tilde{\omega}^{W}\|^2.
\eeq
Recall that $\omega'$ the $i$-th row of $\Omega$.
Since $(\omega^W)'$ and $(\tilde{\omega}^W)'$ are the first rows of $\Sigma^{V,V}$ (see Lemma~\ref{lemma:firstcolumn}) and $\hat{\Sigma}^{V,V}$, respectively, it follows from the triangular inequality that
\beq \label{rho2-temp2}
\|\tilde{\omega}^{W}\| \geq \biggl(\sum_{j\in (S_0\setminus \hs^{(i)}_{m-1})}\omega^2(j)\biggr)^{1/2} - \|\tilde{\omega}^W-\omega^W\|,
\eeq
where
\beq \label{rho2-temp4}
\|\tilde{\omega}^W-\omega^W\|\leq \|(\hat{\Sigma}^{W,W})^{-1}-\Sigma^{W,W}\|.
\eeq
Let $B_2$ be the event that for any $U\subset\{1,\cdots, p\}$ and $|U|\leq 4c_0^{-6}s^2$,
\[
\|\hat{\Sigma}^{U,U}-\Sigma^{U,U}\|\leq Cs\sqrt{\log(p)/n}, \qquad \|(\hat{\Sigma}^{U,U})^{-1}-(\Sigma^{U,U})^{-1}\|\leq Cs\sqrt{\log(p)/n}.
\]
By Lemma~\ref{lemma:stoch}, there exists a constant $C>0$ that only depends on $c_0$ such that $P(B_2^c)=O(p^{-4})$. On the event $B_2\cap \{\hat{k}=k_0\}$,
\[
|V|\leq m(k_0-1)+m_0\leq (k_0-1) + 4c_0^{-6}\sum_{k=1}^{k_0}(s_0-k_0+1)\lesssim 2c_0^{-6}s^2.
\]
By the definition of $B_2$ and that $\Sigma\in\mathcal{M}_p^*(s,c_0)$,
\begin{align} \label{rho2-temp3}
& \|(\hat{\Sigma}^{V,V})^{-1}-(\Sigma^{V_m,V_m})^{-1}\|\leq Cs\sqrt{\log(p)/n}, \cr
& \lambda_{\min}(\hat{\Sigma}^{V,V}) \geq \mu^{(1)}_{N}(\Sigma) - Cs\sqrt{\log(p)/n} \gtrsim c_0, \cr
& \lambda_{\max}(\hat{\Sigma}^{V,V}) \leq \mu^{(2)}_{N}(\Sigma) + Cs\sqrt{\log(p)/n} \lesssim c_0^{-1}.
\end{align}
The claim follows from \eqref{rho2-temp2}-\eqref{rho2-temp3} and that $\bigl(\sum_{j\in (S_0\setminus \hs^{(i)}_{m-1})}\omega^2(j)\bigr)^{1/2}\geq \tau_p^*\gg s\sqrt{\log(p)/n}$.
\qed

%%%%%%%%%%%%%%%
%%%%%%%%%%%%%%%
\subsubsection{Proof of Lemma~\ref{lem:noiselevel}} \label{subsubsec:noiselevel}

Over the event $\{\hat{k}=\infty\}\cap \{S_0\subset\hs^{(i)}_{m}\}$, by Lemma~\ref{lemma:firstcolumn},
\[
\rho_{ij}(\hs^{(i)}_m)=0, \qquad \mbox{for all } j\notin(\{i\}\cup \hs^{(i)}_m).
\]
Therefore, it suffices to show that with probability at least $1-o(p^{-3})$, for any $S\subset(\{1,\cdots,p\}\setminus\{i\})$ and $|S|\leq 4c_0^{-6}s^2-2$,
\beq \label{LemA.5-pf1}
\max_{j\notin(\{i\}\cup S)}|\hat{\rho}_{ij}(S)-\rho_{ij}(S)|\leq c_1 s\sqrt{2\log(p)/n}.
\eeq

Now, we show \eqref{LemA.5-pf1}.
By Lemma~\ref{lemma:stoch} and that $\Sigma\in \mathcal{M}_p^*(s,c_0)$, there is a constant $C>0$ which only depends on $c_1$ such that $P(B_3^c)=o(p^{-3})$ for the following event $B_3$: for any $U\subset\{1,\cdots,p\}$ and $|U|\leq 5c_0^{-6}s^2$,
\[
\|\hat{\Sigma}^{U,U}-\Sigma^{U,U}\|\leq Cs\sqrt{\log(p)/n}, \quad \|(\hat{\Sigma}^{U,U})^{-1}-(\Sigma^{U,U})^{-1}\|\leq Cs\sqrt{\log(p)/n}.
\]
For any fixed $S$ and $j$, let $V=\{i,j\}\cup S$ such that $i$ and $j$ are the first two indices listed in $S$; note that $|V|\leq 4c_0^{-6}+2\leq 5c_0^{-6}$. Write $\hat{A}=(\hat{\Sigma}^{V,V})^{-1}$ and $A=(\Sigma^{V,V})^{-1}$. By definition,
\[
\rho_{ij}(S)= \frac{A(1,2)}{\sqrt{A(1,1)A(2,2)}}, \qquad \hat{\rho}_{ij}(S)= \frac{\hat{A}(1,2)}{\sqrt{\hat{A}(1,1)\hat{A}(2,2)}}
\]
Over the event $B_3$,
\[
\|\hat{A}-A\|_{\max}\leq \|(\hat{\Sigma}^{V,V})^{-1}-(\Sigma^{V,V})^{-1}\|\leq C s\sqrt{\log(p)/n}.
\]
Moreover, $|A(1,2)|\leq \lambda_{\max}(A)\leq c_0^{-1}$, since $|V|\ll N(p,n)$ and $\Sigma\in \mathcal{M}_p^*(s,c_0)$; similarly, $A(1,1)\geq c_0$ and $A(2,2)\geq c_0$. It follows that over the event $B_3$,
\[
|\hat{\rho}_{ij}(S)-\rho_{ij}(S)|\lesssim c_0^{-2}\cdot Cs\sqrt{\log(p)/n},
\]
for all $(S,j)$ such that $|S|\leq 4c_0^{-6}s^2$ and $j\notin(\{i\}\cup S)$.
By taking $c_1\geq 2c_0^{-2}C$, we prove \eqref{LemA.5-pf1}.
%%%%%%%%%%%%%%%%%%%%%
%%%%%%%%%%%%%%%%%%%%%
\subsection{Proof of Theorem~\ref{thm:estimation}}

Recall that $\hat{\Omega}^*$ is the estimator given by PCS without symmetrization. Denote by $\hat{\omega}_i'$ and $\omega_i'$ the $i$-th row of $\hat{\Omega}^*$ and $\Omega$, respectively. It suffices to show that for each $1\leq i\leq p$, with probability at least $1-o(p^{-3})$,
\beq \label{thm2-goal1}
\mbox{$\hat{\omega}_i$ and $\omega_i$ have the same support};
\eeq
and
\beq \label{thm2-goal2}
\|\hat{\omega}_i-\omega_i\|_\infty\leq C\sqrt{\log(p)/n},
\eeq
where $\|\cdot\|_\infty$ denotes the entry-wise max norm for vectors.

Consider \eqref{thm2-goal1}. Write for short $S_0=S^{(i)}(\Omega)$, $\hat{S}^{(i)}_*=\hs_*$ which is the set of recruited nodes in the Screen step of PCS, and $\hat{S}^{(i)}_{**}=\hs_{**}$ which is the support of $\omega_i$.
Let $j_1,\cdots,j_M$ be the nodes in $\hat{S}_*$ and define $W=\{i,j_1,\cdots,j_M\}$ where $i$ is the first index. Denote by $\hat{\eta}'$ the first row of $(\hat{\Sigma}^{W,W})^{-1}$. By definition,
\beq \label{thm2-temp4}
\hs_{**}=\{j_k: |\hat{\eta}(k+1)|>t_q^*\}.
\eeq
By Theorem~\ref{thm:main}, with probability at least $1-o(p^{-3})$,
\[
|\hat{S}_*|\leq Cs^2, \qquad \mbox{and}\qquad S_0\subset\hat{S}_*.
\]
Since $(\{i\}\cup S_0)\subset W$, Lemma~\ref{lemma:firstcolumn} implies that the first row of $(\Sigma^{W,W})^{-1}$ is equal to $(\omega^W)'$. As a result,
\beq \label{thm2-temp1}
\|\hat{\eta} - \omega^W\|_\infty \leq \|(\hat{\Sigma}^{W,W})^{-1}-(\Sigma^{W,W})^{-1}  \|_{\max}\leq  \|(\hat{\Sigma}^{W,W})^{-1}-(\Sigma^{W,W})^{-1}  \|.
\eeq
Note that $|W|\leq Cs^2+1\leq N(p,q)$ and $\Sigma\in \mathcal{M}_p^*(s,c_0)$. It follows from Lemma~\ref{lemma:stoch} that with probability at least $1-o(p^{-3})$,
\beq \label{thm2-temp2}
\| (\hat{\Sigma}^{W,W})^{-1}-(\Sigma^{W,W})^{-1}  \|\leq c_1s\sqrt{2\log(p)/n},
\eeq
for a properly large constant $c_1$ that only depends on $c_0$. Without loss of generality, the constant $c_1$ here is assumed to be the same as that in Lemma~\ref{lem:noiselevel}; if not, we take the maximum of them.
Combining \eqref{thm2-temp1}-\eqref{thm2-temp2}, we find that
\begin{align*}
|\hat{\eta}(k+1)|\left\lbrace
\begin{array}{lcl}
\geq \tau_p^* - c_1s\sqrt{2\log(p)/n}, && j_k\in \{i\}\cup S_0,\cr
\leq c_1 s\sqrt{2\log(p)/n}, &&j_k\in \hat{S}_*\setminus S_0.
\end{array}\right.
\end{align*}
%Take $c_1$ to be the maximum of $C$ here and the constant in Lemma~\ref{lem:noiselevel}.
Then, by the choice of $q$ and that $\tau_p^*\gg s\sqrt{\log(p)/n}$, we have $|\hat{\eta}(k+1)|\gg t_q^*$ for $j_k\in (\{i\}\cup S_0)$ and $|\hat{\eta}(k+1)|\leq t_q^*/2$ for $j_k\in (S_0\setminus \hat{S}_*)$. Plugging these into \eqref{thm2-temp4} gives that with probability at least $1-o(p^{-3})$,
\beq \label{thm2-temp3}
\hat{S}_{**}= S_0.
\eeq
Then, \eqref{thm2-goal1} follows directly.

Second, consider \eqref{thm2-goal2}.
For each $j\in S_0$, introduce the $2 \times 2$ matrices
\begin{align*}
\hat{A} & = \mbox{the ($\{1,k\},\{1,k\}$)-block of $(\hat{\Sigma}^{\{i\}\cup S_0,\{i\}\cup S_0})^{-1}$},\cr
A &=  \mbox{the ($\{1,k\},\{1,k\}$)-block of $(\Sigma^{\{i\}\cup S_0,\{i\}\cup S_0})^{-1}$},
\end{align*}
where we assume $i$ is the first index listed in $\{i\}\cup S_0$ and $j$ is the $k$-th index listed in $\{i\}\cup S_0$.
By Lemma~\ref{lemma:firstcolumn}, $A(1,1) = \omega_i(i)$, $A(1,2) = A(2,1) = \omega_i(j)$, and $\max\{A(2,2), 1/A(2,2)\} \leq C$. By definition and \eqref{thm2-temp3}, with probability at least $1-o(p^{-3})$,
\beq \label{thm2-temp5}
|\hat{\omega}_i(j)- \omega_i(j)| = |\hat{A}(1,2) - A(1,2)|.
\eeq
We apply \cite[Theorem 3.4.6]{Mardia} to the matrix $\hat{\Sigma}^{\{i\}\cup S_0,\{i\}\cup S_0}$ and obtain
\[
\hat{A}^{-1}\sim \mbox{Wishart distribution } W_2(A^{-1}, n - s_0 + 1).
\]
By elementary Random Matrix Theory, with probability at least $1 - o(p^{-3})$,
\beq  \label{thm2-temp6}
|\hat{A}(1,2) - A(1,2)|\leq C\|\hat{A}-A\|_{\max}\leq C\sqrt{\log(p)/n}.
\eeq
Combining \eqref{thm2-temp5}-\eqref{thm2-temp6} gives \eqref{thm2-goal2}.
\qed

%%%%%%%%%%%%%%%%%%%%%
%%%%%%%%%%%%%%%%%%%%%
\subsection{Proof of Theorem~\ref{thm:greedy}}

Fix $1\leq i\leq p$ and write $S_0=S^{(i)}(\Omega)$, $s_0 = |S_0|$, and $\omega = \omega_i$ for short. Note that $s_0 \leq s$ for $s$ in ${\cal M}_p^*(s, c_0)$. The key to our proofs is to relate the Screen step of PCS to a linear regression model.
As before, write $X = [x_1, x_2, \ldots, x_p]$ so that $x_j$ is the $j$-th column of $X$.
As in \eqref{regression}, we can formulate a linear regression model as follows:
\beq \label{lemweak-ls}
x_i = - \sum_{j\in S_0} \frac{\omega(j)}{\omega(i)} x_j + z, \qquad z\sim N(0, \frac{1}{\omega(i)}I_n),
\eeq
where $z$ is independent of $\{x_j:j\in S_0\}$.
For any $S\subset\{1,\cdots,p\}$, let $X_S$ be the
$n \times |S|$ sub-matrix of $X$ such that the $k$-th column of $X_S$ is the $j_k$-th column of $X$, $1 \leq k \leq |S|$.
Suppose $i \notin S$. It is known that the least-squares solution of regressing $x_i$ to $\{x_j: j\in S\}$ is the $s$-dimensional vector
\[
(X_S' X_S)^{-1} X_S' x_i.
\]
We expand this vector to a $p$-dimensional vector $\heta(S)$
by filling $0$ in all coordinates not in $S$:
\begin{equation} \label{DefinehetaS}
\heta(S)(m) =
\left\{
\begin{array}{ll}
\mbox{$k$-th entry of $(X_S' X_S)^{-1} X_S' x_i$}, & \mbox{if $m = j_k$ for some $1 \leq k \leq s$}, \\
0, &  \mbox{if $m \notin S$}.
\end{array}
\right.
\end{equation}
Let $H_S$ be the projection matrix from $R^n$ to $\{x_j: j \in S\}$.
By basic algebra,
\[
x_j'(x_i-X\hat{\eta}(S))=x_j'(I-H_S)x_i.
\]
Combining this with Lemma~\ref{lemma:objective} gives
\beq \label{lemweak-connectLS}
\hat{\rho}_{ij}(S) = \frac{x_j'(x_i-X\hat{\eta}(S))}{\sqrt{x_i'(I-H_S)x_i}\sqrt{x_j'(I-H_S)x_j}}.
\eeq

For preparations, we need two lemmas. The following lemma is the direct result of \cite{ZhangGreedy}.
%%%%%%%%%%%%%%%
%%%%%%%%%%%%%%%
%%%%%%%%%%%%%%%
%%%%%%%%%%%%%%%
%%%%%%%%%%%%%%%
\begin{lemma}  \label{lem:Zhang-LS}
Suppose $y=X\bar{\beta}+N(0,\sigma^2I_n)$ and $S_0 =\mathrm{supp}(\bar{\beta})$, where $X\in R^{n,p}$ and $y\in R^n$.
Let $\hat{\beta}^*=\hat{\beta}^*(S_0)$ be the least-squares solution associated with $S_0$ (defined as a $p$-dimensional vector in a similar fashion to that in (\ref{DefinehetaS})), and define $\theta_{S_0}(X)=\lambda_{\min}(X_{S_0}'X_{S_0})/[\max_{j\notin S_0}\|x_j\|^2]$ and
$\kappa_{S_0}(X)=\max_{j\notin S_0}\|(X_{S_0}'X_{S_0})^{-1}X_{S_0}'x_j\|_1$, where $x_j$ is the $j$-th column of $X$.
\begin{itemize}
\item[(a)] For any $\beta$ whose support $S$ is a strict subset of $S_0$,
\[
\max_{j\in (S_0\setminus S) } \left\{\frac{|x_j'(y-X\beta)|}{\|x_j\|}\right\} \geq \sqrt{n}\theta_{S_0}(X) \frac{\| \beta - \hat{\beta}^*\|}{\sqrt{|S_0\setminus S|}},
\]
where with probability at least $1 - O(p^{-4})$, the right hand side is lower bounded by
\[
\theta_{S_0}(X)\cdot |S_0 \setminus S|^{-1/2}  \sqrt{n}\|\bar{\beta}^{S_0\setminus S}\|  - \sqrt{\theta_{S_0}(X)}\max_{j\notin S_0}\|x_j\| \cdot \sigma \sqrt{10\log(p)}.
\]
\item[(b)] For any $\beta$ whose support is a subset of $S_0$ and any $j \notin S_0$,
\[
 |x_j'(y-X\beta)|\leq  |x_j'(y-X\hat{\beta}^*)| + \kappa_{S_0}(X)\max_{j\in S_0}|x_j'(y-X\beta)|.
\]
\item[(c)] With probability at least $1-O(p^{-4})$,
\[
\max_{j\notin S_0} \left\{\frac{|x_j'(y-X\hat{\beta}^*)|}{ \sqrt{x_j (I - H_{S_0}) x_j}}\right\} \leq   \sigma\sqrt{10\log(p)}.
\]
\end{itemize}
\end{lemma}
Here, $(\beta,S)$ can be either non-random or not; see \cite{ZhangGreedy} for details.
In this lemma, (b) is a slight modification of \cite[Lemma 11]{ZhangGreedy}, and (c) follows from elementary statistics.
As for (a), the first part follows from adapting the proof of \cite[Lemma 7]{ZhangGreedy} and applying \cite[Page 566, third equation]{ZhangGreedy}. For the second part, since $\beta$ has all $0$'s for entries in $S_0\setminus S$,
\beq \label{greedypf1}
\|\beta - \hat{\beta}^*\|  \geq \| (\hat{\beta}^*)^{S_0 \setminus S}\| \geq \|\bar{\beta}^{S_0 \setminus S}\| - \|(\hat{\beta}^*  - \bar{\beta})^{S_0 \setminus S}\|.
\eeq
At the same time, note that
\[
\hat{\beta}^* - \bar{\beta}  \sim N(0, \sigma^2 (X_{S_0}' X_{S_0})^{-1}).
\]
and so with probability at least $1-O(p^{-4})$,
\[
\|(\hat{\beta}^* - \bar{\beta})^{S_0 \setminus S} \| \leq |S_0\setminus S|^{1/2}  \cdot \|(\hat{\beta}^* - \bar{\beta})^{S_0 \setminus S} \|_\infty    \leq \big[|S_0\setminus S|\cdot  10\sigma^2 h^2 \log(p)/n\big]^{1/2}.
\]
where $h$ is the square-root of the maximum diagonal of $(X_{S_0}' X_{S_0})^{-1}$ and $h^2\leq n\lambda^{-1}_{\min}(X_{S_0}'X_{S_0})$.
Inserting it into \eqref{greedypf1} gives the claim.

When we apply Lemma \ref{lem:Zhang-LS} to model (\ref{lemweak-ls}) for each $i$, the quantities $\theta_{S_0}(X)$ and $\kappa_{S_0}(X)$ translate into:
\[
\hat{\theta}(i;X,\Sigma)= [\max_{j\notin S_0}\hat{\Sigma}(j,j)]^{-1}\cdot \lambda_{\min}(\hat{\Sigma}^{S_0,S_0})
\]
and
\[
\hat{\kappa}(i;  X, \Sigma) =    \max_{j \notin (\{i\} \cup S_0)} \|(\hsig^{S_0,S_0})^{-1} \hsig^{S_0, \{j \}}\|_1.
\]
Recalling the definition of $\gamma^*(\Sigma)$, $\theta^*(\Sigma)$ and $\kappa^*(\Sigma)$, we have the following lemma.
%%%%%%%%%%%
%%%%%%%%%%%
%%%%%%%%%%%
\begin{lemma}  \label{lem:greedyconstants}
Under conditions of Theorem~\ref{thm:greedy}, for each $1\leq i\leq p$ and $S_0\equiv S^{(i)}(\Omega)$, with probability at least $1 - O(p^{-4})$,
\begin{itemize}
\item [(i)]   $\max_{k\in S_0,j\notin(\{i\}\cup S_0)} \bigl\{ [x_j' (I - H_{S_0}) x_j]^{-1/2}\|x_k\|  \bigr\}  \leq [\gamma^*(\Sigma)]^{-1} + C \sqrt{\log(p) / n}$,
\item [(ii)] $\hat{\kappa}(i; X, \Sigma) \leq  \kappa^*(\Sigma) + C s \sqrt{\log(p)/n}$,
\item[(iii)] $\hat{\theta}(i;X,\Sigma)\geq \theta^*(\Sigma)-C\sqrt{s\log(p)/n}$.
\end{itemize}
\end{lemma}
For this lemma, the proof of (i) is similar to that of \eqref{thm2-temp6}; (iii) is due to Lemma~\ref{lemma:stoch}, where we have that with probability at least $1-O(p^{-4})$,
\[
\|\hat{\Sigma}^{S_0,S_0}-\Sigma^{S_0,S_0}\|\leq C\sqrt{s\log(p)/n}.
\]
As for (ii), we note that for any vector $\xi \in R^s$, $\|\xi\|_1 \leq  \sqrt{s} \|\xi\|$. By elementary Random Matrix Theory, with probability at least $1 - o(p^{-4})$,
\[
\|\hat{\kappa}(i; X, \Sigma) - \kappa(i; \Sigma)\| \leq C \sqrt{s}\cdot \|  \hat{\Sigma}^{\{j\}\cup S_0,\{j\}\cup S_0} - \Sigma^{\{j\}\cup S_0,\{j\}\cup S_0} \|\leq C s\sqrt{\log(p)/n},
\]
and the claim follows.

We now proceed to show Theorem \ref{thm:greedy}. Consider the Screen step of PCS applied to row $i$. Recall that $S_0 = S^{(i)}(\Omega)$ and $s_0 = |S_0|$.  The proof contains two parts.
\begin{itemize}
\item [(A)] In the first part, we show that provided that the algorithm is not stopped, no noise nodes have been recruited and not all signals have been recruited, then the node we select in the next step
must be a signal node (note that this does not rule out the case that the algorithm stops before it recruits all signal nodes).
\item [(B)]
In the second part, we further check that at each stage $m$:
\begin{itemize}
\item [(B1)]   When $m \leq s_0-1$,  there is a node $j$ the partial correlation associated with which falls above the threshold $t_q^*(p,n)$, so PCS would not stop in less than $s_0$ steps.
\item [(B2)]  When $m = s_0$, we can not find a node the partial correlation associated with which falls above the threshold $t_q^*(p,n)$, and PCS terminates immediately.
\end{itemize}
\end{itemize}

Consider (A). Fix $m \geq 1$. Suppose the Screen step of PCS has not yet stopped by stage $(m-1)$, $\hs^{(i)}_{m-1}\subset S_0$ and $S_0\setminus \hs^{(i)}_{m-1}\neq \emptyset$.
For short, write  $\hat{S} = \hat{S}_{m-1}^{(i)}$ and $\tilde{X}$ as the sub-matrix of $X$ formed by removing the
$i$-th column. Applying item (b) of Lemma \ref{lem:Zhang-LS} to the linear regression model \eqref{lemweak-ls},
with $y = x_i$, $X=\tilde{X}$ and $\sigma^2 = 1/\omega(i)$ and $\beta=\hat{\eta}(\hat{S})$,
it follows that for each $j\notin (\{i\}\cup S_0)$,
%%%%%%%%%
%%%%%%%%%
%%%%%%%%%
\[
|x_j' (x_i - X \heta(\hat{S}))|   \leq  |x_j' (x_i -  X\heta(S_0))|   + \hat{\kappa}(i;X,\Sigma) \max_{k\in (S_0\setminus \hs)}   |x_k' (x_i - X\heta(\hat{S}))|,
\]
where we have replaced $\tilde{X}$ by $X$ according to \eqref{DefinehetaS}.
Combining this with (\ref{lemweak-connectLS}), it follows from basic algebra and Lemma~\ref{lem:greedyconstants} that
\begin{align} \label{greedypf4}
|\hrho_{ij}(\hs)| \leq &
[x_i' (I - H_{\hs}) x_i]^{-1/2} \frac{|x_j' (x_i - X \heta(S_0))| }{\sqrt{x_j'(I-H_{\hat{S}})x_j}} \cr
+ \left[\kappa^*(\Sigma)\right. &\left. +Cs\sqrt{\log(p)/n}\right] \max_{k\in (S_0\setminus \hs)}  \left\{|\hat{\rho}_{ik}(\hs)| \sqrt{ \frac{(x_k' (I - H_{\hs}) x_k)}{ (x_j' (I - H_{\hs}) x_j)}} \right\}.
\end{align}
Now, on one hand, note that for any $j \notin  (\{i\}\cup S_0)$ and $k  \in (S_0 \setminus \hs)$, using Lemma~\ref{lem:greedyconstants} and the assumption that $\hs \subset S_0$ gives that with probability at least $1 - O(p^{-4})$,
\begin{equation} \label{greedypf51}
\frac{x_k' (I - H_{\hs}) x_k}{x_j' (I - H_{\hs}) x_j} \leq \frac{\|x_k\|^2}{x_j' (I - H_{S_0})  x_j} \leq [\gamma^*(\Sigma)]^{-2} + C \sqrt{\log(p)/n}.
\end{equation}
On the other hand, applying item (c) of Lemma~\ref{lem:Zhang-LS} and noting that $\hs\subset S_0$, with probability at least $1 - O(p^{-4})$,
\begin{equation} \label{greedypf52}
\frac{|x_j' (x_i - X\heta(S_0))| }{\sqrt{x_j'(I-H_{\hat{S}})x_j}} \leq \frac{|x_j' (x_i - X\heta(S_0))| }{\sqrt{x_j'(I-H_{S_0})x_j}}\leq \sqrt{10 \omega(i)^{-1}  \log(p)}.
\end{equation}
Inserting (\ref{greedypf51})-(\ref{greedypf52}) into (\ref{greedypf4}) gives \begin{align}  \label{greedypf3}
\max_{j\notin (\{i\}\cup S_0)}  |\hrho_{ij}(\hs)|  & \leq  [x_i' (I - H_{\hs}) x_i]^{-1/2}    \sqrt{10 \omega(i)^{-1}  \log(p)}     \nonumber  \\
  + &  \left[ \kappa^*(\Sigma)/\gamma^*(\Sigma)  + o(1)\right]  \max_{j\in (S_0\setminus \hs)}  |\hat{\rho}_{ij}(\hs)|.
\end{align}

At the same time,  applying item (a) of Lemma \ref{lem:Zhang-LS} to the linear regression model \eqref{lemweak-ls}, there exists $j\in (S_0\setminus \hat{S})$ such that
\begin{align*}
\|x_j\|^{-1} |x_j' & (x_i-\tilde{X}\hat{\eta}(\hs))| \geq  \hat{\theta}(i;X,\Sigma)\cdot |S_0\setminus\hs|^{-1/2} \sqrt{n}\|\eta^{S_0\setminus\hs}\| \cr
 & -  \sqrt{\hat{\theta}(i;X,\Sigma)} \cdot \max_{k\notin S_0}\|x_k\| \sqrt{10\omega(i)^{-1} \log(p)/n},
\end{align*}
where $\eta\equiv \omega(i)^{-1}\omega$, recalling that $\omega'$ is the $i$-th row of $\Omega$. Combining it with with \eqref{lemweak-connectLS}, using  $x_j'(I-H_{\hs})x_j\leq \|x_j\|^2$ and that
\[
\|\eta^{S_0\setminus \hs}\|\geq  |S_0\setminus\hs|^{1/2} \omega(i)^{-1} \tau_p^*,
\]
and applying Lemma~\ref{lem:greedyconstants}, it follows that with probability at least $1-O(p^{-4})$,
\begin{align} \label{greedypf21}
\max_{j\in (S_0\setminus\hs)}& |\hrho_{ij}(\hs)| \geq [x_i'(I-H_{\hat{S}})x_i]^{-1/2}\left\{ [\theta^*(\Sigma)+o(1)] \omega(i)^{-1} n^{1/2}\tau_p^*\right.\cr
&\left. -  [\sqrt{\theta^*(\Sigma)}+o(1)]\sqrt{10\omega(i)^{-1}\log(p)} \right\}.
\end{align}
Recall that $\tau_p^*\geq r\sqrt{2\log(p)/n}$ and by the choice of $r$,
\[
\theta^*(\Sigma)\omega(i)^{-1}r  - \sqrt{\theta^*(\Sigma)}\sqrt{5\omega(i)^{-1}} \geq 2\delta^{-1}\sqrt{5\omega(i)^{-1}}.
\]
Therefore,
\beq \label{greedypf2}
\max_{j\in (S_0\setminus\hs)} |\hrho_{ij}(\hs)| \geq [ 2\delta^{-1}+o(1)]\cdot  [x_i'(I-H_{\hat{S}})x_i]^{-1/2} \sqrt{10\omega(i)^{-1}\log(p)}.
\eeq

Write for short $x=\max_{j\notin (\{i\}\cup S_0)}  |\hrho_{ij}(\hs)|$, $y=\max_{j \in (S_0\setminus\hs)}  |\hrho_{ij}(\hs)|$ and $x_0=[x_i'(I-H_{\hat{S}})x_i]^{-1/2}\sqrt{10\omega(i)^{-1}\log(p)}$. We now combine \eqref{greedypf3} and \eqref{greedypf2}.
Noting that $\kappa^*(\Sigma)/\gamma^*(\Sigma)\leq 1-\delta$, it follows that for sufficiently large $p$,
\[
x \leq  x_0 + (1-\delta) y,  \qquad \mbox{and}  \qquad   y > \delta^{-1}  x_0.
\]
Then we must have $x > y$; otherwise, $y \leq x < x_0 + (1-\delta) y$, and a contradiction follows. Now, we have shown that with with probability at least $1 - O(p^{-4})$,
\[
\max_{j\notin (\{i\}\cup S_0)}  |\hrho_{ij}(\hs)|   <   \max_{j\in (S_0\setminus \hs)}  |\hat{\rho}_{ij}(\hs)|.
\]
This implies that if the algorithm has not stopped and we have not recruited all signal nodes, then we can always find a signal node whose associated partial correlation is larger than the partial correlations associated with all noise nodes.
Since PCS is a greedy algorithm, it must select a signal node in the next step.
This proves the first part.

Next, consider  (B1).  When $m \leq s_0 -1$, by (\ref{greedypf21}), there is at least a signal node $j \in (S_0 \setminus\hs)$ such that
\[
|\hat{\rho}_{ij}(\hs)| \gtrsim \theta^*(\Sigma)\omega(i)^{-1}r \sqrt{2\log(p)/n},
\]
where we have used $(1/n)x_i'(I-H_{\hat{S}})x_i\leq \hat{\Sigma}(i,i)\leq 1+C\sqrt{\log(p)/n}$ with probability at least $1-O(p^{-4})$. By the choice of $q$, the right hand side $> t_q^*(p, n)$ by definition. Together with (A), this implies that the algorithm will not stop but recruit a signal node in the next step.

Last, consider (B2). By the above arguments, we must have $\hs = S_0$ when $m = s_0$.  Using item (c) of Lemma \ref{lem:Zhang-LS}, Lemma \ref{lem:greedyconstants} and (\ref{lemweak-connectLS}), with probability at least $1 - O(p^{-4})$,
\[
\max_{j \notin (\{i\} \cup S_0)} |\hrho_{ij}(S_0)| \leq [x_i'(I-H_{S_0})x_i]^{-1/2}  \sqrt{10 \omega(i)^{-1} \log(p)} \lesssim \sqrt{10 \log(p)/n},
\]
where the last inequality follows from the observation that $n[x_i'(I-H_{S_0})x_i]^{-1}$ is the first diagonal of $(\hat{\Sigma}^{\{i\}\cup S_0, \{i\}\cup S_0})^{-1}$, which is no large than $\omega(i)+o(1)$ with an overwhelming probability, by Lemma~\ref{lemma:firstcolumn} and basics in Random Matrix Theory.
By the way $t_q^*(p,n)$ is chosen, all these coefficients fall below $t_q^*(p,n)$ and the algorithm stops immediately. \qed

%%%%%%%%%%
%%%%%%%%%%

\bibliographystyle{asa}
\bibliography{Jiashun}

\begin{thebibliography}{41}
\newcommand{\enquote}[1]{``#1''}
\expandafter\ifx\csname natexlab\endcsname\relax\def\natexlab#1{#1}\fi

\bibitem[{An et~al.(2014)An, Huang, Yao, and Zhang}]{ZhangYao}
An, H., Huang, D., Yao, Q., and Zhang, C.-H. (2014), \enquote{Stepwise
  searching for feature variables in high-dimensional linear regression,}
  \textit{Manuscript}.

\bibitem[{Anonymous(2006 (retrieved))}]{Elephant}
Anonymous (2006 (retrieved)), \enquote{Elephant and the blind men,}
  \textit{Jain Stories}.

\bibitem[{Arias-Castro et~al.(2011)Arias-Castro, Candes, and Plan}]{Plan}
Arias-Castro, E., Candes, E., and Plan, Y. (2011), \enquote{Global testing
  under sparse alternatives: ANOVA, multiple comparisons and the higher
  criticism,} \textit{Ann. Statist.}, 39, 2533--2556.

\bibitem[{Bi et~al.(2003)Bi, Bennett, Embrechts, Breneman, and Song}]{Bi2003}
Bi, J., Bennett, K., Embrechts, M., Breneman, C., and Song, M. (2003),
  \enquote{Dimensionality reduction via sparse support vector machines,}
  \textit{J. Mach. Learn. Res.}, 3, 1229--1243.

\bibitem[{Bickel and Levina(2008)}]{BLT}
Bickel, P.~J. and Levina, E. (2008), \enquote{Regularized estimation of large
  covariance matrices,} \textit{Ann. Statist.}, 36, 199--227.

\bibitem[{Breiman(2001)}]{RF}
Breiman, L. (2001), \enquote{Random forests,} \textit{Mach. Learn.}, 24, 5--32.

\bibitem[{Buhlmann and van~de Geer(2011)}]{BuhlmannBook}
Buhlmann, P. and van~de Geer, S. (2011), \textit{Statistics for
  High-Dimensional Data: Methods, Theory and Applications}, Springer.

\bibitem[{Burges(1998)}]{SVM}
Burges, C. (1998), \enquote{A tutorial on support vector machines for pattern
  recognition,} \textit{Data Min. Knowl. Discov.}, 2, 121--167.

\bibitem[{Cai et~al.(2011)Cai, Liu, and Luo}]{CLIME}
Cai, T., Liu, W., and Luo, X. (2011), \enquote{A constrained $l_{1}$
  minimization approach to sparse precision matrix estimation,} \textit{J.
  Amer. Statist. Assoc.}, 106, 594--607.

\bibitem[{Cawley and Talbot(2010)}]{Cawley2010}
Cawley, G.~C. and Talbot, N.~L. (2010), \enquote{On over-fitting in model
  selection and subsequent selection bias in performance evaluation,}
  \textit{J. Mach. Learn. Res.}, 11, 2079--2107.

\bibitem[{Dettling and Buhlmann(2003)}]{Buhlmann}
Dettling, M. and Buhlmann, P. (2003), \enquote{Boosting for tumor
  classification with gene expression data,} \textit{Bioinformatics}, 19,
  1061--1069.

\bibitem[{Donoho and Jin(2008)}]{DJ08}
Donoho, D. and Jin, J. (2008), \enquote{Higher Criticism Thresholding: Optimal
  feature selection when useful features are rare and weak,} \textit{Proc.
  Natl. Acad. Sci.}, 105, 14790--14795.

\bibitem[{Donoho and Jin(2014)}]{DJ14}
--- (2014), \enquote{Higher Criticism for large-scale inference: especially for
  rare and weak effects,} \textit{Manuscript}.

\bibitem[{Donoho et~al.(2012)Donoho, Tsaig, Drori, and Starck}]{StOMP}
Donoho, D., Tsaig, Y., Drori, I., and Starck, J.-L. (2012), \enquote{Sparse
  solution of undetermined systems of linear equations by stagewise orthogonal
  matching pursuit,} \textit{IEEE Trans. Inform. Theory}, 58, 1094--1121.

\bibitem[{Edelman(1998)}]{Edelman}
Edelman, A. (1998), \enquote{Eigenvalues and condition number of random
  matrices,} \textit{SIAM J. Matrix Anal. Appl.}, 9, 543--560.

\bibitem[{Efron(2004)}]{EmpNull}
Efron, B. (2004), \enquote{Large-scale simultaneous hypothesis testing: the
  choice of a null hypothesis,} \textit{J. Amer. Statist. Assoc.}, 99, 96--104.

\bibitem[{Fan and Fan(2008)}]{FAIR}
Fan, J. and Fan, Y. (2008), \enquote{High-dimensional classification using
  features annealed independent rules,} \textit{Ann. Statist.}, 36, 2605--2637.

\bibitem[{Fan et~al.(2012)Fan, Feng, and Tong}]{ROAD}
Fan, J., Feng, Y., and Tong, X. (2012), \enquote{A road to classification in
  high dimension space: the regularized optimal affine discriminant,}
  \textit{J. Roy. Statist. Soc.}, 74, 745--771.

\bibitem[{Fan and Lv(2008)}]{FanLv}
Fan, J. and Lv, J. (2008), \enquote{Sure independence screening for ultrahigh
  dimensional feature space,} \textit{J. Roy. Statist. Soc. Ser. B}, 70,
  849--911.

\bibitem[{Fan et~al.(2013)Fan, Jin, and Yao}]{FJY}
Fan, Y., Jin, J., and Yao, Z. (2013), \enquote{Optimal classification in sparse
  Gaussian graphic model,} \textit{Ann. Statist.}, 41, 2537--2571.

\bibitem[{Friedman et~al.(2007)Friedman, Hastie, and Tibshirani}]{glasso}
Friedman, J., Hastie, T., and Tibshirani, R. (2007), \enquote{Sparse inverse
  covariance estimation with the graphical lasso,} \textit{Biostatistics}, 9,
  432--441.

\bibitem[{Hall and Jin(2010)}]{HJ09}
Hall, P. and Jin, J. (2010), \enquote{Innovated higher criticism for detecting
  sparse signals in correlated noise,} \textit{Ann. Statist.}, 38, 1686--1732.

\bibitem[{Horn and Johnson(1990)}]{Horn}
Horn, R.~A. and Johnson, C.~R. (1990), \textit{Matrix Analysis}, Cambridge
  University Press.

\bibitem[{Ingster et~al.(2009)Ingster, Pouet, and Tsybakov}]{Ingster}
Ingster, Y., Pouet, C., and Tsybakov, A. (2009), \enquote{Classification of
  sparse high-dimensional vectors,} \textit{Phil. Trans. R. Soc. A}, 367,
  4427--4448.

\bibitem[{Jager and Wellner(2007)}]{Wellner}
Jager, L. and Wellner, J. (2007), \enquote{Goodness-of-fit tests via
  phi-divergences,} \textit{Ann. Statist.}, 35, 2018--2053.

\bibitem[{Jin and Ke(2014)}]{JK}
Jin, J. and Ke, Z. (2014), \enquote{Rare and weak effects in large-scale
  inference: methods and phase diagrams,} \textit{Manuscript}.

\bibitem[{Jin et~al.(2014)Jin, Zhang, and Zhang}]{JZZ}
Jin, J., Zhang, C.-H., and Zhang, Q. (2014), \enquote{Optimality of Graphlet
  Screening in high dimensional variable selection,} \textit{J. Mach. Learn.
  Res. To appear}.

\bibitem[{Ke et~al.(2014)Ke, Jin, and Fan}]{KJ}
Ke, Z., Jin, J., and Fan, J. (2014), \enquote{Covariance Assisted Screening and
  Estimation,} \textit{Ann. Statist. To appear}.

\bibitem[{Li and Zhu(2012)}]{Runze}
Li, R. and Zhu, L. (2012), \enquote{Feature screening via distance correlation
  learning,} \textit{J. Amer. Statist. Assoc.}, 107, 1129--1139.

\bibitem[{Mardia et~al.(2003)Mardia, Kent, and M.}]{Mardia}
Mardia, K.~V., Kent, J.~T., and M., B.~J. (2003), \textit{Multivariate
  Analysis}, New York: Academic Press.

\bibitem[{Mazumder and Hastie(2012)}]{glassocomplexity}
Mazumder, R. and Hastie, T. (2012), \enquote{Exact covariance thresholding into
  connected components for large-scale graphical lasso.} \textit{J. Mach.
  Learn. Res.}, 13, 723--736.

\bibitem[{Ravikumar et~al.(2011)Ravikumar, Wainwright, Raskutti, and
  Yu}]{BinYu}
Ravikumar, P., Wainwright, M.~J., Raskutti, G., and Yu, B. (2011),
  \enquote{High-dimensional covariance estimation by minimizing
  $l_{1}$-penalized log-determinant divergence,} \textit{Electron. J.
  Statist.}, 5, 935--980.

\bibitem[{Seber and Lee(2003)}]{Seber}
Seber, G. and Lee, A. (2003), \textit{Linear Regression Analysis}, New Jersey:
  Wiley.

\bibitem[{Spiegelhalter(2014)}]{Science}
Spiegelhalter, D. (2014), \enquote{The future lies in uncertainty,}
  \textit{Science}, 345, 264.

\bibitem[{Sun and Zhang(2012)}]{slasso}
Sun, T. and Zhang, C.-H. (2012), \enquote{Scaled sparse linear regression,}
  \textit{Biometrika}, 99, 879--898.

\bibitem[{Vershynin(2010)}]{Vershynin}
Vershynin, R. (2010), \enquote{Introduction to the non-asymptotic analysis of
  random matrices,} \textit{arXiv:1011.3027}.

\bibitem[{Wasserman and Roeder(2009)}]{Wasserman}
Wasserman, L. and Roeder, K. (2009), \enquote{High dimensional variable
  selection,} \textit{Ann. Statist.}, 37, 2178--2201.

\bibitem[{Yousefi et~al.(2010)Yousefi, Hua, Sima, and Dougherty}]{liver}
Yousefi, M., Hua, J., Sima, C., and Dougherty, E. (2010), \enquote{Reporting
  bias when using real data sets to analyze classification performance,}
  \textit{Bioinformatics}, 26, 68--76.

\bibitem[{Zhang(2009)}]{ZhangGreedy}
Zhang, T. (2009), \enquote{On the consistency of feature selection using greedy
  least squares regression,} \textit{J. Mach. Learn. Res.}, 10, 555--568.

\bibitem[{Zhang(2011)}]{ZhangFoBa}
--- (2011), \enquote{Adaptive Forward-Backward greedy algorithm for learning
  sparse representations,} \textit{IEEE Trans. Inf. Theory}, 57, 4689--4708.

\bibitem[{Zhong et~al.(2013)Zhong, Chen, and Xu}]{Chen}
Zhong, P., Chen, S.-X., and Xu, M. (2013), \enquote{Test alternative to higher
  criticism for high dimensional means under sparsity and column wise
  dependence,} \textit{Ann. Statist.}, 41, 2820--2851.

\end{thebibliography}

\end{document}